\documentclass[12pt]{article}
\usepackage[top=1in, bottom=1in, left=1in, right=1in]{geometry}
\usepackage{amsfonts}
\usepackage{amsmath}
\usepackage[ngerman,british]{babel}
\usepackage{natbib}
\setcitestyle{aysep={}} 
\usepackage{url}
\usepackage{amssymb}
\usepackage{pdflscape}
\usepackage{placeins} 
\usepackage{subfigure}

\usepackage{array}
\usepackage[flushleft]{threeparttable}
\usepackage{rotating,graphicx}
\usepackage{setspace}
\usepackage{color}
\usepackage{mathtools}
\usepackage[dvipsnames]{xcolor}
\definecolor{CardinalRed}{cmyk}{0,1,0.65,0.34} 
\usepackage[colorlinks,linkcolor=CardinalRed,citecolor=CardinalRed,urlcolor = CardinalRed]{hyperref}
\usepackage[labelsep=period, font=sc, skip=0pt,justification=centering]{caption}
\usepackage{pdfpages}
\usepackage{framed} 
\usepackage{titletoc}

\DeclareMathOperator*{\argmin}{arg\,min}

\usepackage{bm}
\usepackage{bbm}

\usepackage{changepage}

\usepackage{accents}

\usepackage{etoolbox}
\newcommand{\zerodisplayskips}{%
  \setlength{\abovedisplayskip}{5pt}%
  \setlength{\belowdisplayskip}{5pt}%
  \setlength{\abovedisplayshortskip}{5pt}%
  \setlength{\belowdisplayshortskip}{5pt}}
\appto{\normalsize}{\zerodisplayskips}
\appto{\small}{\zerodisplayskips}
\appto{\footnotesize}{\zerodisplayskips}  
\newcommand{\indep}{\perp \!\!\! \perp}

\usepackage{titling}
\usepackage[flushmargin]{footmisc}
\usepackage{titlesec}
\titlelabel{\thetitle.\ }
\startcontents[sections]

\usepackage[noend]{algpseudocode}
\usepackage{enumitem}

\let\oldcenter\center
\let\oldendcenter\endcenter
\renewenvironment{center}{\setlength\topsep{0pt}\oldcenter}{\oldendcenter}

\providecommand{\U}[1]{\protect \rule{.1in}{.1in}}

\newcommand{\E}{\mathbb{E}}

\graphicspath{{./graph/}}%

\begin{document}

\onehalfspacing

\title{\Large\bf A Practical Guide to Counterfactual Estimators for Causal Inference with Time-Series Cross-Sectional Data%
\thanks{Licheng Liu, Graduate Student, Department of Political Science, Massachusetts Institute of Technology. Email: \url{liulch@mit.edu}. Ye Wang, Assistant Professor, Department of Political Science, University of North Carolina at Chapel Hill, Chapel Hill, NC 27514. Email: \url{yewang@unc.edu}. Yiqing Xu, corresponding author, Assistant Professor, Department of Political Science, Stanford University. Email: \url{yiqingxu@stanford.edu}. We thank Neal Beck, Bernie Black,  Naoki Egami, Avi Feller, Dan de Kadt, Erin Hartman, Chad Hazlett, Danny Hidalgo, Apoorva Lal, Kosuke Imai, In Song Kim, Jeff Lewis, Marc Ratkovic, and seminar participants at NYU, UCSD, UCLA, and MIT, as well as participants of PolMeth 2019 for helpful comments. We are grateful for the constructive comments by editors of AJPS and three anonymous reviewers, which we believe have helped us improve the paper significantly. We thank Ziyi Liu for superb research assistance.}}

\author{Licheng Liu\and Ye Wang\and Yiqing Xu}

\date{
  \vspace{-0.5em}\hspace{1.2em}(MIT)\hspace{4.8em}(UNC)\hspace{3.5em}{(Stanford)} \\\vspace{1em}
  First version: 11th July 2019\\
  This version: 2nd August 2022\vspace{2em} 
}

\maketitle

\vspace{-3em}
\begin{abstract}
\noindent This paper introduces a simple framework of counterfactual estimation for causal inference with time-series cross-sectional data, in which we estimate the average treatment effect on the treated by directly imputing counterfactual outcomes for treated observations. We discuss several novel estimators under this framework, including the fixed effects counterfactual estimator, interactive fixed effects counterfactual estimator, and matrix completion estimator. They provide more reliable causal estimates than conventional twoway fixed effects models when treatment effects are heterogeneous or unobserved time-varying confounders exist. Moreover, we propose a new dynamic treatment effects plot, along with several diagnostic tests, to help researchers gauge the validity of the identifying assumptions. We illustrate these methods with two political economy examples and develop an open-source package, \texttt{fect}, in both \texttt{R} and \texttt{Stata} to facilitate implementation.

\end{abstract}

\thispagestyle{empty}  
\clearpage

\doublespace
\setcounter{page}{1}

\section{Introduction}

The linear twoway fixed effects (TWFE) model is one of the most commonly used statistical routines in the social sciences to establish causal relationships using observational time-series cross-sectional (TSCS) data, or long panel data. Such models are a popular choice because they can potentially control for a large set of unobserved unit- and time-invariant confounders. However, recent research points to several important drawbacks of FE models \citep{blackwell2018make, ImaiKim2019}. First, the strict exogeneity assumption they rely on is often unrealistic---it not only requires the absence of time-varying confounders, but also rules out the possibility that past outcomes directly affect current treatment assignment (\emph{no feedback}). It is well known that violations of strict exogeneity lead to biases in the causal estimates, yet methods for relaxing it or diagnosing its failure remain limited. 

Second, TWFE models involve rigid functional form assumptions. When the treatments are dichotomous, TWFE models often assume their effects to be constant (\emph{constant treatment effect}) and they affect the contemporaneous outcome only, not future outcomes (\emph{no carryover effects}). Violation of the former will likely result in biased estimates even when strict exogeneity is satisfied, a problem receiving much attention in the literature recently. For example, \citet{de_Chaisemartin2018-iw} show that TWFE estimates are weighted averages of individualistic treatment effects, or treatment effects on each cell under the treatment condition. Because the weights can sometimes be negative due to differential treatment timing and heterogeneous treatment effects, TWFE estimates may not even be convex combinations of the individualistic effects. In a staggered adoption setting where a treatment never switches back once it is on, \citet{Goodman-Bacon2018} shows that negative weights are caused by temporal changes in the treatment effects of early treatment adaptors. Several papers aim to address this issue. For example, \citet{strezhnev2018} and \citet{callaway2020difference} suggest that under staggered adoption, researchers can instead estimate the average treatment effects for units that adopt the treatment at the same time, which they call the cohort average treatment effect; \citet{de_Chaisemartin2018-iw} propose to use observations only one period before or after the treatment’s onset or exit, which leads to an estimator they call DID$_{M}$ ($M$ for ``multiple''). However, these approaches either have limited applicability by requiring a staggered adoption design or are statistically inefficient due to dropping many observations. Researchers so far have paid little attention to the no carryover effects assumption, though it is often testable by data. 

In this paper, we introduce a simple framework that ameliorates these problems. We focus on TSCS data with dichotomous treatments, but they are allowed to switch back and forth (we call it a general panel treatment structure). Estimators under this framework take observations under the treatment condition as missing, use data under the control condition to build models, and impute counterfactuals of treated observations based on the estimated models. We call them \emph{counterfactual estimators}.  This framework has several benefits. First, by not using the treated observations at the modeling stage and by imposing uniform weights on individualistic treatment effects on treated observations, it avoids the aforementioned negative weights problem and corrects biases induced by treatment effect heterogeneity. Second, it accommodates a variety of models, some of which can potentially relax the conventional strict exogeneity assumption. Third, it makes diagnostics and visualization much easier than with traditional TWFE models.

We discuss three methods under this framework, including (1) the fixed effects counterfactual (FEct) estimator, of which difference-in-differences (DID) is a special case; (2) the interactive fixed effects counterfactual (IFEct) estimator; and (3) the matrix completion (MC) estimator. Both IFEct and MC have recently emerged in the literature---see, e.g., \citet{GobillonMagnac2016} and \citet{xu2017generalized} for the former and \citet{kidzinski2018longitudinal} and \citet{athey2018matrix} for the latter. They are designed to construct a lower-rank approximation of the outcome data matrix using information of untreated observations to account for potential time-varying confounders but differ in their ways of regularizing latent factors. Although FEct can be seen as a special case of IFEct, it is uniquely important because it provides a simple solution to the aforementioned weighting problem with the TWFE estimator. In addition to us, \citet{borusyak2020revisiting} and \citet{gardnertwo} have independently proposed it as an improvement over TWFE. They call it the efficient estimator---because it is shown to be the most efficient among a class of linear unbiased estimators for the ATT---and the two-stage DID estimator, respectively.  

Moreover, this paper aims to provide researchers with a set of diagnostic tools when making causal claims using TSCS data. A popular practice among researchers to evaluate the validity of the identifying assumptions is to draw a plot of the so-called ``dynamic treatment effects,'' which are coefficients of a series of interactions between a dummy variable indicating the treatment group---units that are exposed to the treatment for at least one period during the observed time window---and a set of time dummies indicating the time period relative to the onset of the treatment using a TWFE model. If these coefficients exhibit a monotonic trend leading toward the onset of the treatment, or a ``pretrend,'' the assumptions are deemed problematic. However, this method relies on parametric assumptions and the statistical tests derived from it are informal, often underpowered, or even misleading \citep{Roth2020,sun2020estimating}. Taking advantage of the counterfactual estimation framework, we improve the practice of estimating and plotting the dynamic treatment effects, or the average treatment effects on the treated (ATT) over different periods, without assuming treatment effect homogeneity of any kind. 

In addition to visual inspections, we propose a set of statistical tests to help researchers evaluate the validity of the identifying assumptions. The core of these tests is based on a panel \emph{placebo test}, in which we hide a few periods of observations right before the onset of the treatment for the treated units and use a model trained using the rest of the untreated observations to predict the untreated outcomes of those held-out periods. If the identifying assumptions are valid, the average differences between the observed and predicted outcomes in those periods should be close to zero. If, on the contrary, these differences are significantly different from zero, we obtain a piece of evidence that either the functional form assumption or strict exogeneity is likely invalid. 

We then use this basic idea to construct two additional tests, a test for \emph{no pretrend} and a test for \emph{no carryover effects}. With the former, instead of hiding a few periods right before the treatment begins, we use a leave-one-period-out approach to consecutively hide one pretreatment period (relative to the timing of the treatment) and repeatedly conduct a placebo test on observations in that period. By doing so, we have a more holistic view of whether the identifying assumptions will likely hold. The test for no carryover effects, on the other hand, is the mirror opposite of the placebo test in that it hides a few periods right after the treatment ends. If carryover effects do not exist, the average differences between the observed and predicted outcomes in those periods should be close to zero. This test is infeasible for the staggered adoption treatment structure, in which the treatment never switches back. However, under staggered adoption, potential carryover effects may not be concerns for researchers who care about the overall cumulative effects of the treatment over an extended period of time.  

For all three tests, we use both a conventional difference-in-means (DIM) approach, which tests against the null of no difference, and an equivalence approach, which flips the null and tests against a prespecified difference. Consistent with the literature on equivalence tests in cross-sectional settings \citep{hartman2018equivalence,Hartman2020-jb}, we show that the equivalence approach has advantages over the DIM approach when limited power is a concern. This is because as researchers collect more data, under valid identifying assumptions, it should be easier for them to declare equivalence by rejecting null in an equivalence test, not harder. \citet{Bilinski2018-eh}, \citet{Dette2020-hg}, and \citet{Egami2020} propose similar tests recently in a DID setting. We recommend researchers consider the equivalence approach when data are limited. 

This paper makes two main contributions to the literature. First, it introduces a counterfactual estimation framework to TSCS analysis that covers a variety of novel estimators. This new imputation approach addresses the weighting issue of TWFE models that causes concern for many researchers, and the new estimators introduced here can potentially control for decomposable time-varying confounders in a general panel data setting. Our second contribution is to develop a set of visualization and diagnostic tools to assist researchers in gauging the validity of the identifying assumptions and choosing the most suitable model for their applications. 

This paper builds on earlier work on counterfactual estimation (or imputation methods) for causal inference. \citet{heckman1997matching, heckman1998matching} first noted that, to identify the ATT, one only needs to impute counterfactuals for observations in the treatment group. This perspective has motivated a series of studies that try to predict the counterfactual in cross-sectional studies using various methods, such as regression \citep{lin2013agnostic}, the Oaxaca-Blinder estimator \citep{kline2011oaxaca}, and machine learning algorithms \citep{kunzel2019metalearners}. The synthetic control method (SCM) first adopts the counterfactual approach in a panel setting \citep{ADH2010}, but it is limited to comparative case studies, a specialized user case. We introduce it to systematically analyzing panel/TSCS data. 

We also contribute to an emerging literature on causal inference with panel/TSCS data and our approach has advantages over existing methods under various circumstances. Compared with existing factor-augmented methods (e.g., \citealt{GobillonMagnac2016,xu2017generalized}), which also use imputation methods, our framework can accommodate more complex TSCS designs, such as treatment reversal. Compared with TSCS methods based on matching and reweighting (e.g., \citealt{Abadie2005,ImaiKim2019,strezhnev2018,callaway2020difference,de_Chaisemartin2018-iw}), our approach can accommodate more complex data structure and incorporate covariates more conveniently, and is often more efficient. It can also serve as a building block for doubly robust estimators, such as the augmented SCM \citep{BFR2019}. 

This approach, of course, has limitations. First, the strict exogeneity assumption, which corresponds to baseline randomization, may be unrealistic in many applied settings, in which case researchers should consider methods based on sequential ignorability \citep{blackwell2018make, imai2018matching,hazlettxu2018}. See \cite{Xu2022-hw} for a detailed discussion on the two identification regimes. With a general panel treatment structure, our method only allows limited carryover effects. Second, although we provide flexible modeling options, such as IFEct and MC, they are no panacea for all TSCS applications. The factor-augmented approach is more likely to suffer from biases due to model dependency and misspecification. Researchers have recently made efforts to alleviate this concern by proposing doubly robust estimators \citep[e.g.,][]{BFR2019,Arkhangelsky2019-lz}; this paper does not incorporate these innovations because doing so would limit the applicability of our methods (e.g., by not allowing treatment reversal). Last but not least, the equivalence test approach requires users to specify an equivalence range, which may leave room for post hoc justification. Despite these drawbacks, we believe that the counterfactual imputation approach is a promising framework for TSCS analysis and can be extended to support a wide range of models.

\section{Counterfactual Estimators}\label{sc:estimators}

We first introduce the framework and the overall estimation strategy, and then discuss three novel estimators as examples. 

\subsection{A Simple Framework}

\paragraph{Setup.} Though our approach can accommodate both balanced and imbalanced panels, we consider a balanced panel with $N$ units and $T$ periods for notational convenience. Denote $D_{it}$ the treatment status. Denote $Y_{it}(1)$ and $Y_{it}(0)$ the potential outcomes of unit $i$ in period $t$ when $D_{it} = 1$ and $D_{it} = 0$, respectively. Denote $\mathbf{X}_{it}$ a vector of the exogenous covariates, $\mathbf{U}_{it}$ the unobservable attributes, and $\varepsilon_{it}$ the idiosyncratic error term. Without loss of generality, we can define $\delta_{it} = Y_{it}(1) - Y_{it}(0)$ for unit $i$ in period $t$. We assume the following class of outcome models for the untreated potential outcome:

\bigskip\noindent\textbf{Assumption 1} \textit{(Functional form)} $Y_{it}(0) = f(\mathbf{X}_{it}) + h(\mathbf{U}_{it}) + \varepsilon_{it},$ in which $f(\cdot)$ and $h(\cdot)$ are known, parametric functions. 

\medskip Note that Assumption 1 requires additive separability of the four right-hand side terms. This class of models is scale dependent \citep{AtheyImbens2006}, e.g., transforming the outcome from levels to logarithms may render the identification assumptions discussed below invalid. It is easy to see that the classic two-group two-period DID approach assumes a model that is a special case in Assumption 1: 
\begin{center}
$Y_{it}(0) = U_{it} + \varepsilon_{it} = \alpha_i + \xi_t + \varepsilon_{it},\quad t = 1, 2$
\end{center}
in which $\alpha_i$ and $\xi_{t}$ are unit and period fixed effects. Hence, TWFE models' ability to control for unobserved confounders rests on the functional form assumption.

The setup, together with Assumption 1, rules out both temporal and spatial interference \citep{Wang2021-xo}, including potential anticipation effects and carryover effects. \citet{borusyak2020revisiting} show that the presence of anticipation effects will cause underidentification of the causal effects; the same logic applies to carryover effects.  In a staggered adoption design, however, carryover effects are allowed because we can interpret $\delta_{it}$ as a combination of instant effect of the current treatment and cumulative carryover effects of past treatments on a treated unit relative to its potential outcome history under the never-treated condition---see Figure A3 on page 6 in Supporting Information (SI) for a graphic illustration. 

\medskip\noindent\textbf{Estimands.} The primary causal quantity of interest is the average treatment effect on the treated units, whose treatment status has changed at least once during the observed time window, i.e., 
$$ATT = \E[\delta_{it}|D_{it}=1, C_{i} = 1]$$
in which $\delta_{it} = Y_{it}(1) - Y_{it}(0)$ by definition; and $C_{i} = 1$ if $\exists t,t' \text{ s.t. } D_{it} = 0, D_{it'}=1$; otherwise, $C_{i} = 0$. For units that have never been exposed to the treatment condition, it is difficult to compute their treated potential outcomes without strong structural assumptions. Similarly, it is difficult to estimate causal effects on units that are always treated, and we drop them from the sample at the preprocessing stage. 

In empirical work, researchers may be also interested in the average treatment effect on the treated at $s$th ($s>0$) periods since the treatment's onset:
$$ATT_{s} = \E[\delta_{it}| D_{i,t-s}=0, \underbrace{D_{i,t-s+1} = D_{i,t-s+2} = \cdots =  D_{it} = 1}_{s \text{ periods}}, C_i = 1],\quad s>0.$$
For the purpose of the diagnostic tests we will introduce later, we define $ATT_{s} = 0, \forall s\leq0$. An alternative estimand used by \citet{de_Chaisemartin2018-iw} is the average instant treatment effect of changes in the treatment, i.e.,
$$AITC =  \E[\delta_{it}| (D_{i,t-1}=0, D_{i,t} = 1) \text{ or } (D_{i,t}=1, D_{i,t+1} = 0)].$$
It has the benefit of relaxing the no-carryover-effect assumption, but it is less applicable in many empirical applications because the effect of a treatment often takes time to manifest itself. For that reason, it is not the main focus of this paper. Our software package \texttt{fect} provides support for all the above estimands.

\bigskip\noindent\textbf{Assumption 2} ({\it Strict exogeneity})
$\varepsilon_{it} \indep \{D_{js}, \mathbf{X}_{js}, \mathbf{U}_{js}\}$, for all $i,j \in \{1,2,\dots,N\}$ and $s,t \in \{1,2,\dots,T\}$.\bigskip

Together with Assumption 1, Assumption 2 corresponds to baseline quasi-randomization conditional on $\mathbf{X}$ and $\mathbf{U}$, that is, $Y_{is}(0)\indep D_{it}\ |\mathbf{X}_{i,1...T}, \mathbf{U}_{i,1...T}$, $\forall i,s,t$, in which $\mathbf{X}_{i,1...T}$ and $\mathbf{U}_{i,1...T}$ are the time series of $\mathbf{X}_{it}$ and $\mathbf{U}_{it}$, respectively. When $h(\mathbf{U}_{it}) = \alpha_{i} + \xi_{t}$ (as in DID), Assumption 2 implies the parallel trends assumption, i.e., 
$$\E[Y_{it}(0)|\mathbf{X}_{it}]-\E[Y_{is}(0)|\mathbf{X}_{is}] = \E[Y_{jt}(0)|\mathbf{X}_{jt}]-\E[Y_{js}(0)|\mathbf{X}_{js}],\quad \forall i, j, \forall t, s,$$
which states that, by expectation, the untreated potential outcome of all units follow parallel paths. When $\mathbf{U}_{it}$ is of a more general form, Assumption 2 implies 
$$
\E[Y_{it}(0)|\mathbf{X}_{it}, \mathbf{U}_{it}]-\E[Y_{is}(0)|\mathbf{X}_{is}, \mathbf{U}_{is}] = \E[Y_{jt}(0)|\mathbf{X}_{jt}, \mathbf{U}_{jt}]-\E[Y_{js}(0)|\mathbf{X}_{js},\mathbf{U}_{js}],\quad \forall i, j, \forall t, s
$$
which states that conditional on the observed exogenous covariates and unobserved attributes (if we can extract them), the average changes in untreated potential outcome from period $s$ to period $t$ is the same between unit $i$ and unit $j$. This leads to the third assumption.

\bigskip\noindent\textbf{Assumption 3}  ({\it Low-dimensional decomposition}) There exists a low-dimensional decomposition of $h(\mathbf{U}_{it})$: $h(\mathbf{U}_{it}) = L_{it}$, and $rank(\mathbf{L}_{N \times T}) \ll \min\{N, T\}$. For example, $\mathbf{L} = \mathbf{\Lambda}\mathbf{F}$, in which $\mathbf{\Lambda}$ is a $(N\times r)$ matrix of factor loadings and $\mathbf{F}$ is a $(r\times T)$ matrix of factors and $r \ll \min\{N, T\}$.\bigskip

Assumption 3 allows us to condition on $\mathbf{U}_{it}$. To give a concrete example, if $\mathbf{U}_{it} = f_{t} \cdot \lambda_{i}$ is one dimensional, we can understand it as the impact of a common time trend $f_{t}$ having a heterogeneous impact on each unit, whose heterogeneity is captured by $\lambda_{i}$. Moreover, when $f_{t}$ is constant, $\mathbf{U}_{it}$ reduces to a set of unit fixed effects; when $\lambda_{i}$ is constant, it reduces to time fixed effects. Hence, additive fixed effects in DID models obviously satisfy this assumption. When unobserved confounders $\mathbf{U}_{it}$ exist, treatment assignment is dependent on observed untreated outcomes, thus, we are operating under a special case of missing not at random \citep{rubin1976inference}. Assumption 3 allows us to break this dependency by controlling for $\mathbf{U}_{it}$ approximated using data and can be understood as a feasibility assumption.

\begin{figure}[!ht]\caption{A DAG Illustration\label{fg.dag}}
\begin{center}
\begin{minipage}{0.85\linewidth}
\begin{center}\vspace{-0.5em}
\includegraphics[width = 0.9\textwidth]{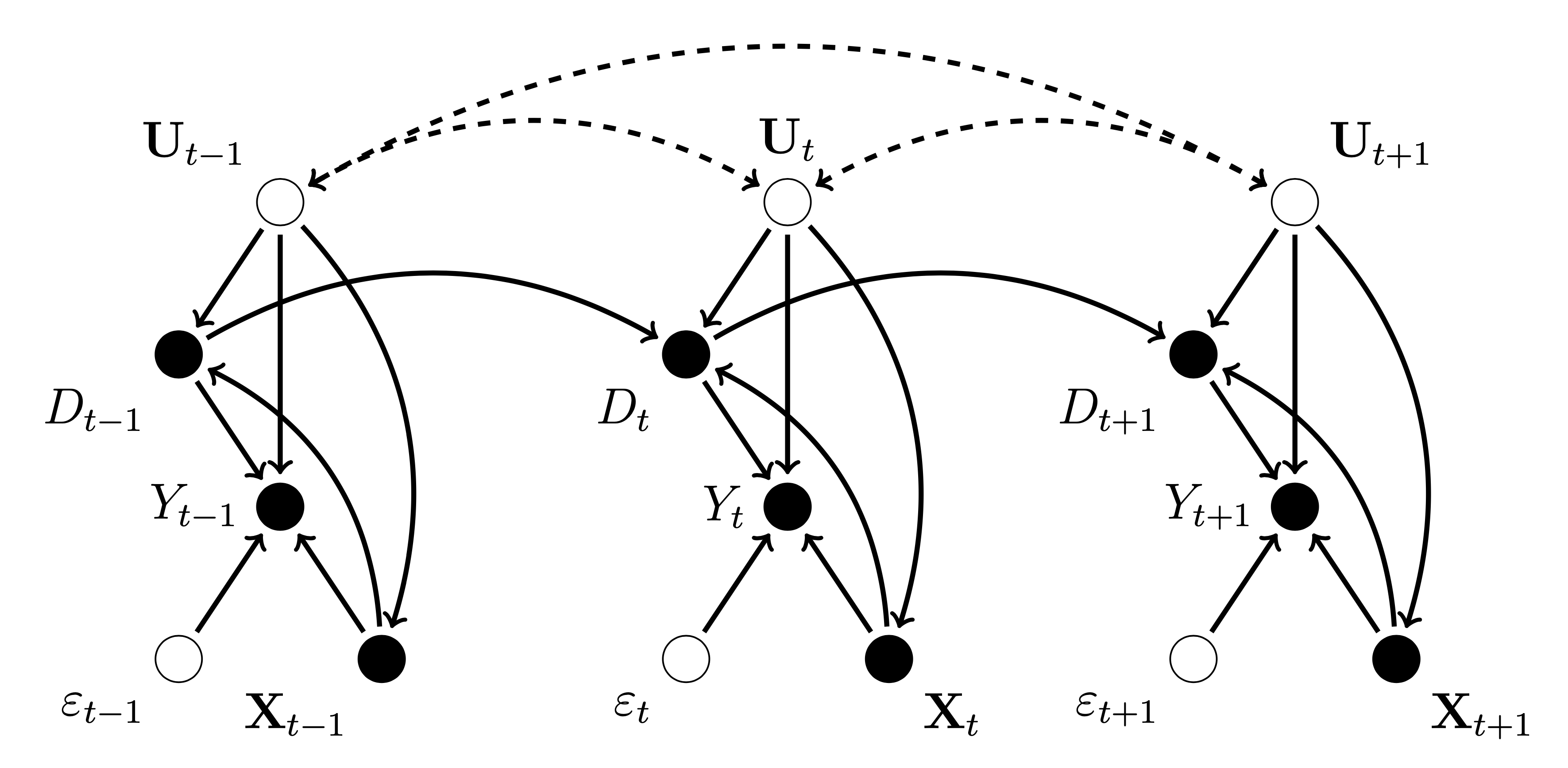}
\end{center}
\footnotesize\textbf{Note:} The above figure presents a DAG consistent with Assumptions 1-3. Unit indices are dropped for simplicity. $Y$, $D$, $\mathbf{X}$, $\varepsilon$ represent the outcome, treatment, covariates, and error term, respectively.
\end{minipage}
\end{center}\vspace{-1em}
\end{figure}

In Figure~\ref{fg.dag}, we illustrate what the identifying assumptions entail using a directed acyclic graph (DAG). It shows that Assumptions~1-2  rule out anticipation effects (e.g., no arrows from $D_{t}$ to $Y_{t-1}$ or $Y_{t+1}$), feedback (e.g., no arrow from $Y_{t-1}$ to $D_{t}$), and lagged dependent variables (no arrow from $Y_{t-1}$ to $Y_{t}$); it also shows that the treatment effects of $D_{it}$ on $Y_{it}$ are separable from the influences of $\mathbf{U}_{it}$ and $\mathbf{X}_{it}$. This setup nests many existing models for TSCS data analyses, including TWFE and IFE models, although these models usually assume constant treatment effect, i.e, $\delta_{it} = \delta$. If these assumptions are unsatisfied, research may turn to methods under sequential ignorability. See more discussion in \citet{blackwell2018make} and \citet{ImaiKim2019} on the potential tradeoffs.

\paragraph{Estimation strategy.} We define the observations under control and treatment conditions as $\mathcal{O} = \{(i,t)| D_{it} = 0\}$ and $\mathcal{M} = \{(i,t)| i\in\mathcal{T}, D_{it} = 1\}$, respectively, in which $\mathcal{O}$ stands for ``observed'' and $\mathcal{M}$ stands for  ``missing.'' Although the outcome model researchers choose to employ may vary, estimation proceeds in a similar fashion with the following steps:

\begin{adjustwidth}{0pt}{0pt}

\noindent\textbf{Step 1.} On the subset of untreated observations ($\mathcal{O}$), fit a model of the response surface $Y_{it}$, obtaining $\hat f$ and $\hat h$. This step relies on the functional form assumptions on $f(\mathbf{X}_{it})$ and $h(\mathbf{U_{it}})$, as well as a lower-rank representation of $\mathbf{U}$. 

\noindent\textbf{Step 2.} Predict the counterfactual outcome $Y_{it}(0)$ for each treated observation using $\hat f$, $\hat h(\mathbf{U})$, i.e., $\hat Y_{it}(0) = \hat f(\mathbf{X}_{it}) + \hat h(\mathbf{U}_{it})$, for all $(i,t)\in\mathcal{M}$.

\noindent\textbf{Step 3.} Estimate the individualistic treatment effects $\delta_{it}$ using $\hat \delta_{it} = Y_{it} - \hat Y_{it}(0)$ for each treated observation $(i,t)\in\mathcal{M}$. 

\noindent\textbf{Step 4.} Take averages of $\hat\delta_{it}$ to produce estimates for the quantities of interest. For example, for the ATT,
  $\widehat{ATT} = \frac{1}{|\mathcal{M}|} \sum_{\mathcal{M}} \hat \delta_{it}$; for the ATT at time period $s$ since the treatment occurred $\widehat{ATT}_{s} = \frac{1}{|\mathcal{S}|} \sum_{(i,t)\in\mathcal{S}}\hat{\delta}_{it},$ in which $\mathcal{S} = \{(i,t)| D_{i,t-s}=0,D_{i,t-s+1} = D_{i,t-s+2} = \cdots =
  D_{it} = 1\}$. $|\mathcal{A}|$ denotes the number of elements in set $\mathcal{A}$.
\end{adjustwidth}

Because treated observations of early treatment adopters never serve as controls for late treatment adopters, we prevent the negative weights problem from its root cause \citep{Goodman-Bacon2018,de_Chaisemartin2018-iw}. Compared with DID$_{M}$, our method is more efficient because it uses most available data without imposing stronger functional form assumptions.

\subsection{Three Novel Estimators as Examples} 

In this subsection, we review three estimators as examples of this framework. They are conceptually similar because they follow the same identification strategy laid out above.

\paragraph{a) The fixed effects counterfactual estimator.} We start by introducing a counterfactual estimator in which $Y_{it}(0)$ is imputed based on a TWFE model, i.e.:
$$Y_{it}(0) = \mathbf{X}_{it}'\beta + \alpha_{i} + \xi_{t} +
  \varepsilon_{it},\quad\text{for all } (i, t).$$
In other words, we assume $f(\mathbf{X}_{it}) =  \mathbf{X}_{it}'\beta$ and $h(\mathbf{U}_{it}) = \alpha_{i} + \xi_{t}$. A linear constraint over the fixed effects, $\sum_{D_{it}=0}\alpha_{i} = \sum_{D_{it}=0}\xi_{t}$, is imposed to achieve identification. This constraint also makes the grand mean parameter redundant.

It is easy to see that in a classic DID setup with two groups, two periods and no covariates, the FEct estimator is the DID estimator. How does FEct address the weighting issue with a general panel treatment structure? \citet{arkhangelsky2021double} show that with additive unit and time fixed effects, any estimator that aims at identifying a convex combination of $\delta_{it}$ can be written as a weighted average of $Y_{it}$, where the weights $\{w_{it}\}_{1\leq i \leq N, 1 \leq t \leq T}$ must satisfy the following four conditions: (1) $\frac{1}{NT}\sum_{i=1}^N \sum_{t=1}^T w_{it}D_{it}= 1$; (2) $\sum_{t=1}^T w_{it} = 0$ for any $i$; (3) $\sum_{t=1}^N w_{it} = 0$ for any $t$; and (4) $w_{it}D_{it} \geq 0$ for any $(i, t)$. Weights from both a TWFE model and FEct meet conditions (1)-(3). However, the former violates the last condition while latter does not; in fact, FEct imposes $w_{it: D_{it}=1} = \frac{1}{|\mathcal{M}|}$, which guarantees the identification of the causal quantities, such as $ATT$ and $ATT_{s}$. We can therefore rewrite FEct as a weighting estimator; that is, each treated observation is matched with its predicted counterfactual $\hat{Y}_{it}(0) = \mathbf{W}^{(it)'} \mathbf{Y}_{\mathcal{O}}$, which is the weighted sum of all untreated observations. Comparison within each matched pair removes the biases caused by improper weighting that plague conventional FE models. We provide all the proofs for in Section B in SI (pp. 12--15). 

\paragraph{Proposition 1} ({\it Unbiasedness and consistency of FEct}) Under Assumptions 1-3, as well as some regularity conditions,
\begin{center}
    $\E[\widehat{ATT}_{s}] = ATT_{s}$; $\E[\widehat{ATT}] = ATT$;\\
    $\widehat{ATT}_{s} - ATT_{s} \overset{p}{\to} 0$; and $\widehat{ATT} -  ATT \overset{p}{\to} 0$ as $N\to\infty$.
\end{center}\vspace{-1ex}
\paragraph{Proposition 2} ({\it FEct as a weighting estimator}) Under Assumptions~1-3 and when there are no covariates, we have 
   \begin{center}
    $\widehat{ATT}_{s} = \frac{1}{|\mathcal{S}|}
                        \sum_{(i,t) \in \mathcal{S}} [Y_{it} - \mathbf{W}^{(it)'} \mathbf{Y}_{\mathcal{O}}]$,
  \end{center} 
  where $\mathbf{W}^{(it)'} = \left(\dots, W_{js}^{(it)}, \dots\right)_{(j,s) \in \mathcal{O}}$ is a vector of weights that satisfy
  \begin{align*}
  \sum_{(s: (i, s)\in\mathcal{O})}W_{is}^{(it)} = 1, \sum_{(j: (j, t)\in\mathcal{O})}W_{jt}^{(it)} = 1, \sum_{(j: s \neq t, (j, s)\in\mathcal{O})}W_{js}^{(it)} = \sum_{(s: j \neq i, (j, s)\in\mathcal{O})}W_{js}^{(it)} = 0.
  \end{align*}

\medskip\paragraph{b) The interactive fixed effects counterfactual estimator.} FEct estimates will be biased when unobserved time-varying confounders exist. A couple of authors have proposed using factor-augmented models to relax the strict exogeneity assumption \citep{Bai2009,GobillonMagnac2016,xu2017generalized,BaiNg2020}. IFEct models the response surface of untreated potential outcomes using a factor-augmented model:
\begin{center}
  $Y_{it}(0) = \mathbf{X}_{it}'\beta + \alpha_{i} + \xi_{t} + \lambda_{i}'f_{t} + \varepsilon_{it}$, for all $(i, t)$.
\end{center}
\medskip\noindent In other words, $f(\mathbf{X}_{it}) =  \mathbf{X}_{it}'\beta$ and $h(\mathbf{U}_{it}) = \alpha_{i} + \xi_{t} + \lambda_{i}'f_{t}$. When the model is correctly specified, IFEct is consistent. 

\medskip\noindent\textbf{Proposition 3} ({\it Consistency of IFEct}) Under Assumptions~1-3, as well as some regularity conditions, $\widehat{ATT} \overset{p}{\to}  ATT \text{ as } N, T\to\infty.$

\paragraph{c) The matrix completion estimator.} \citet{athey2018matrix} introduce the MC method from the computer science literature as a generalization of factor-augmented models. Similar to FEct and IFEct, it treats a causal inference problem as a task of completing a $(N \times T)$ matrix with missing entries, where missing occurs when $D_{it} = 1$. Mathematically, MC assumes that the ($N\times T$) matrix of $[h(\mathbf{U}_{it})]_{i=1,2,\hdots,N, t=1,2,\hdots,T}$ can be approximated by a lower-rank matrix $\mathbf{L}_{(N\times T)}$, i.e.,
\begin{center}
 $\mathbf{Y(0)} = \mathbf{X}\beta + \mathbf{L} + \bm{\varepsilon}$,
\end{center}
in which $\mathbf{Y}$ is a $(N\times T)$ matrix of untreated outcomes; $\mathbf{X}$ is a $(N\times T\times k)$ array of covariates; and  $\bm{\varepsilon}$ represents a $(N\times T)$ matrix of idiosyncratic errors. As with IFEct, $\mathbf{L}$ can be expressed as the product of two $r$-dimension matrices: $\mathbf{L}  = \mathbf{\Lambda} \mathbf{F}$. Unlike IFEct, however, the MC estimator does not explicitly estimate $\mathbf{F}$ and $\mathbf{\Lambda}$; instead, it seeks to directly estimate $\mathbf{L}$ by solving the following minimization problem:
\begin{equation*}
\mathbf{\widehat{L}} = \argmin_{\mathbf{L}} \left[ \sum_{(i,t) \in \mathcal{O}} \frac{(Y_{it} - L_{it})^2}{|\mathcal{O}|} + \lambda_{L} \|\mathbf{L}\| \right],
\end{equation*}
in which $\mathcal{O} = \{(i,t)| D_{it} = 0\}$, $\|\mathbf{L}\|$ is the chosen matrix norm of $\mathbf{L}$, and $\lambda_{L}$ is a tuning parameter. \citet{athey2018matrix} propose an iterative algorithm to obtain $\mathbf{\widehat{L}}$ and show that $\mathbf{\widehat{L}}$ is an asymptotically unbiased estimator for $\mathbf{L}$. We summarize the algorithms for both IFEct and MC in SI (pp. 2--3). 

\paragraph{Remark 1: The difference between IFEct and MC.} The main difference between IFEct and MC lies in the way they regularize the singular values when decomposing the residual matrix. IFEct uses a ``best subset'' approach that selects the $r$ biggest singular values, in which $r$ is a fixed number and $r < \min\{N, T\}$, while MC imposes an $L_1$ penalty on all singular values with a tuning parameter $\lambda_{L}$ (Figure~\ref{fg.impute}). In the machine learning literature, they are referred to as hard impute and soft impute, respectively.
\begin{figure}[!ht]
\caption{Hard Impute (IFEct) vs. Soft Impute (MC)}\label{fg.impute}
\begin{center}
\begin{tabular}{cc}
{\tiny$\begin{pmatrix}
    {\sigma_{1}}       & 0 & 0 & \cdots & 0 \\
    0  & {\sigma_{2}} & 0 & \cdots & 0 \\
    0  & 0 & 0 & \dots & 0 \\
    \vdots  & \vdots & \vdots & \ddots & \vdots \\
    0  & 0 & 0 & \cdots & 0 \\
    \vdots  & \vdots & \vdots & \vdots & \vdots \\
    0  & 0 & 0 & \cdots & 0
\end{pmatrix}_{N\times T}$} &
{\tiny$\begin{pmatrix}
    { |\sigma_{1}-\lambda_{L}|_{+}}       & 0 & 0 & \cdots & 0 \\
    0  & {|\sigma_{2}-\lambda_{L}|_{+}} & 0 & \cdots & 0 \\
    0  & 0 & {|\sigma_{3}-\lambda_{L}|_{+}} & \dots & 0 \\
    \vdots  & \vdots & \vdots & \ddots & 0 \\
      &  &  &  & {|\sigma_{T}-\lambda_{L}|_{+}} \\
    0  & 0 & 0 & \cdots & 0 \\
    \vdots  & \vdots & \vdots & \vdots & \vdots \\
    0  & 0 & 0 & \cdots & 0
\end{pmatrix}_{N\times T}$}\\
{\small Hard Impute} & {\small Soft Impute} \\
\end{tabular}
\end{center}
{\footnotesize{\textbf{Note:}} The above figure, adapted from \citet{athey2018matrix}, illustrates how regularization works with IFEct---which selects two factors in this case---and MC. It also shows that they are fundamentally similar ideas. $|a|_{+} = \max(a, 0)$}.
\end{figure}

Whether IFEct or MC performs better depends on context. In Section D.2 in SI (pp. 20--21), we provide Monte Carlo evidence to show that when the factors are strong and sparse, IFEct outperforms MC; otherwise, MC performs better. In practice, researchers may choose between the two models based on how they behave under the diagnostic tests we introduce in the next section. When $r = 0$ or when $\lambda_{L}$ is bigger than the biggest singular value of the residual matrix, no factors are included in the model; as a result, IFEct or MC reduces to FEct. 

The IFEct estimator was first proposed by \citet{GobillonMagnac2016} in a DID setting where the treatment takes place at the same time for a subset of units. It is also closely related to the generalized synthetic control method  \citep{xu2017generalized}, in which factors are estimated using the control group data only. In this paper, we accommodate with panel treatment structure, which allows treatment reversal. In other words, the generalized synthetic control method can be seen as a special case of IFEct when the treatment does not switch back.

\paragraph{Remark 2: Choosing the tuning parameters.} In order to choose $r$ for IFEct, we repeat Step 2 on a training set of untreated observations until $\hat{\beta}$ converges. The optimal $r$ is then chosen based on a prespecified model performance metric, such as mean squared prediction error, using a k-fold cross-validation scheme. To preserve temporal correlations in the data, the test set consists of a number of triplets (three consecutive untreated observations of the same unit) from the treatment group. Similarly, for the MC estimator, we use k-fold cross-validation to select the $\lambda_{L}$. The test set is constructed in the same way as in IFEct. 

\paragraph{Remark 3: Inferential methods.} We rely on nonparametric block bootstrap and jackknife---both clustered at the unit level---to obtain uncertainty estimates for the treatment effect estimates. Our simulation results, reported in SI (pp. 16--17), suggest that both inferential methods work well with reasonable sample sizes (e.g. $T = 20, N = 50$). In practice, we recommend researchers use jackknife when the number of treated units is small.

\section{Diagnostics}\label{sc:diag}

In this section, we introduce a set of diagnostic tools to assist researchers probing the validity of the identifying assumptions. These assumptions should be considered collectively because strict exogeneity (Assumption~2) hinges on a correct functional form (Assumption~1) and bias removal is only possible when the feasibility condition (Assumption~3) is met. We first introduce a plot for dynamic treatment effects based on counterfactual estimators. We then propose several statistical tests for the implications of the identifying assumptions, including a placebo test, a test for no pretrend, and a test for no carryover effects. The latter two can be seen as extensions of the placebo test.  

\subsection{A Plot for Dynamic Treatment Effects}

In applied research with TSCS data, researchers often plot the so-called ``dynamic treatment effects,'' which are coefficients of the interaction terms between the treatment indicator and a set of dummy variables indicating numbers of periods relative to the onset of the treatment (lags and leads)---for example, $s = -4, -3, \cdots, 0, 1, \cdots, 5$ with $s=1$ representing the first period a unit receives the treatment---while controlling for unit and time fixed effects. Researchers then gauge the plausibility of the strict exogeneity assumption by eyeballing whether the coefficients in the pretreatment periods (when $s \leq 0$) exhibit an upward or a downward trend---often known as a ``pretrend''---or are statistically significant from zero. The magnitudes of the coefficients and corresponding {\it p}-values often depend on the baseline category researchers choose, which varies from case to case.

\begin{figure}[!ht]
\caption{Estimating the Dynamic Treatment Effects}\label{fg:treat}
\centering
\begin{minipage}{1\linewidth}{
\centering
\hspace{0em}
\subfigure[Staggered Adoption]{\includegraphics[width = 0.47\textwidth]{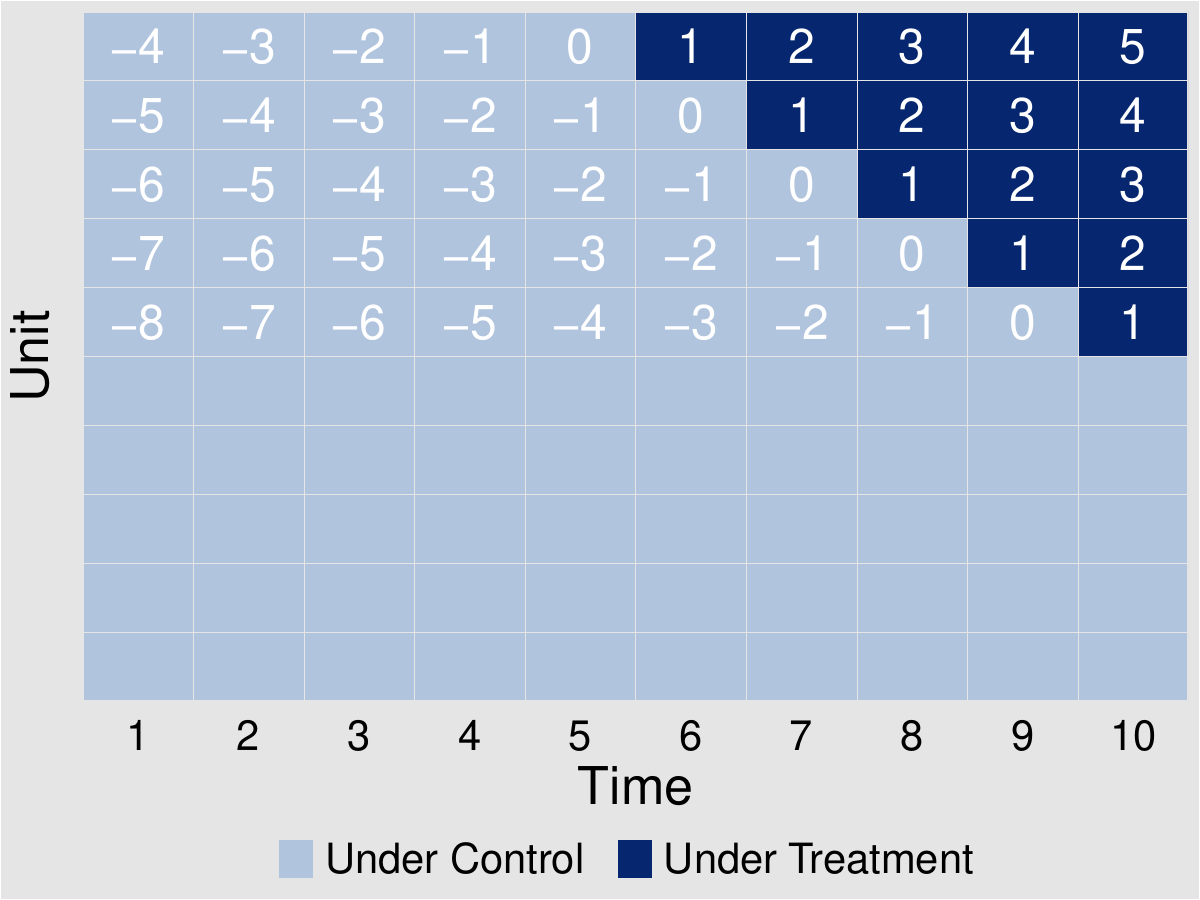}}\hspace{1em}
\subfigure[General Pattern]{\includegraphics[width = 0.47\textwidth]{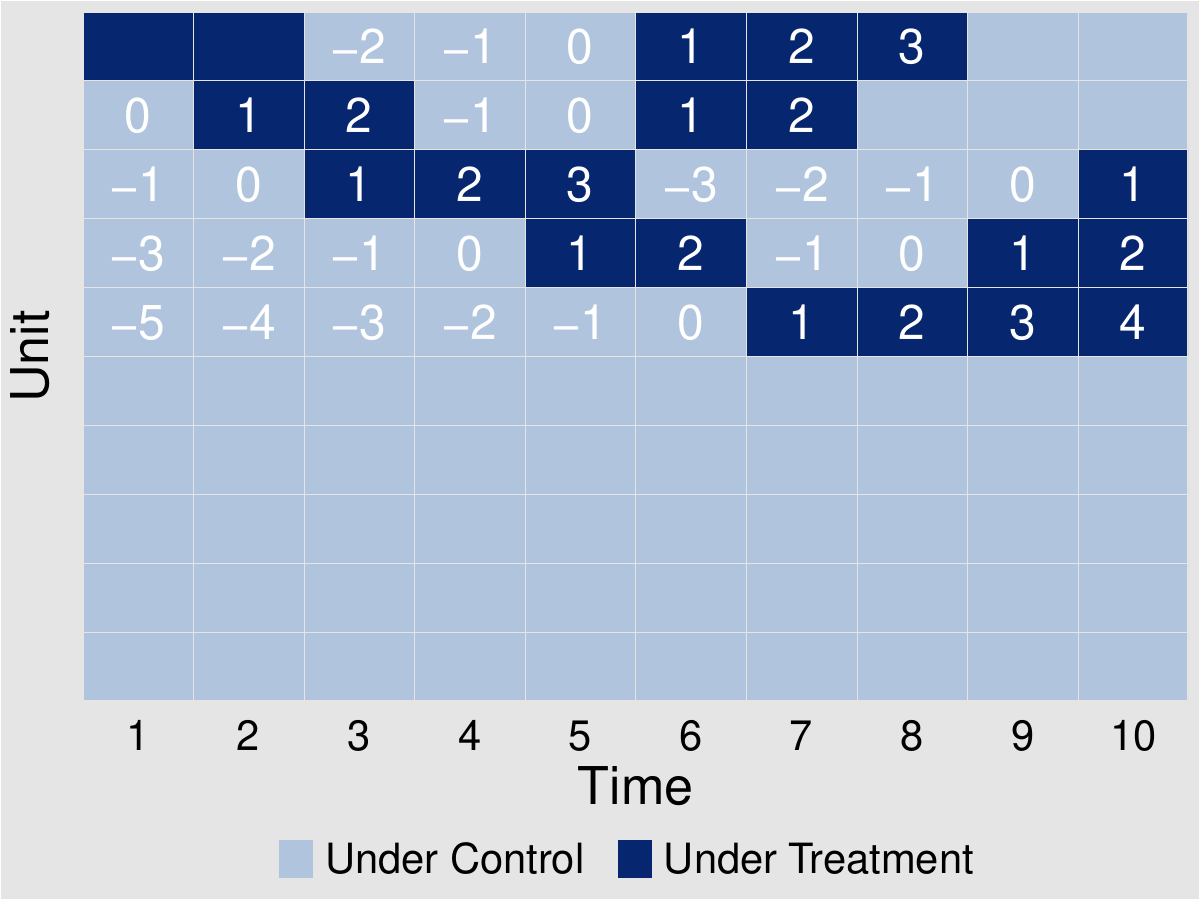}}
}\\
\footnotesize\textbf{Note:} The above figure shows the treatment status with two hypothetical examples: (a) staggered adoption and (b) a general panel treatment structure. Numbers correspond to time relative to the onset of a treatment. Several cells in (b) are not assigned numbers because left or right censorship of data makes their relative positions to a treatment uncertain.
\end{minipage}\vspace{-0.5em}
\end{figure}

We improve the dynamic treatment effect plot by taking advantage of the counterfactual estimators. Instead of plotting the interaction terms, we plot the averages of the differences between $Y_{it}$ and $\hat{Y}_{it}(0)$ for units in the treatment group ($C_i = 1$), re-indexed based on the time relative to the onset of the treatment. Specifically, we define $\hat{\delta}_{it} = Y_{it} - \hat{Y}_{it}(0)$, for all $t, C_i = 1$. When the identifying assumptions are correct, it is easy to see that average pretreatment residuals will converge to zero, i.e., $\widehat{ATT}_{s}\overset{p}{\to}0$ for all $s\leq0$.\footnote{With some abuse of the terminology, we call the residual averages $\widehat{ATT}_{s}$ when $s\leq 0$.} Therefore, we should expect pretreatment residual averages to be bouncing around zero, i.e., no strong pretrend. Figure~\ref{fg:treat} illustrates how we takes averages of $\hat\delta_{it}$ based on the timing relative to the next closest treatment. 

This method has two main advantages over the traditional approach. First, it relaxes the constant treatment effect assumption. Even though the conventional dynamic treatment effect plot allows the treatment effects to be different across time, it assumes a constant effect for all treated units in a given time period (relative to the start of the next treatment).\footnote{\citet{sun2020estimating} show that, under a staggered adoption design, if the dynamic treatment effects differ across cohorts, a spurious pretrend may arise even when the parallel trends assumption is valid.} Second, because a unit's untreated average has already been subtracted from $\hat{\delta}_{it}$, it is no longer necessary for researchers to choose a base category; to put it differently, the base category is set at a unit's untreated average after the time effects are partialed out. The dynamic treatment effects plot is an intuitive ``eyeball'' test that can help researchers detect data and modeling issues instantly. However, it cannot differentiate the specific reasons why the identifying assumption may have failed, such as the anticipation effect, the presence of time-varying confounders, or feedback from past outcomes.

We illustrate the plot using a simulated panel dataset of 200 units and 35 time periods based on the following data generating process (DGP) with two latent factors, $f_{1t}$ and $f_{2t}$:
\[
Y_{it} = \delta_{it} D_{it} + 5 + 1 \cdot X_{it,1} + 3 \cdot X_{it,2} + \lambda_{i1}\cdot f_{1t} + \lambda_{i2}\cdot f_{2t}+ \alpha_{i} + \xi_{t} + \varepsilon_{it},
\]
where the heterogeneous individualistic treatment effects are governed by $\delta_{it}= 0.4 s_{t} + e_{it}$ when $D_{it} = 1$, in which $s_{t}$ represents the number of periods since the latest treatment's onset  and $e_{it}$ is i.i.d. $N(0,0.16)$; and $\delta_{it}= 0$ when $D_{it} = 1$. This means the expected value of the treatment effect gradually increases as a unit takes up the treatment and there is no carryover effect. $f_{1t}$ is a linear trend plus white noise and $f_{2t}$ is an i.i.d. $N(0,1)$ white noise. For each unit, the treatment may switch on and off. The probability of getting the treatment is dependent on the treatment status in the previous period as well as the interactive and additive fixed effects (see page 18 in SI for details; this DGP satisfies Assumptions~1-3). As a result, failure to adjust for these factors will lead to biases in the causal estimates. 

\begin{figure}[!ht]
\caption{Dynamic Treatment Effect for the Simulated Example}\label{fg:gap}
\centering
\begin{minipage}{1\linewidth}{\centering
\includegraphics[width = 1\textwidth]{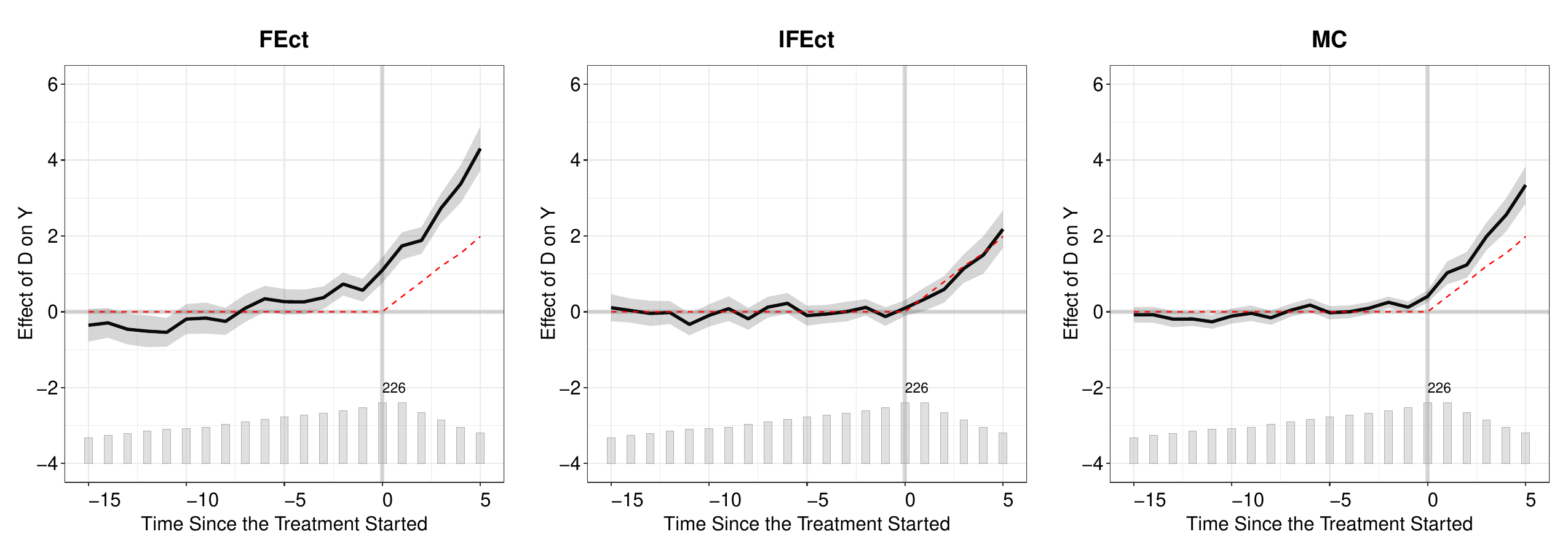}\\}
 \footnotesize\textbf{Note:} The above figure shows the dynamic treatment effects estimates from the simulated data using three different estimators: FEct, IFEct, and MC. The bar plot at the bottom of each panel illustrates the number of treated units at the given time period relative to the onset of the treatment (the number decreases as time goes by because there are fewer and fewer units that are treated for a sustained period of time). The red dashed lines indicate the true ATT.  
\end{minipage}
\end{figure}

Figure~\ref{fg:gap} shows the estimated dynamic treatment effects with 95\% confidence intervals based on block bootstraps of 1,000 times using the aforementioned counterfactual estimators. They are benchmarked against the true ATTs, which we depict with red dashed lines. 
From the left panel of Figure~\ref{fg:gap}, we see that using the FEct estimator, (1) a strong pretrend leads towards the onset of the treatment and multiple ``ATT''  estimates (residual averages) in the pretreatment periods are significantly different from zero; and (2) there are sizable positive biases in the ATT estimates in the posttreatment periods. We see a similar pattern in the posttreatment periods in the right panel where the MC estimator is applied, though with smaller biases. However, when using the IFEct estimator, the ATT estimates in both pretreatment and posttreatment periods are very close to the truth. This is expected because the DGP is generated by an IFE model with two latent factors and our cross-validation scheme picks the correct number of factors. To help researchers gauge the effective sample size, we plot the number of treated units at a given time period beneath the corresponding ATT estimate.

The dynamic treatment effects plot displays the temporal heterogeneity of treatment effects in an intuitive way. It is also a powerful visual tool for researchers to evaluate how plausible the identifying assumptions are. Next, we introduce several statistical procedures that formally test the implications of these assumptions. We start with a placebo test.

\subsection{A Placebo Test}

The basic idea for the placebo test is straightforward: we assume that the treatment starts $S$ periods earlier than its actual onset for each unit in the treatment group ($C_i = 1$) and apply the same counterfactual estimator to obtain estimates of $ATT_s$ for $s = -(S-1), \hdots, -1, 0$. We can also estimate the overall ATT for the $S$ pretreatment periods. If Assumptions~1-3 hold, we should expect the magnitude of this fake ``ATT'' estimate is close to zero. If this ``ATT'' estimate is statistically different from zero, we obtain a piece of evidence that some or all of the identifying assumptions are likely to be invalid.\footnote{In practice, $S$ should not be set too large because the larger $S$ is, the fewer pretreatment periods will remain for estimating the model. If both $S$ and $N_{tr}$ are too small, however, the test may be underpowered. In this and the following examples, we set $S = 3$.} For example, if a feedback effect from past outcome to current treatment exists (e.g., $Y_{t-1}$ and $D_{t}$ are positively correlated in Figure~\ref{fg.dag}), which is a failure of the strict exogeneity assumption, it is likely to be detected by the placebo test given sufficient data. 

Because a placebo test is a test for equivalence, as \citet{hartman2018equivalence} point out, a simple difference-in-means approach may suffer from limited power; that is, when the number of observations is small, failing to reject the null of the zero placebo effect does not mean equivalence holds. To address this concern, we introduce a variant of the equivalence test, where the null hypothesis is reversed:
\begin{center}
$ATT^{p} < -\theta_2$ or $ATT^{p} > \theta_1$,
\end{center}
in which $-\theta_{2} < 0 < \theta_1$ are prespecified parameters, or equivalence thresholds. Rejection of the null hypothesis implies the opposite holds with a high probability, i.e., $-\theta_2 \leq ATT^{p} \leq \theta_1$. In other words, if we collect sufficient data and show that the fake ``ATT'' falls within a prespecified narrow range, we obtain a piece of evidence to support the validity of the identifying assumptions. $[-\theta_{2}, \theta_{1}]$ is therefore called the \emph{equivalence range}. 
We use the two one-sided tests (TOST) to check the equivalence of $ATT^{p}$ to zero. Following \citet{hartman2018equivalence}, we set $\theta_1 = \theta_2 = 0.36\hat\sigma_{\varepsilon}$, in which $\hat\sigma_{\varepsilon}$ is the standard deviation of the residualized untreated outcome;\footnote{Specifically, we run a twoway fixed effects model with time-varying covariates using untreated data only and calculate the standard deviation of the residuals. The literature maps it at a moderate effect size.} alternatively, researchers may set the equivalence range based on an effect size they deem reasonable. 

\begin{figure}[!ht]
\caption{Placebo Tests for the Simulated Example}\label{fg:placebo}
\centering
\begin{minipage}{1\linewidth}{\centering
\includegraphics[width = 1\textwidth]{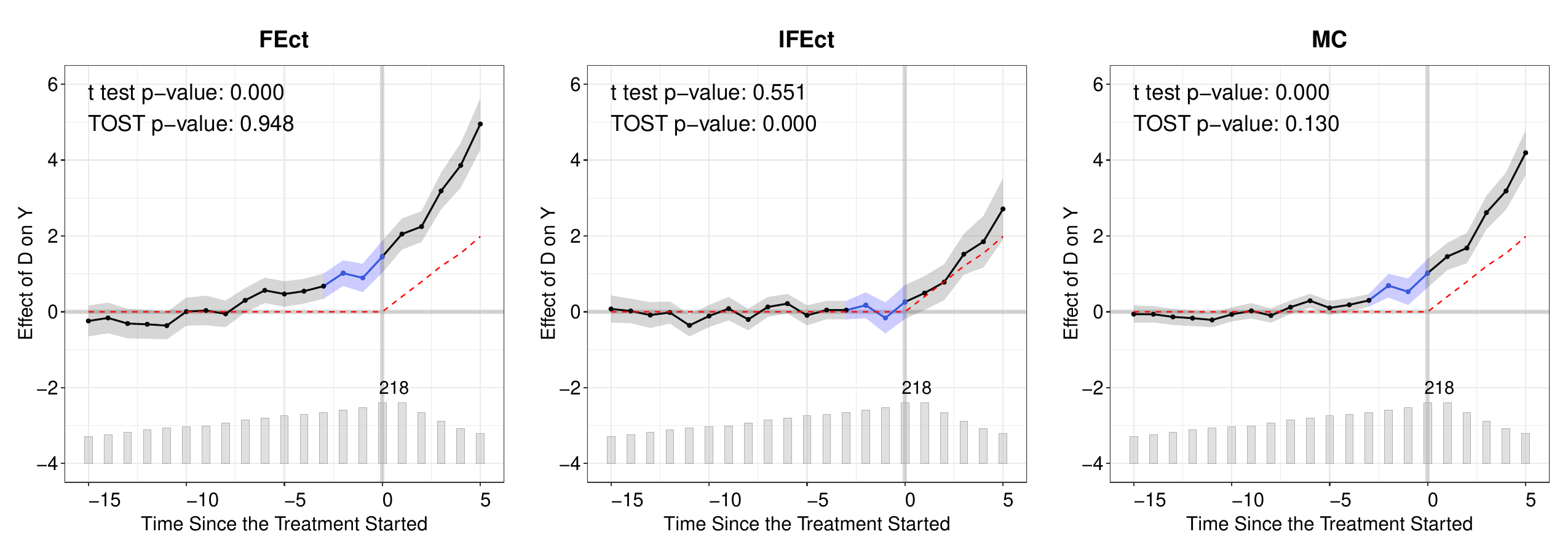}\\}
 \footnotesize\textbf{Note:} The above figure shows the results of the placebo tests based on three different estimators: FEct, IFEct, and MC. The bar plot at the bottom of each panel illustrates the number of treated units at the given time period relative to the onset of the treatment. The red dashed lines indicate the true ATT. Three pretreatment periods ($s = -2, -1, 0$) serving as the placebo are rendered in blue. The $p$-values for the $t$ test of the placebo effect and for the TOST are shown at the top-left corner of each panel. The equivalence range is set as $[-0.36\hat\sigma_{\varepsilon}, 0.36\hat\sigma_{\varepsilon}]$.
\end{minipage}
\end{figure}

One advantage of the placebo test is that it is robust to model misspecification and immune from over-fitting because it relies on out-of-sample predictions of $Y(0)$ in the placebo periods. Figure~\ref{fg:placebo} shows the results from the placebo tests based on the three counterfactual estimators. We see that for FEct and MC, we can reject the null that the placebo effect is zero under the DIM test but cannot reject the null that the effect is outside the equivalence range---hence, equivalence does not hold---while IFEct behaves in the exact opposite way: the placebo effect is statistically indistinguishable from zero ($p = 0.534$), and we can reject the null hypothesis that the placebo effect is bigger than the true ATT ($p = 0.000$). Although the MC method fits the pretreatment periods well, it does not pass the placebo test using either the DIM approach ($p = 0.000$) or the equivalence approach ($p = 0.131$). 

The main shortcoming of the equivalence approach is that researchers need to prespecify the equivalence range. $[0.36\hat{\sigma}_{\varepsilon}, 0.36\hat{\sigma}_{\varepsilon}]$ may be too lenient when the effect size is small relative to the variance of the residualized outcome. An alternative the literature suggests is to benchmark the minimum range against a reasonable guess of the effect size based on previous studies (e.g., \citealt{weiens2001}). However, such information is often unavailable. Because the ATT estimates from a TSCS analysis can be severely biased due to failures of the identification assumptions, unlike in experimental settings, they cannot provide valuable information for the true effect size, either. Moreover, setting the equivalence range in a post hoc fashion can lead to problematic results \citep{campbell2018}.  The best practice would be for researchers to preregister a plausible effect size and use it to set the equivalence range before analyzing data, as is a common practice in clinical trials. 

\subsection{Two Extensions}

We now extend the placebo test to testing (1) whether a pretrend exists, especially when it takes place a few periods before the treatment starts and cannot be detected by the placebo test; and (2) whether the treatment has carryover effects.

\paragraph{A test for no pretrend.} When a potential time-varying confounder is cyclical or does not present itself right before the treatment's onset, the placebo test may not be able to pick it up. Under this circumstance, we need a more global test for no pretrend. A natural approach is to jointly test a set of null hypotheses that the average of residuals for any pretreatment period is zero, i.e., $ATT_{s} = 0$ for all $s\leq 0$ using an $F$ test (see SI Section A.3, pp. 4--5, for details). However, because the test for no pretrend is also a test for equivalence, we develop an equivalence test with the following null:
\begin{center}
$ATT_s < -\theta_2$ or $ATT_s > \theta_1,\quad\forall s \leq 0$,
\end{center}
in which $[-\theta_{2}, \theta_1]$ is the equivalence range. In other words, the null is considered rejected (hence, equivalence holds) only when the tests for all pretreatment periods generate significant results. This is clearly a conservative standard, as we are simultaneously testing multiple hypotheses; as a result, the Type-I error will be smaller than the test size (e.g., 0.05).\footnote{Because the goal of an equivalence test is to control the Type-I error, multiple testing, which makes the test more conservative, is not a major concern. See \citet{Hartman2020-jb} (footnote 11) for a discussion.}  The equivalence approach has an additional advantage over the $F$ test in that when the sample size is large, a small confounder (or a few outliers) that only contributes to a neglectable amount of bias in the causal estimates will almost always cause rejection of the null hypothesis of joint zero means using the $F$ test. The equivalence test avoids this problem. 

Building upon the basic idea of the placebo test, we use a leave-one-period-out approach to obtain an average out-of-sample prediction error for each period before the treatment's onset as long as data permits. Given a prespecified equivalence range, each of the TOST rejects the null of inequivalence when the bootstrapped one-sided confidence interval of pretreatment $ATT_s$ (average prediction error in period $s$) falls within the range. In addition, we also calculate the \emph{minimum range}, the smallest symmetric bound within which we can reject the null of inequivalence using our sample. In other words, the minimal range is determined by the largest absolute value of the range of the 90\% confidence intervals of $\widehat{ATT}_{s, s\leq 0}$ in the pretreatment periods if we control the size $\alpha = 0.05$ \citep{hartman2018equivalence}. A rule of thumb is that when the minimum range is within the equivalence range, the test is considered passed. In Section D.3 in SI (pp. 21--22), we compare the performance of the $F$ test and the equivalence test using simulations.%
\begin{figure}[!ht]
\caption{Tests for No Pretrend: The Simulated Example}\label{fg:equivalence}
\centering
\begin{minipage}{0.95\linewidth}{\centering
\includegraphics[width = 1\textwidth]{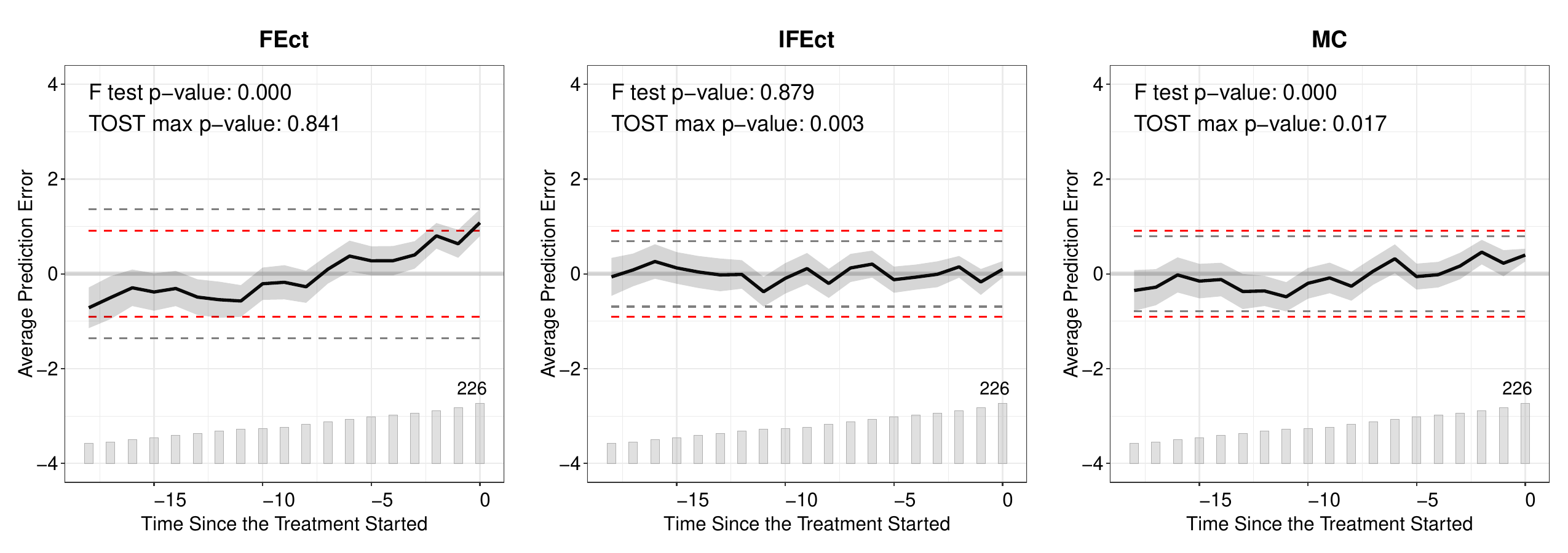}\\}
\footnotesize\textbf{Note:} The above figure shows the results of the equivalence tests based on three different estimators: FEct, IFEct, and MC. Pretreatment average prediction errors and their 90\% confidence intervals are drawn. The red dashed lines mark the equivalence range, while the gray dashed lines mark the minimum range. The bar plot at the bottom of each panel illustrates the number of treated units at the given time period relative to the onset of the treatment. 
\end{minipage}
\end{figure}
Figure~\ref{fg:equivalence} demonstrates the results of the equivalence test based on FEct, IFEct, and MC using the simulated dataset. With FEct, the trend leading towards the onset of the treatment goes beyond the equivalence range and results in a wide minimum range. Therefore, we cannot reject the null that the pretreatment average prediction errors are beyond a narrow range---in other words, we cannot say that equivalence holds with high confidence. However, both IFEct and MC pass the test. The 90\% confidence intervals of the pretreatment prediction error averages are within the equivalence range and the minimum range is narrower than the equivalence range. Note that the $F$ test $p$-value for MC is $0.000$, which points to potential model misspecification. 

\begin{figure}[!ht]
\caption{Tests for No Carryover Effects Using the Simulated Example}\label{fg:carryover}
\centering
\begin{minipage}{1\linewidth}{\centering
\includegraphics[width = 1\textwidth]{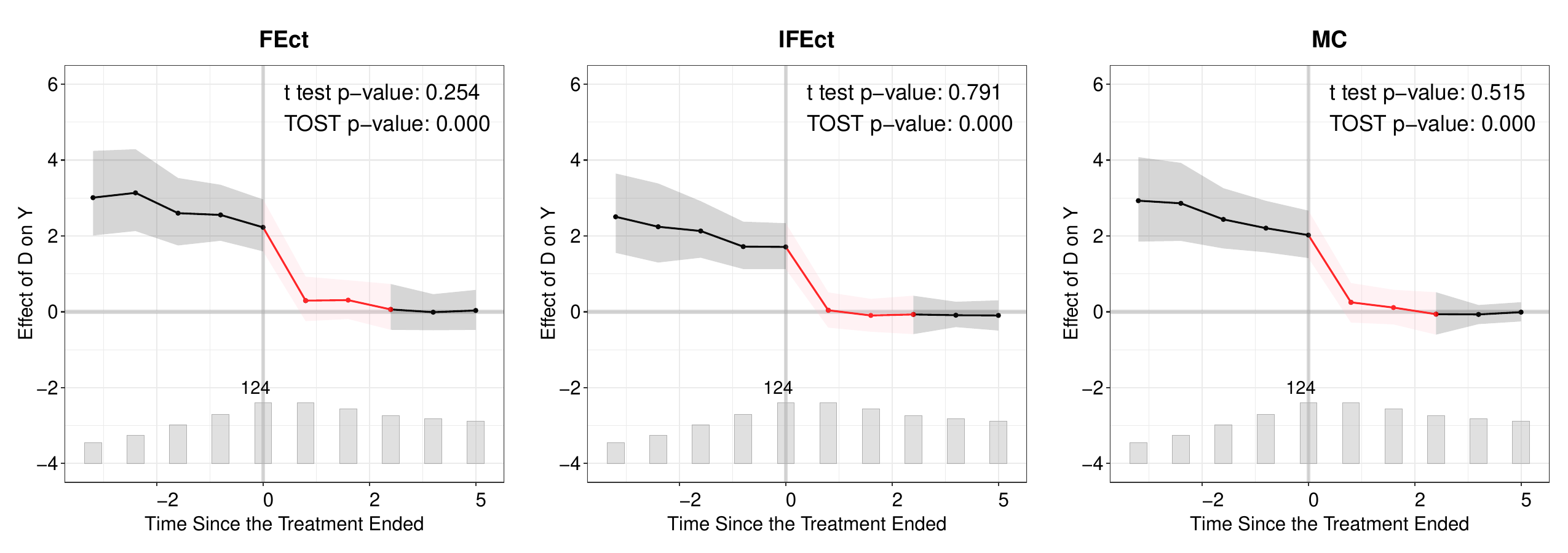}\\}
 \footnotesize\textbf{Note:} The above figure shows the results of the tests for no carryover effects based on three different estimators: FEct, IFEct, and MC. The bar plot at the bottom of each panel illustrates the number of treated units at the given time period relative to the end of the treatment. Three periods  after the treatment ends are rendered in pink. The $p$-values for the $t$ test of the carryover effects and for the TOST are shown at the top-right corner of each panel. 
\end{minipage}
\end{figure}

\paragraph{A test for no carryover effects.} We extend the idea of the placebo test to testing the presence of carryover effects. Instead of hiding a few periods right before the treatment starts, we hide a few periods right after the treatment ends and predict $Y_{it}(0)$ in those periods. If carryover effects do not exist, we would expect the average prediction error in those periods to be close to zero. Once again, we use both the DIM approach and the equivalence approach. Figure~\ref{fg:carryover} shows the results from applying this test to the simulated sample. Different from the dynamic treatment effects plot, the x-axis is now realigned based on the timing of the treatment's exit, not onset, e.g., 1 represents one period after the treatment ends. The results show that carryover effects do not seem to exist no matter which estimator or test is used, which is consistent with the DGP.

It is worth noting that the failure of the no carryover effects assumption does not necessarily invalidate our counterfactual estimation approach. If, by employing the proposed test, researchers find that the treatment effect persists after the treatment ends but in a limited time window, one strategy to proceed is to leave \emph{a sufficient number} of periods after the end of the treatment as ``treated'' and estimate the effects over these periods (as we do in the proposed test). Alternatively, researchers can change the definition of the treatment to ``$D_{it} = 1$ if a unit has \emph{ever} been under the treatment conditions, and $D_{it} = 0$ if otherwise,'' which essentially converts treatment assignment to a staggered adoption process, thus making the assumption for the no carryover effects unnecessary. 

We summarize the diagnostic tests in Table~\ref{tb:tests}. To lend support to the identifying assumptions, researchers can use either the DIM approach, if power is not a big concern, or the equivalence approach, if they have prior knowledge about the approximate effect size. No matter which approach researchers choose to use, visual inspection is always the first line of defense against erroneous causal claims based on invalid identifying assumptions.

\begin{table}[htbp]
  \caption{Diagnostic Tests Summary\label{tb:tests}}\footnotesize
  \resizebox{1\textwidth}{!}{
    \begin{tabular}{p{11em}cccccccc}\hline\hline
    \multicolumn{1}{r}{} & \multicolumn{2}{p{15em}}{\centering Placebo test} &       & \multicolumn{2}{p{13em}}{\centering Testing (no) pretrend} &
    & \multicolumn{2}{p{13em}}{\centering Testing (no) carryover effects}\\ \cline{2-3}\cline{5-6}\cline{8-9}
    \multicolumn{1}{r}{} & \multicolumn{1}{c}{$t$ test} &  \multicolumn{1}{c}{TOST}    &   & \multicolumn{1}{c}{$F$ test} &  \multicolumn{1}{c}{TOST}  &    & \multicolumn{1}{c}{$t$ test} &  \multicolumn{1}{c}{TOST}\\ \hline\\
    \multicolumn{1}{l}{Null} &  $ATT^{p} = 0$      &   $|ATT^{p}| > \theta $    &     &   $ATT_{s} = 0, \forall s\leq 0$      & $|ATT_{s}| > \theta, \exists s\leq 0$  &     &   $ACOE = 0$      & $|ACOE| > \theta$ \\ \\
    If Rejecting the Null & \multicolumn{1}{p{7.5em}}{\centering Invalidate\\ Assumptions} & \multicolumn{1}{p{7.5em}}{\centering Support\\ Assumptions} &  & \multicolumn{1}{p{6.5em}}{\centering Invalidate\\ Assumptions} & \multicolumn{1}{p{6.5em}}{\centering Support\\  Assumptions} & & \multicolumn{1}{p{6.5em}}{\centering Invalidate\\ No Carryover} & \multicolumn{1}{p{6.5em}}{\centering Support\\  No Carryover} \\
    Equivalence threshold $\theta$ &       & $0.36\hat\sigma_{\varepsilon}$ or eff &       &       & $0.36\hat\sigma_{\varepsilon}$ or eff  & & & $0.36\hat\sigma_{\varepsilon}$ or eff\\ \hline 
    \end{tabular}}\medskip

   {\footnotesize\textbf{Note:} Both the $t$ and $F$ tests are conventional difference-in-means tests, testing against the null of no difference. ``Assumptions'' refers to Assumptions 1-3 as a whole. $\hat\sigma_{\varepsilon}$ is the standard deviation of the residuals after twoway fixed effects are partialled out using untreated data only.  $ATT^p$ denotes the average placebo treatment effect on the treated. $ACOE$ denotes the average carryover effect. ``eff'' represents an effect size that researchers deem reasonable.} 
\end{table}

\section{Empirical Examples}\label{sc:examples}

We now apply the counterfactual estimators, as well as the diagnostics tools, to two empirical examples in political economy. The first example has a staggered adoption treatment structure while in the second one, the treatment switches back and forth. We start with FEct. If the results from FEct pass both the ``eyeball'' test and the diagnostic tests, there is little need for more complex methods except for potential efficiency gains. If, however, the visual inspection or the tests suggest the identifying assumptions are unlikely to be true, we apply IFEct and MC and run diagnostics again. In both applications, we set $S = 3$ in the placebo tests. All uncertainty estimates are obtained using clustered bootstrap at the unit level 1,000 times. 

\paragraph{Direct democracy and naturalization rates.} \citet{hainmueller2015does} study whether switching from direct democracy to indirect democracy increases naturalization rates for minority immigrants in Swiss municipalities using a generalized DID design. The outcome variable is minorities' naturalization rate in municipality $i$ during year $t$.  The treatment is a dummy variable indicating whether naturalization decisions are made by popular referendums. The dataset consists of 1,211 Swiss municipalities over 19 years, from 1991 to 2009. The authors report that the naturalization rate increases by  1.339\% on average (with a standard error of $0.161$) after a municipality shifts the decision-making power from popular referendums to elected officials using a twoway fixed effects model. 
\begin{figure}[!ht]
\caption{The Effect of Indirect Democracy on Naturalization}\label{fg:HH2015}
\centering
\begin{minipage}{1\linewidth}{
\centering
\hspace{-1em}
\subfigure[Dynamic Treatment Effects]{\includegraphics[width = 0.32\textwidth]{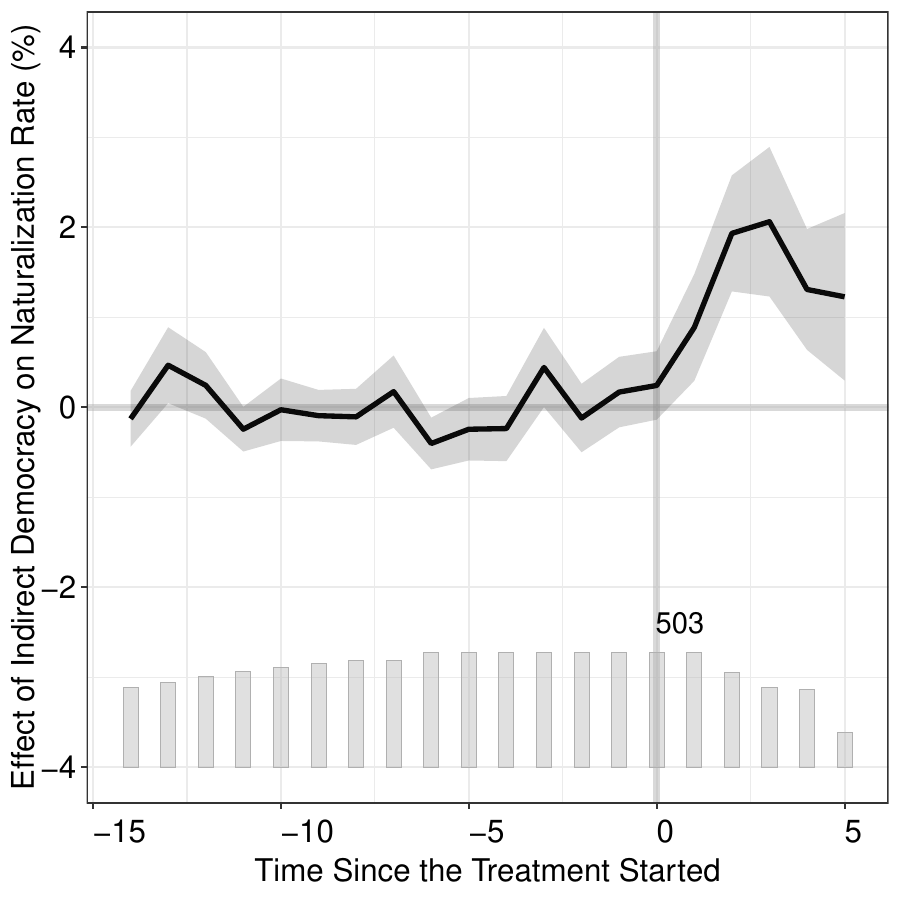}}
\subfigure[Placebo Test]{\includegraphics[width = 0.32\textwidth]{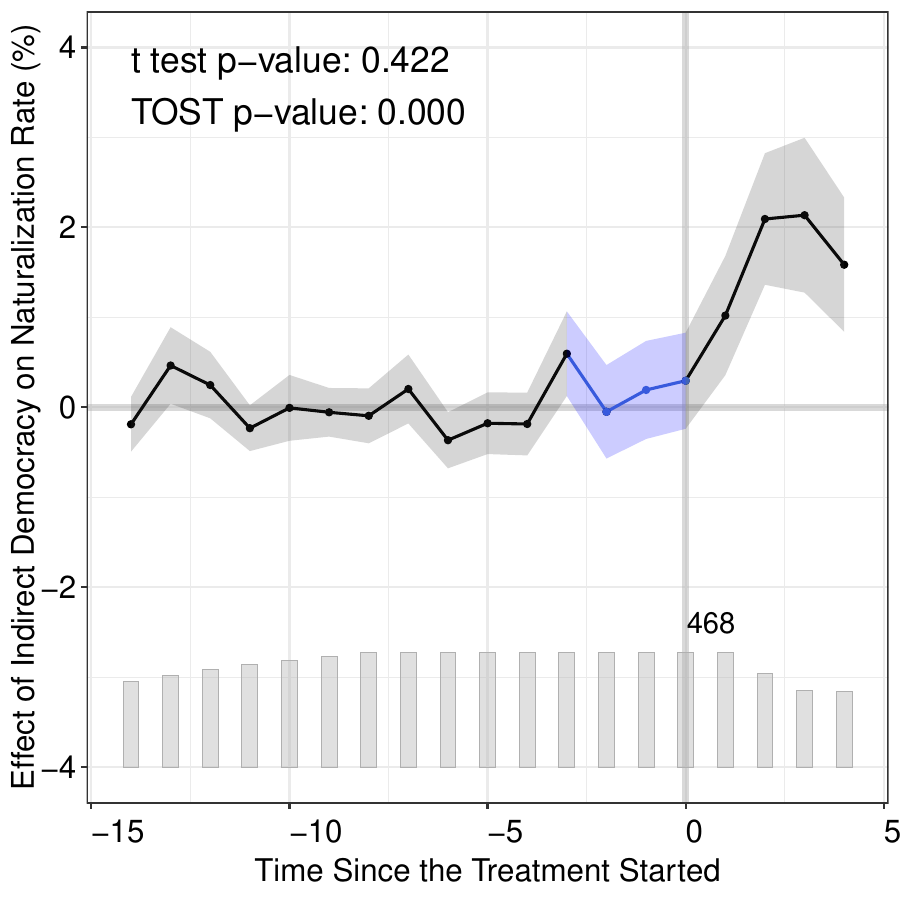}}
\subfigure[Testing No Pretrend]{\includegraphics[width = 0.32\textwidth]{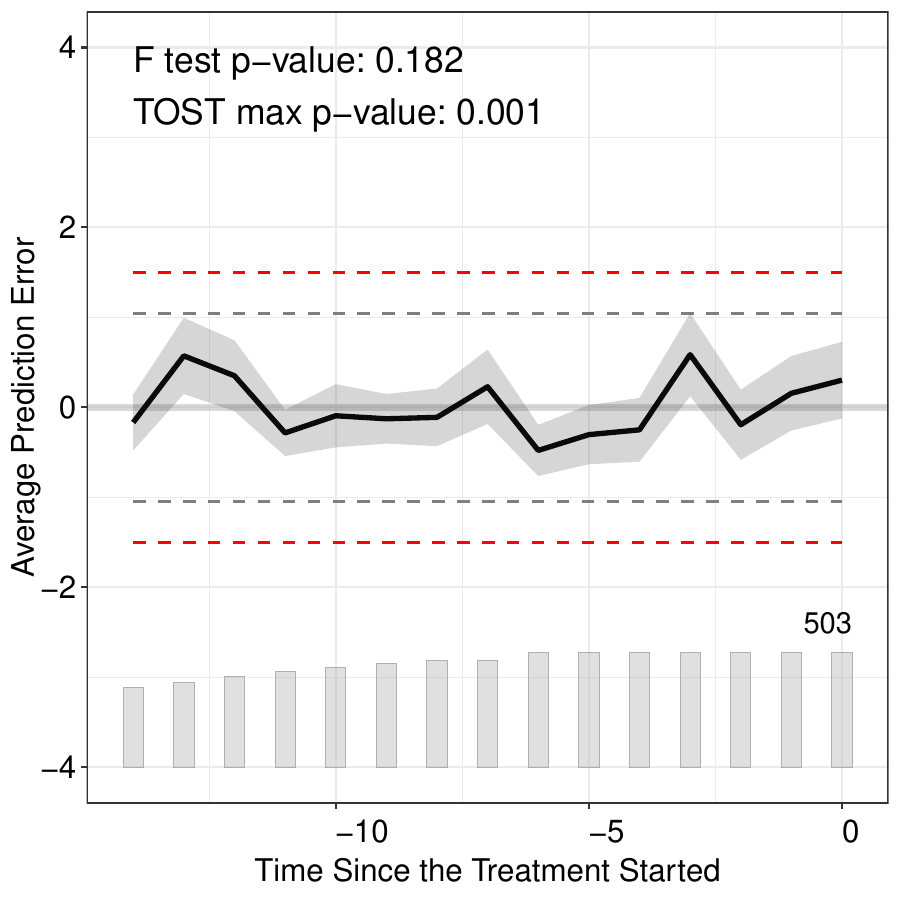}}\\
}
\footnotesize\textbf{Note:} The above figure shows the results from applying FEct to data from \citet{hainmueller2015does}. The left panel shows the estimated dynamic treatment effects using FEct. The middle panel shows the results from a placebo test using the ``treatment'' in three pretreatment periods as a placebo. The right panel shows the results of an equivalence test for no pretrend, in which the red and gray dashed lines mark the equivalence range and the minimum range, respectively. The bar plot at the bottom of each panel illustrates the number of treated units at a given time period relative to the onset of the treatment.
\end{minipage}\vspace{-0.5em}
\end{figure}

We then apply FEct and obtain an estimate of 1.767 (with a standard error of 0.197), even larger than the original estimate. Plots for the dynamic treatment effects and placebo test are shown in Figure~\ref{fg:HH2015}. We find that, first, the residual averages in the pretreatment periods are almost flat and around zero and the effect gradually takes off after the treatment begins. Second, with the placebo test, we cannot reject the null of zero placebo effect ($p = 0.425$), while we can reject the null whose magnitude is bigger than the default equivalence threshold ($p = 0.000$). Third, although the FEct estimator does not pass the $F$ test at the 5\% level ($p = 0.022$), it passes the TOST ($p = 0.001$).\footnote{With a large sample size, any small disturbances in data, such as outliers or an inconsequential confounder, can cause rejection of the $F$ test. The equivalence approach is immune to this problem.} The test for carryover effects is not applicable because of the staggered adoption treatment structure. We also apply both IFEct and MC estimators to this example. It turns out that the cross-validation schemes find zero factors, in the case of IFEct, and a tuning parameter bigger than the first singular value of the residual matrix, in the case of MC, both of which imply maximum regularization (no factors). Hence, both methods reduce to FEct and give the exact same estimates as FEct. 

In short, results from FEct are substantively the same as those from conventional TWFE models. However, counterfactual estimators like FEct allow us to check the validity of the identifying assumptions in a more convenient and transparent way. 

\paragraph{Partisan alignment and grant allocation.} Our second example is based on \citet{FM2015-yy}, in which the authors investigate whether partisan alignment between local councils in England and the central government bring localities more grants. The outcome of interest is the logarithm of specific grants per capita allocated to a local council. The treatment is a dummy variable indicating whether the government party controls the majority of local councils. The dataset spans 466 local councils from 1992 to 2012. The authors add locality-specific linear time-trends to a TWFE specification and find that partisan alignment increases specific grants allocated to a council---the increase peaks three years after alignment (see page 24 in SI for the original figure). A TWFE model without the locality-specific trends, however, returns negative estimates for the effect of partisan alignment.  

\begin{figure}[!ht]
\caption{The Effect of Partisan Alignment on Grant Allocation}\label{fg:FM2015}
\centering
\begin{minipage}{0.9\linewidth}{
\centering
\subfigure[Dynamic Treatment Effects]{\includegraphics[width = 1\textwidth]{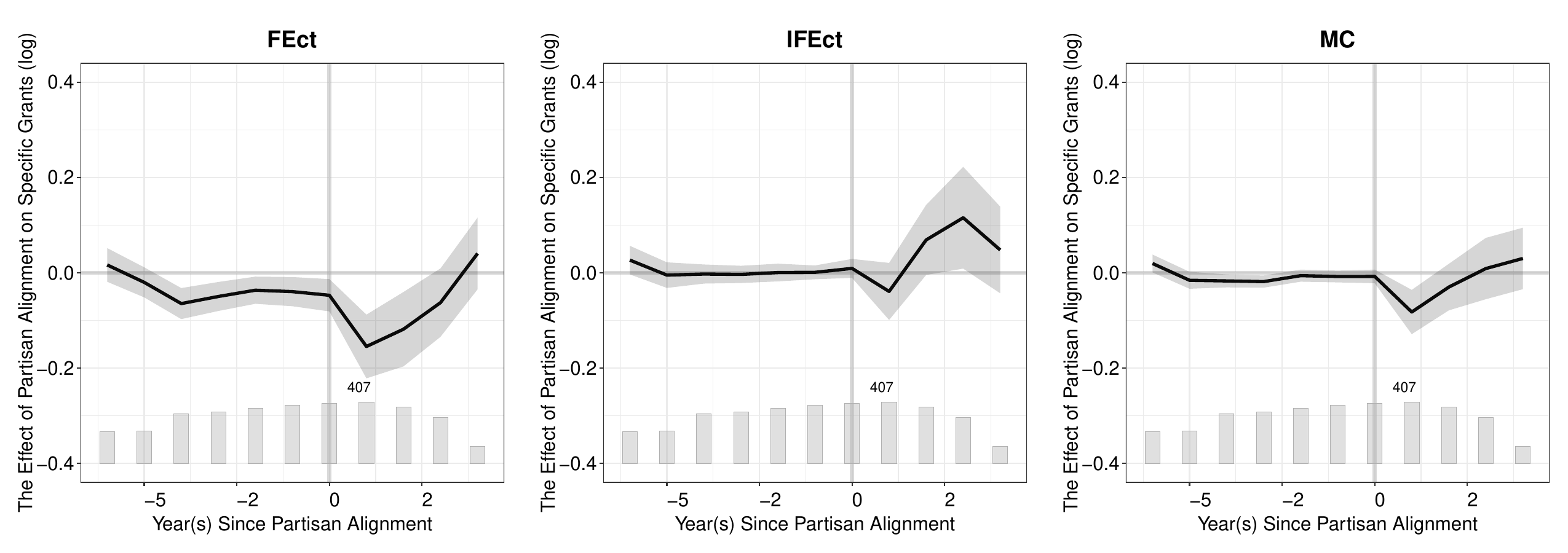}}\\
\subfigure[Placebo Test]{\includegraphics[width = 1\textwidth]{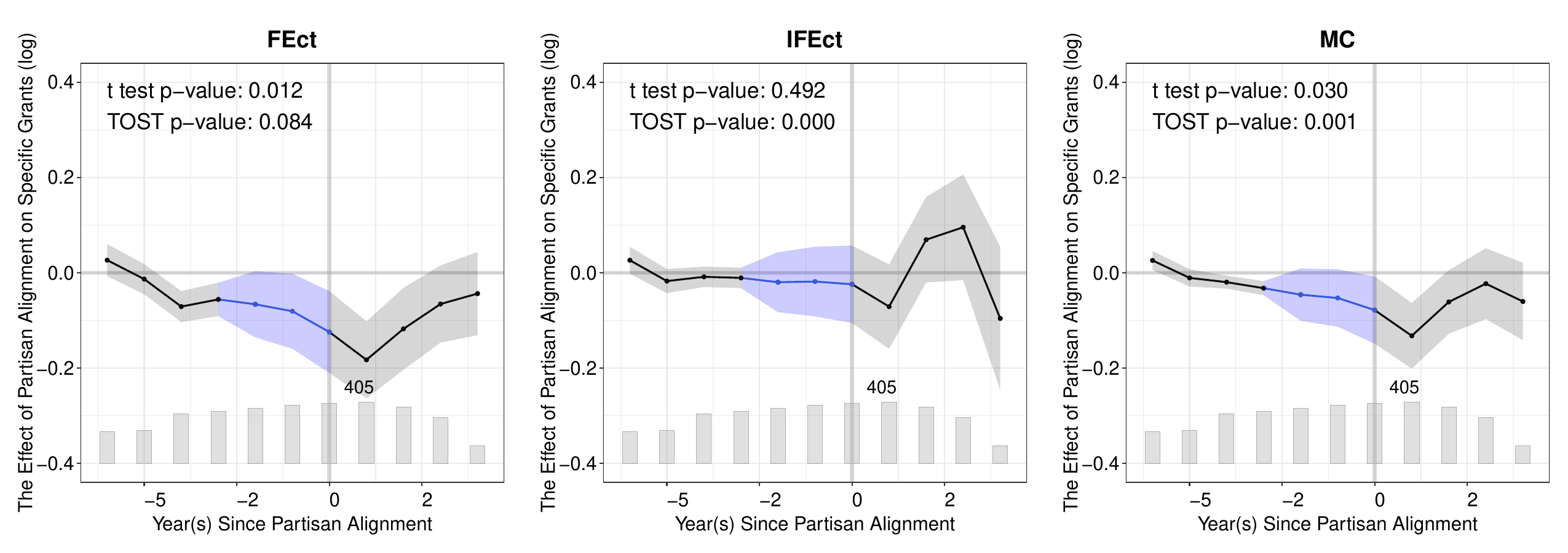}}
\subfigure[Test for Carryover Effects]{\includegraphics[width = 1\textwidth]{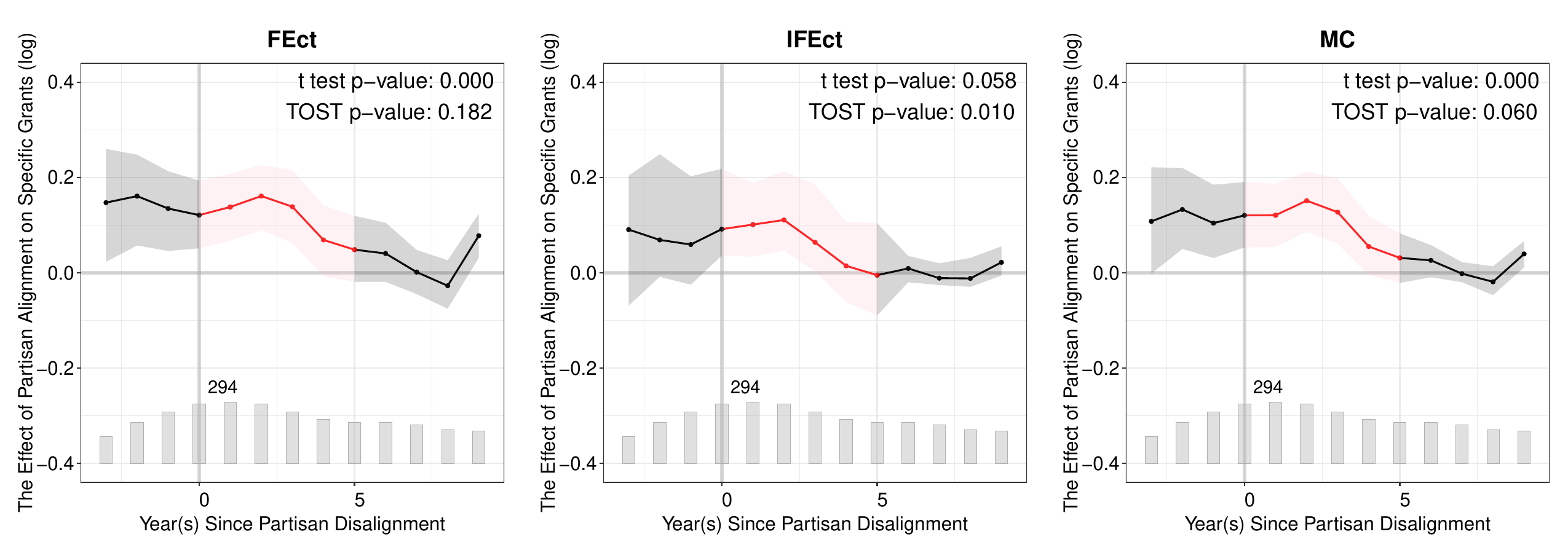}}
}
\footnotesize\textbf{Note:} The blue dots in (b) represent the periods used in the placebo tests. The red dots in (c) represent the periods used in the tests for no carryover effects. 
\end{minipage}
\end{figure}

We apply the three estimators to the data and plot the estimated dynamic treatment effects in Figure~\ref{fg:FM2015}(a). It shows that, with FEct, the pretreatment residual averages consistently deviate from zero, suggesting potential violations of the identifying assumptions. With IFEct and MC, however, these averages are very close to zero. Figure~\ref{fg:FM2015}(b) shows the results from the placebo test. With FEct, we cannot reject the null hypothesis of a non-zero placebo effect at the 5\% level. With either IFEct or MC, both the DIM ($t$ test) and the equivalence tests suggest the placebo effect is close to zero; however, IFEct seem to approximate the data better than MC as the pretrend looks almost completely flat and the estimated treatment effects are close to those in Figure~\ref{fg:FM2015}(a).\footnote{The results from the equivalence tests are not greatly informative and are reported in Figure A14 in in SI (p. 25).} Finally, we report the results from the test for carryover effects in Figure~\ref{fg:FM2015}(c), in which we test the carryover effects up to five years after partisan alignment ends. Based on the result from IFEct, the test suggests that there are positive carryover effects at least three years after partisan alignment ends. As discussed earlier, violation of the no carryover effect assumption does not necessarily invalidate the research design, but suggests that a more flexible estimation strategy is required. After we remove observations in the three periods after the treatment ended in the model-building stage (Step 1 in the algorithm), we re-estimate the ATT and conduct the diagnostic tests again. The new results shown in Figure~A15 in SI (p. 26) suggest that IFEct passes all diagnostic tests and is the most suitable model among the three. The magnitude of the effect remains similar.

\section{Conclusion}

The commonly used TWFE models require strong assumptions to produce interpretable causal estimates; however, they remain highly valuable because of their versatility in accommodating different data structures and high computational efficiency. In this paper, we seek to improve current practices with TWFE models by providing a simple but powerful counterfactual estimation framework, the key to which can be described as  ``fit data in the controls and impute counterfactuals to the treated,'' and by offering easy-to-implement diagnostic tests to assist researchers in probing the validity of the identifying assumptions. 

We discuss three estimators under this framework, including FEct, IFEct, and MC. It is important to note that IFEct and MC are not this paper's invention; they already exist in the literature. However, putting them in the same framework allows us to conduct diagnostics and evaluate their respective assumptions. Table~\ref{tb:methods} compares these estimators and other existing approaches and shows that they have several important advantages: they address the negative weights problem under heterogeneous treatment effects, accommodate general panel treatment structure without discarding data, can flexibly incorporate time-varying covariates, and are amenable for diagnostic tests. In addition, IFEct and MC can account for decomposable time-varying confounders. 

\begin{table}[htbp]
  \caption{Comparison of Methods}\footnotesize\label{tb:methods}
  \resizebox{1\textwidth}{!}{
    \begin{tabular}{lccccccc}\hline\hline
          & DID   & wDID  & DID$_{M}$  & PM & TWFE     & FEct  & IFEct/MC \\
    Accommodate heterogeneous treatment effects & x     & x     &   x  &   x  &           & x     & x \\
    Allow treatment reversal &       &       & x   & x  & x     & x     & x \\
    Condition on time-invariant covariates &       & x     &     & x   &      &       &  \\
    Condition on time-varying covariates &       &       &   & x    & x         & x     & x \\
    Use most available data & x     & x     &  &    &  x        & x     & x \\
    Easy-to-implement diagnostic tests & x     &       &  & x     &  x       & x     & x \\
    Condition on $U_{it} = \lambda_{i}'f_{t}$ &       &    &   &           &       &       & x \\     \hline
    \end{tabular}}\medskip

   {\footnotesize\textbf{Note:} DID, wDID, DID$_{M}$ , PM, TWFE, FEct, and IFEct/MC represent difference-in-differences, weighted difference-in-differences \citep{strezhnev2018,sun2020estimating,callaway2020difference}, multiple difference-in-differences \citep{de_Chaisemartin2018-iw}, panel match \citep{imai2018matching}, two-way fixed effects, fixed effects counterfactual, and interactive fixed effects counterfactual/matrix completion, respectively. $U_{it} = \lambda_{i}'f_{t}$ represents decomposable time-varying confounders.}
\end{table}

We also improve the existing practice of estimating and plotting dynamic treatment effects and develop several statistical tests based on the new plot. These tests are based on out-of-sample predictions of untreated potential outcomes, and thus are immune to model misspecification or overfitting. We recommend researchers use the visual and statistical tests in a holistic manner to gauge the validity of the identifying assumptions, as we do with two empirical examples. Below we provide a checklist as a practical guide to analyzing TSCS data using counterfactual estimators:
\begin{itemize}[leftmargin=*]\itemsep0em
  \item Plot the treatment status of your data and ask whether strict exogeneity assumption is a plausible description of the treatment assignment process; if not, consider using methods based on sequential ignorability.
  \item Plot the outcome variable in a time-series fashion to spot outliers and irregularities; transform the data if necessary. 
  \item Start with the simplest estimator, FEct, draw the dynamic treatment effects plot and perform both visual inspection and diagnostic tests (using either the DIM approach or the equivalence approach). 
  \item If FEct does not pass the placebo test or the test for no pretrend, apply more complex models, such as IFEct and MC, and perform diagnostics again. 
  \item If the chosen method fails the test for no carryover effects, remove several periods after the treatment ends from the model-building stage, then re-apply the method and1746-1763 conduct diagnostics again. 
  \item Optionally, if a treatment effect is detected, perform subgroup analysis to understand which group(s) of units are driving the effect.
  \item Communicate your findings effectively, ideally with figures.
\end{itemize}
We provide two packages, \texttt{panelView} and \texttt{fect}, in both \texttt{R} and \texttt{Stata} to assist researchers in achieving these goals. We hope that this guide, as well as the tools we provide, will contribute to improved practices when researchers analyze TSCS data.  

\section*{Replication Materials}

The data and materials required to verify the computational reproducibility of the results, procedures and analyses in this article are available on the {\it American Journal of Political Science} Dataverse within the Harvard Dataverse Network, at: \url{https://doi.org/10.7910/DVN/ZVC9W5}.

\vspace{2em}

\clearpage
\FloatBarrier
\setstretch{1.5}
\bibliographystyle{apsr}
\bibliography{tscs}
\clearpage

\end{document}


\onehalfspacing
\date{
  \vspace{-0.5em}\hspace{1.2em}(MIT)\hspace{4.8em}(UNC)\hspace{3.5em}{(Stanford)} \\\vspace{1em}
  First version: 11th July 2019\\
  This version: \today\vspace{2em} 
}

\onehalfspacing
\renewcommand\thesection{\Alph{section}}
\setcounter{page}{1}
\setcounter{table}{0}
\setcounter{figure}{0}
\setcounter{equation}{0}
\setcounter{footnote}{0}
\renewcommand\thetable{A\arabic{table}}
\renewcommand\thefigure{A\arabic{figure}}
\renewcommand{\theequation}{A\arabic{equation}}
\renewcommand{\thefootnote}{A\arabic{footnote}}

\setcounter{proposition}{0}

\begin{center}
{\Large\bf Supplementary Information \emph{for}}\\\bigskip
{\large\bf A Practical Guide to Counterfactual Estimators for Causal Inference with Time-Series Cross-Sectional Data}\\\bigskip
{\normalsize Licheng Liu (MIT)\hspace{2em}Ye Wang (UNC)\hspace{2em}Yiqing Xu (Stanford)}
\end{center}
\bigskip

\vspace{2em}

\noindent\hspace{0em}{\large\bf \underline{Table of Contents}}

{\bf
\begin{enumerate}\itemsep0ex
    \item[A.] Algorithms and Tests
    \begin{enumerate}\itemsep0ex\vspace{-0.5em}
    	\item[A.1.] The IFEct algorithm
      \item[A.2.] The MC algorithm
    	\item[A.3.] The difference-in-means tests and equivalence tests
      \item[A.4.] Discussion on the no carryover effects assumption
      \item[A.5.] Procedures for the diagnostic tests
	  \end{enumerate}
    \item[B.] Proofs 
    \begin{enumerate}\itemsep0ex\vspace{-0.5em}
      \item[B.1.] Unbiasedness and consistency of FEct and IFEct
      \item[B.2.] FEct as a weighting estimator
    \end{enumerate}
    \item[C.] Inferential Methods
    \item[D.] Additional Monte Carlo Evidence
    \begin{enumerate}\itemsep0ex\vspace{-0.5em}
    \item[D.1.] Describing the data generating processes
    \item[D.2.] IFEct vs MC
    \item[D.3.] $F$ test vs the equivalence test
    \end{enumerate}
    \item[E.] Additional Information on the Empirical Examples
    \begin{enumerate}\itemsep0ex\vspace{-0.5em}
    \item[E.1.] Replicating Hainmueller and Hangartner (2015)
    \item[E.2.] Replicating Fouirnaies and Mutlu-Eren (2015)
    \end{enumerate}    
\end{enumerate}
}
\clearpage

\small

\newpage

\section{Algorithms and Tests}

\subsection{The IFEct Algorithm}\label{sc:algo}

The IFEct algorithm takes for the following four steps. 

\begin{adjustwidth}{20pt}{20pt}

\noindent\textbf{Step 1.} Assuming in round $h$ we have $\hat{\mu}^{(h)}$, 
$\hat{\alpha}_i^{(h)}$, $\hat{\xi}_t^{(h)}$, $ \hat{\lambda}_i^{(h)} $, 
$ \hat{f}_t^{(h)} $ and $ \hat{\beta}^{(h)} $. Denote 
$\dot{Y}_{it}^{(h)} \coloneqq Y_{it} - \hat{\mu}^{(h)} - \hat{\alpha}_i^{(h)} - 
\hat{\xi}_t^{(h)} - \hat{\lambda}_i^{(h)\prime} \hat{f}_t^{(h)} $ 
for the untreated ($ D_{it} = 0 $):

\noindent\textbf{Step 2a.} Update $\hat{\beta}^{(h+1)}$ using the untreated data only (we can set $\hat{\lambda}_{i}^{(0)} = \mathbf{0}$, $\hat{f}_{t}^{(0)} = \mathbf{0}$ in round 0 and run a twoway fixed effects model to initialize $ \mu $, $ \alpha_i $ and $ \xi_t $):
\begin{align*}
  \hat{\beta}^{(h+1)} = 
  \left(\sum_{D_{it} = 0} \mathbf{X}_{it} \mathbf{X}_{it}^{\prime}\right)^{-1}
  \sum_{D_{it} = 0} \mathbf{X}_{it} \dot{Y}_{it}^{(h)}
\end{align*}
Note that matrix $\left(\sum_{D_{it} = 0} \mathbf{X}_{it} \mathbf{X}_{it}^{\prime}\right)^{-1}$ is fixed and does not need to be updated every time.

\noindent\textbf{Step 2b.} For all $i$, $t$, define
\begin{center}
$
W_{it}^{(h+1)} \coloneqq \left\{
\begin{aligned}
& = Y_{it} - \mathbf{X}_{it}^{\prime}\hat{\beta}^{(h+1)}, \ D_{it} = 0 \\
& = \hat{\mu}^{(h)} + \hat{\alpha}_i^{(h)} +
\hat{\xi}_t^{(h)} + \hat{\lambda}_i^{(h)\prime} \hat{f}_t^{(h)}, \ D_{it} = 1
\end{aligned}
\right.
$
\end{center}
For all untreated observations (i.e., $D_{it} = 0$) , calculate $W_{it}^{(h)}$. 
For all treated observations (i.e., $D_{it} = 1$), calculate its conditional expectation:
\begin{center}
   $\E\left(W_{it}^{(h+1)} \big|\hat{\lambda}_{i}^{(h)},\hat{F}_{t}^{(h)}\right) = 
    \hat{\mu}^{(h)} + \hat{\alpha}_i^{(h)} +
\hat{\xi}_t^{(h)} + \hat{\lambda}_{i}^{(h)} \hat{F}_{t}^{(h)}$ 
\end{center}

\noindent\textbf{Step 2c.} Denote 
$ W_{..}^{(h+1)} = \frac{\sum\limits_{i=1}^{N}\sum\limits_{t=1}^{T}W_{it}^{(h+1)}}
{NT} $, 
$ W_{i.}^{(h+1)} = \frac{\sum\limits_{t=1}^{T}W_{it}^{(h+1)}}{T}, \forall i $, 
$ W_{.t}^{(h+1)} = \frac{\sum\limits_{i=1}^{N}W_{it}^{(h+1)}}{N}, \forall t $ and 
$ \tilde{W}_{it}^{(h+1)} = W_{it}^{(h+1)} - W_{i.}^{(h+1)} - W_{.t}^{(h+1)} 
+ W_{..}^{(h+1)} $.  With restrictions: 
$ \sum_{i=1}^{N} \alpha_i = 0 $, $ \sum_{t=1}^{T} \xi_t = 0 $, 
$ \sum_{i=1}^{N} \lambda_i = \mathbf{0} $ and 
$ \sum_{t=1}^{T} f_t = \mathbf{0} $.

\noindent\textbf{Step 2d.} Update estimates of factors and factor loadings by minimizing the least squares objective function using the complete data of $\mathbf{W}^{(h+1)} = [\tilde{W}_{it}^{(h+1)}]_{\forall i,t}$:
\begin{gather}
  (\hat{\mathbf{F}}^{(h+1)},\hat{\mathbf{\Lambda}}^{(h+1)}) = 
  \mathop{\arg\min}_{(\tilde{\mathbf{F}},\tilde{\mathbf{\Lambda}})}\text{\textbf{tr}}
  \left[
    (\mathbf{W}^{(h+1)}-\tilde{\mathbf{F}}\tilde{\mathbf{\Lambda}}^{\prime})^{\prime}
    (\mathbf{W}^{(h+1)}-\tilde{\mathbf{F}}\tilde{\mathbf{\Lambda}}^{\prime})
  \right] \nonumber  \\
  s.t.
    \qquad \tilde{\mathbf{F}}^{\prime}\tilde{\mathbf{F}}/T = \mathbf{I}_{r} ,
    \tilde{\mathbf{\Lambda}}^{\prime}\tilde{\mathbf{\Lambda}} = \text{diagonal} \nonumber
\end{gather}

\noindent\textbf{Step 2e.} Update estimates of grand mean and twoway fixed effects:
\begin{gather}
\hat{\mu}^{(h+1)} = W_{..}^{(h+1)} \nonumber \\
\hat{\alpha}_i^{(h+1)} = W_{i.}^{(h+1)} - W_{..}^{(h+1)} \nonumber \\
\hat{\xi}_t^{(h+1)} = W_{.t}^{(h+1)} - W_{..}^{(h+1)} \nonumber
\end{gather}

\noindent\textbf{Step 3.} Estimate treated counterfactual, obtaining:
\begin{center}
  $\hat{Y}_{it}(0) = \mathbf{X}_{it}'\hat\beta + \hat\alpha_{i} + \hat\xi_{t} + \hat{\lambda}_{i}'\hat{f}_{t},$ for all $i$, $t$, $D_{it}=1$
\end{center}

\noindent\textbf{Step 4.} Obtain the $ATT$ and $ATT_{s}$ as in FEct.

\end{adjustwidth}

\bigskip

\subsection{The MC Algorithm}

We summarize the algorithm for the matrix completion (MC) method below. First, define $P_{\mathcal{O}}(\mathbf{A})$ and $P^{\bot}_{\mathcal{O}}(\mathbf{A})$ for any matrix $\mathbf{A}$: 
\begin{center}
$P_{\mathcal{O}}(\mathbf{A})=\begin{cases}
    \mathbf{A}_{it}, & \text{if} (i, t) \in \mathcal{O}.\\
    0, & \text{if} (i, t) \notin \mathcal{O}.
  \end{cases}$ \quad  and \quad 
$P^{\bot}_{\mathcal{O}}(\mathbf{A})=\begin{cases}
    0, & \text{if} (i, t) \in \mathcal{O}.\\
    \mathbf{A}_{it}, & \text{if} (i, t) \notin \mathcal{O}.
  \end{cases}$
\end{center}
\noindent Conduct Singular Value Decomposition (SVD) on matrix $\mathbf{A}$ and obtain $\mathbf{A} = \mathbf{S} \mathbf{\Sigma} \mathbf{R^T}$. The matrix shrinkage operator is defined as $\text{shrink}_{\theta}(\mathbf{A}) = \mathbf{S} \mathbf{\tilde{\Sigma}} \mathbf{R^T}$, where $\mathbf{\tilde{\Sigma}}$ equals to $\mathbf{\Sigma}$ with the i-th singular value $\sigma_{i}(A)$ replaced by $\max(\sigma_{i}(A) - \theta, 0)$, which is called ``soft impute'' in the machine learning literature. The MC algorithm takes the following iterative steps:
\begin{adjustwidth}{20pt}{20pt}\small
\noindent\textbf{Step 0.} Given a tuning parameter $\theta$, we start with the initial value $\mathbf{L}_{0}(\theta) = P_{\mathcal{O}}(\mathbf{Y})$.

\noindent\textbf{Step 1.} For $h = 0,1,2,\hdots$, we use the following formula to calculate $\mathbf{L}_{h+1}(\theta)$:
\begin{center}
$\mathbf{L}_{h+1}(\theta) = \text{shrink}_{\theta}\left\lbrace P_{\mathcal{O}}(\mathbf{Y}) + P^{\bot}_{\mathcal{O}}(\mathbf{L}_{h}(\theta)) \right\rbrace$
\end{center}

\noindent\textbf{Step 2.} Repeat Step 1 until the sequence $\left\lbrace \mathbf{L}_{h}(\theta) \right\rbrace_{h \geq 0}$ converges.

\noindent\textbf{Step 3.} Given $\hat{Y}_{it}(0) = \hat{L}_{it}^*$, and $\hat{\delta}_{it} = Y_{it} - \hat{Y}_{it}(0)$, compute $ATT$ and $ATT_{s}$ as before.
\end{adjustwidth}

\noindent If we replace  $\sigma_{i}(A)$ by $\sigma_{i}(A) \mathbf{1}\{\sigma_{i}(A) \geq \theta \}$, which is called ``hard impute,'' the algorithm will produce estimates almost identical to the IFEct algorithm.  

\bigskip
\clearpage

\subsection{The Difference-in-Means Tests and Equivalence Tests}\label{sc:equiv}

\begin{figure}[!th]
\caption{$t$ distribution under the Two Tests}\label{fg:tost}
\centering
\begin{minipage}{0.9\linewidth}{
\begin{center}
\includegraphics[trim=0 .5em 0 .5em,clip,width = 0.8\textwidth]{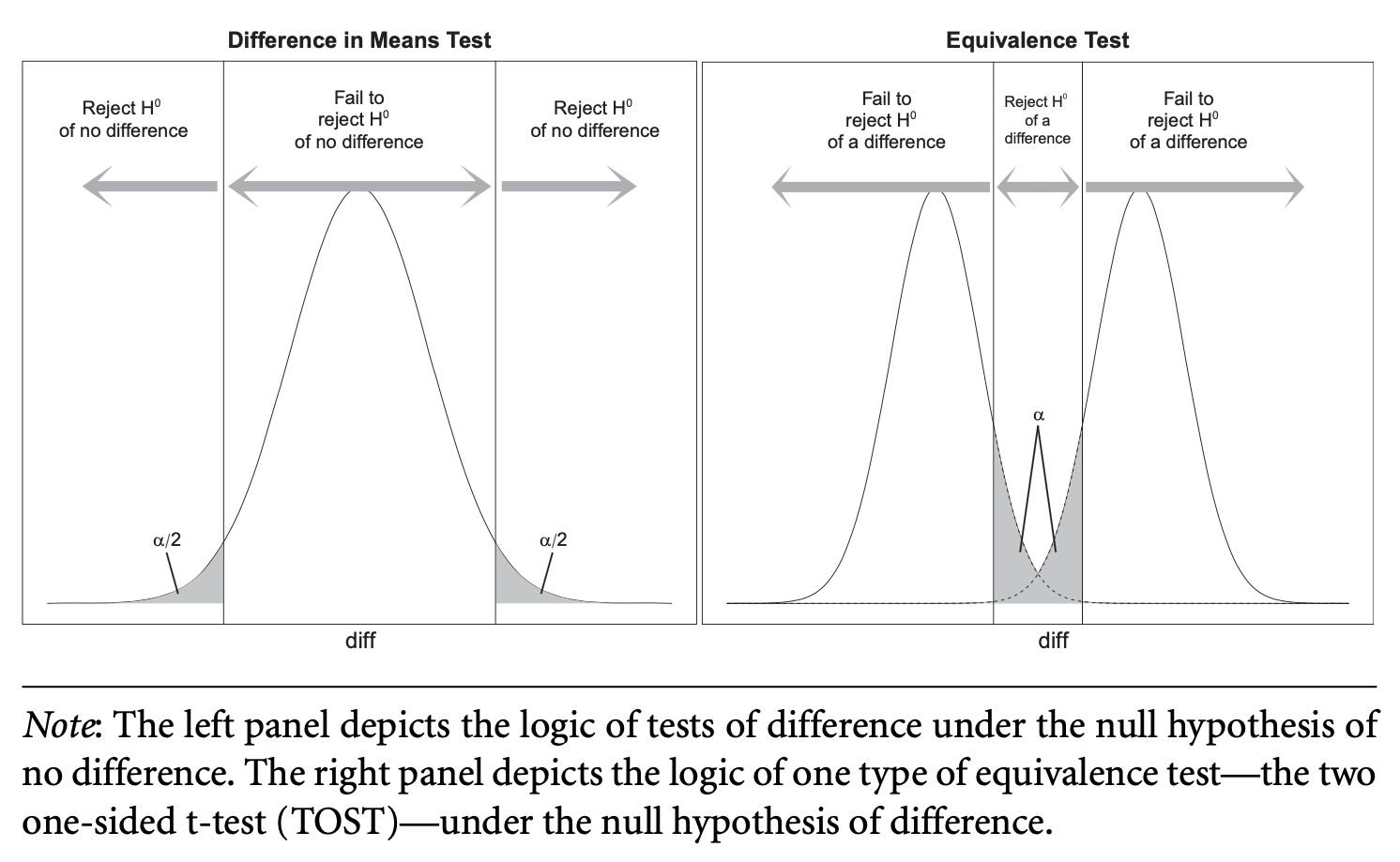}\\
\end{center}
}
\end{minipage}
\end{figure}

The equivalence test we introduce in the main text takes the form of two one-sided tests (TOST) for each pretreatment period $s$. We declare equivalence only when the test rejects the null hypothesis in all the periods of interest. Figure~\ref{fg:tost} is adapted from \citet{hartman2018equivalence} and shows the difference between the usual $t$ test and the TOST. The null hypothesis is rejected when the test statistic's value falls in the shaded region on both panels. But only the shaded region on the right reflects the magnitude of Type-I error that we are trying to control for.

\FloatBarrier

An alternative approach is an equivalence $F$ test, which uses the same statistic as the $F$ test:
$$F = \frac{Ntr(Ntr-m-1)}{(Ntr-1)(m+1)} \bm{\delta}_{(-m:0)}'\Sigma_{\bm{\delta}_{(-m:0)}}\bm{\delta}_{(-m:0)}$$
where $\bm{\delta}_{(-m:0)}= (ATT_{-m},ATT_{-(m-1)},\dots, ATT_{0})'$ and $\Sigma_{\bm{\delta}_{(-m:0)}}$ is the covariance matrix of $\bm{\delta}_{(-m:0)}$. The key difference lies in that we impose a reversed null hypothesis for the equivalence test:
$$H_0: \quad \bm\delta_{(-m:0)}'\Sigma_{\bm\delta_{(-m:0)}}\bm\delta_{(-m:0)} > \kappa,$$
\citet{wellek2010testing} shows that under this hypothesis, the statistic converges to a non-central $F$-distribution $F(m+1, N_{tr}-m-1, N_{tr}\kappa^2)$, where $N_{tr}\kappa^2$ is the distribution's centrality parameter. The null is considered rejected (hence, equivalence holds) when the statistic's value is smaller than the $100\alpha$th percentile of the distribution. When the absolute values of all the $ATT_{s}$ are smaller, there will be a higher chance to reject the null . Based on the discussion in \citet{wellek2010testing} and simulation results, we recommend to set $\kappa = 0.6$.

\begin{figure}[!th]
\caption{F distribution under the Two Tests}\label{fg:fvse}
\centering
\begin{minipage}{0.85\linewidth}{
\begin{center}
\includegraphics[trim=0 1em 0 4em,clip,width = 0.8\textwidth]{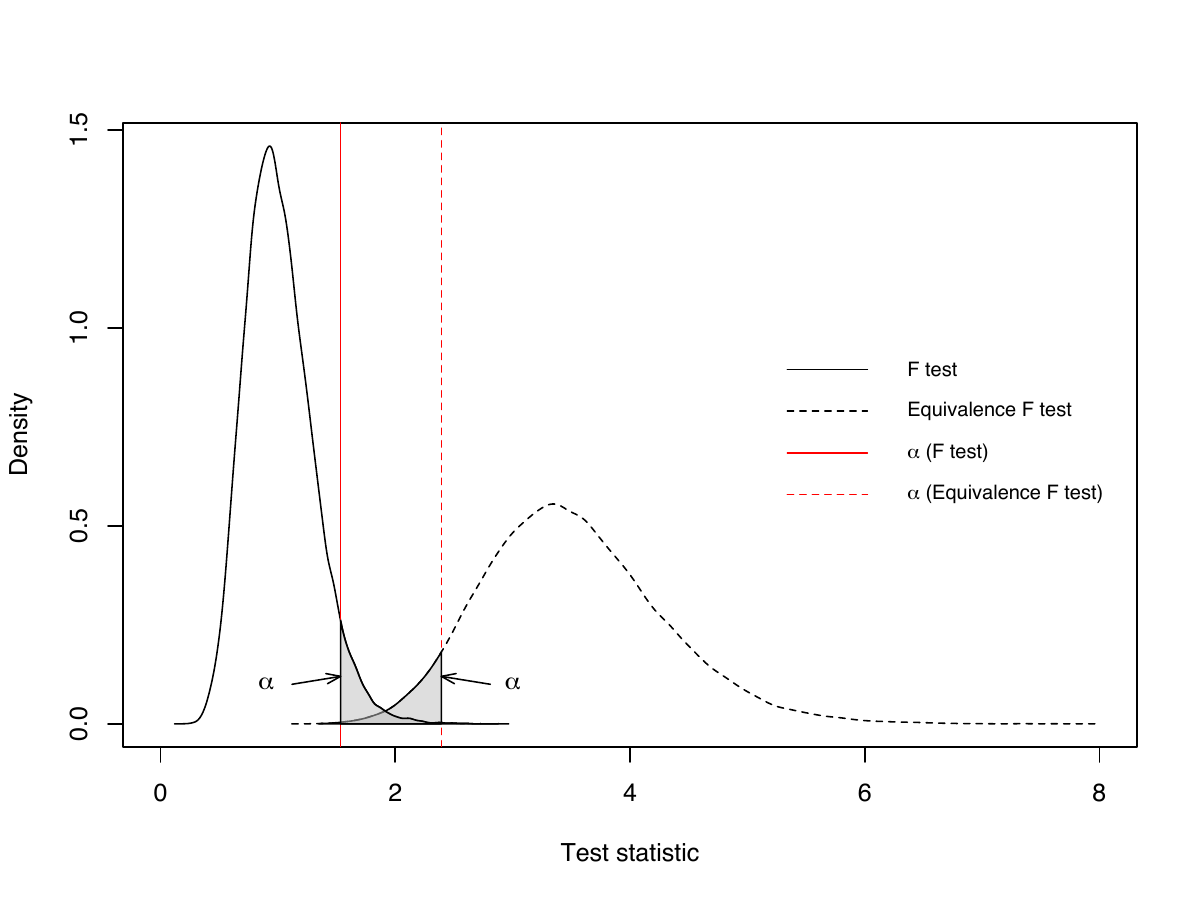}\\
\end{center}
}
\footnotesize\textbf{Note:} The above figure plots the distribution of the test statistic under the null of the $F$ test and its distribution under the null of the equivalence test. The shaded areas represent the size of the two tests ($\alpha = 0.05)$. 
\end{minipage}
\end{figure}

We compare the distribution of the $F$ test with that of the equivalence test in Figure~\ref{fg:fvse} above. The solid black curve represents the distribution of the test statistic under the null of the $F$ test. The dotted black curve represents its distribution under the null of the equivalence test. We reject the null under the former if the value of the test statistic falls on the right side of the solid red line and reject the null under the latter if it falls on the left side of the dotted red line, i.e., the equivalence threshold. 

In the above case (with a chosen equivalence threshold of 0.6), the equivalence test is more lenient than the $F$ test: when the test statistic falls between the two red lines, we reject the null under the $F$ test (suggesting inequivalence), but also reject the null under the equivalence test (declaring equivalence). Therefore, the equivalence test has the same advantages as the TOST does. However, because its threshold is less intuitive and harder to interpret, we choose the TOST as the primary approach to conduct the equivalence test. 

\bigskip
\clearpage

\subsection{Discussion on the No Carryover Effects Assumption}\label{sc:no-co}

The violation of no carryover effects assumption is a violation of the stable unit treatment value assumption (SUTVA) along the temporal dimension, i.e., the potential outcome of unit $i$ in period $t$ could be affected by its own treatment status in earlier periods. Therefore, the presence of carryover effects is equivalent to temporal interference and does not imply failure of the strict exogeneity assumption. 

As we briefly discuss in the main text, violations of the no carryover effects assumption is not a concern under staggered adoption; it is a concern when the treatment switches on and off for some unit. In the latter case, the carryover effects can be seen as a result of special ``time-varying confounders.'' Hence, we can use the placebo test introduced earlier in paper to gauge whether they exist. When the average prediction error in those periods deviate from zero, we obtain a piece of evidence that the assumption is likely invalid (of course, it is also possible it is a result of some temporal shocks unrelated to carryover effects). A potential solution in this scenario is to re-code the treatment such that we label a few periods after the treatment's ending as under treatment to allow the carryover effects to fully present themselves.
\begin{figure}[!ht]
\caption{Illustrating Carryover Effects under Staggered Adoption}\label{fg:carryover}
\vspace{0.5em}
\centering
\begin{minipage}{1\linewidth}{
\begin{center}
\hspace{1em}\includegraphics[width = 0.7\textwidth]{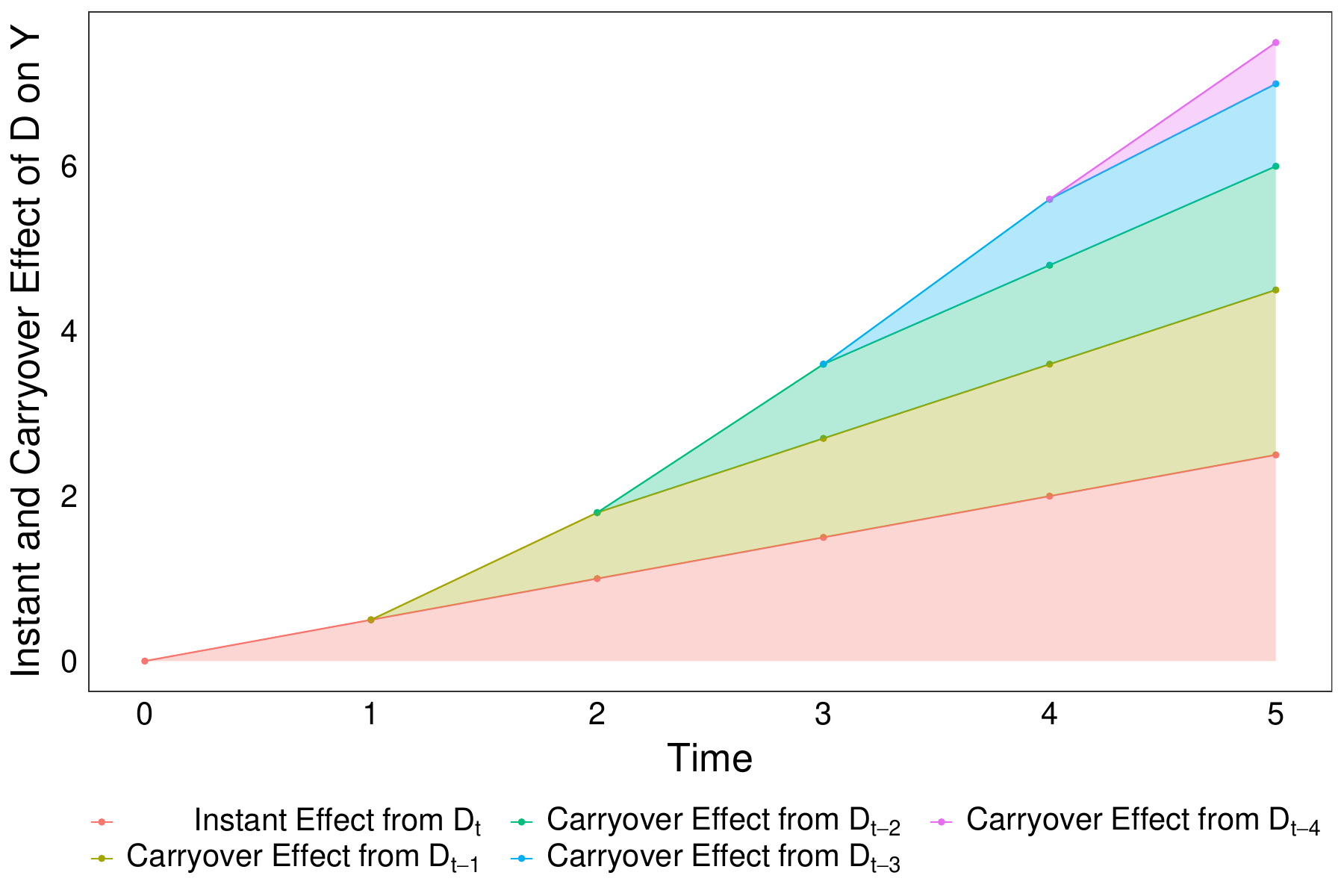}
\end{center}
\footnotesize\textbf{Note:} The above figures demonstrates a decomposition of $\delta_{it}$ in a hypothetical case under staggered adoption when carry over effects exist. The x-axis indicates the time relative the onset of the treatment.}
\end{minipage}\vspace{-0.5em}
\end{figure}

Under staggered adoption, however, we may not be concerned about the carryover effect because we can reinterpret $\delta_{it}$ as a combination of the instant effect of the current treatment (red area in Figure~\ref{fg:carryover}) and cumulative carryover effects of past treatments (other colored areas) on a treated unit $i$ relative to its potential outcome history under the never-treated condition.

\clearpage

\subsection{Procedures for the Diagnostic Tests}

\begin{figure}[!ht]
\caption{Illustrating the Diagnostic Tests}\label{fg:tests}
\centering
\begin{minipage}{0.95\linewidth}
\begin{center}
\hspace{-2em}
\subfigure[Placebo Test]{\includegraphics[width = 0.45\textwidth]{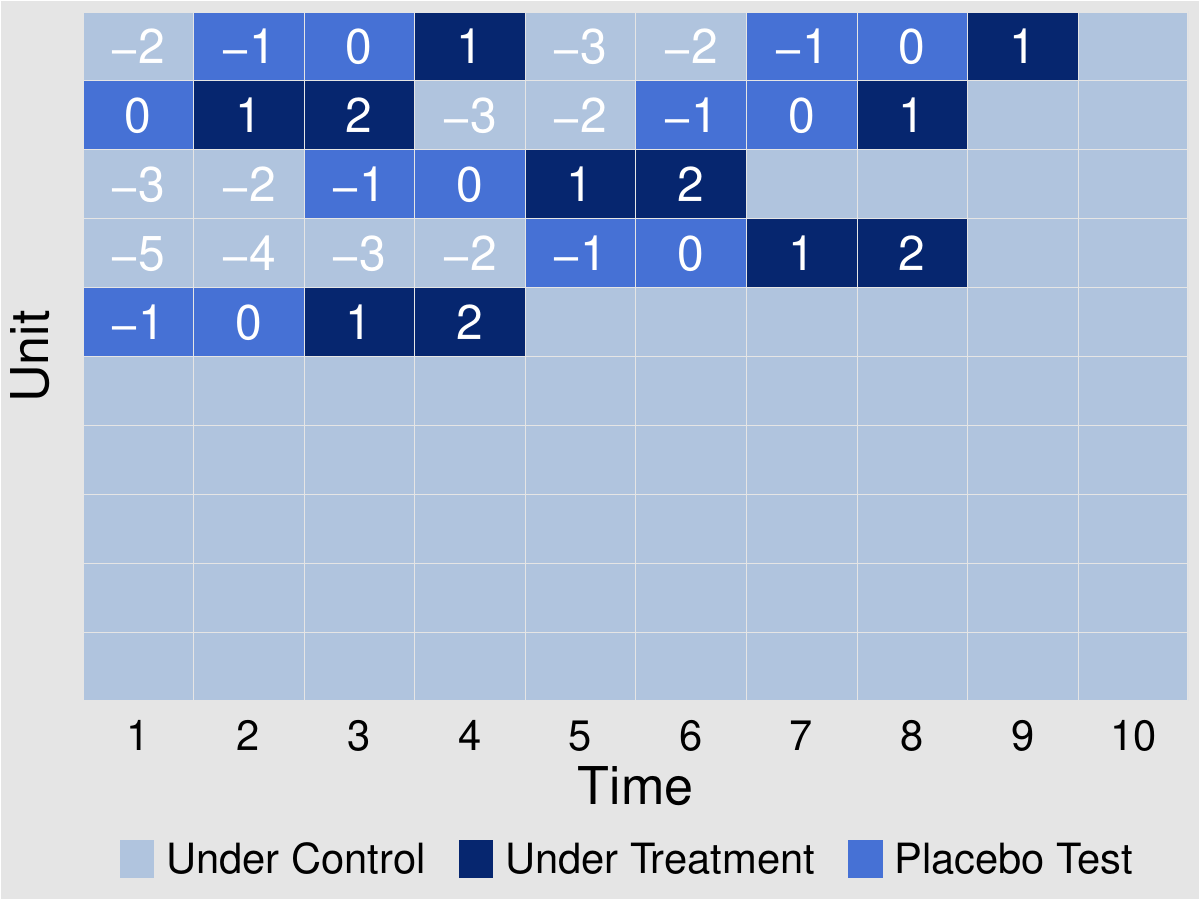}}\hspace{2em}
\subfigure[Test for No Carryover Effects]{\includegraphics[width = 0.45\textwidth]{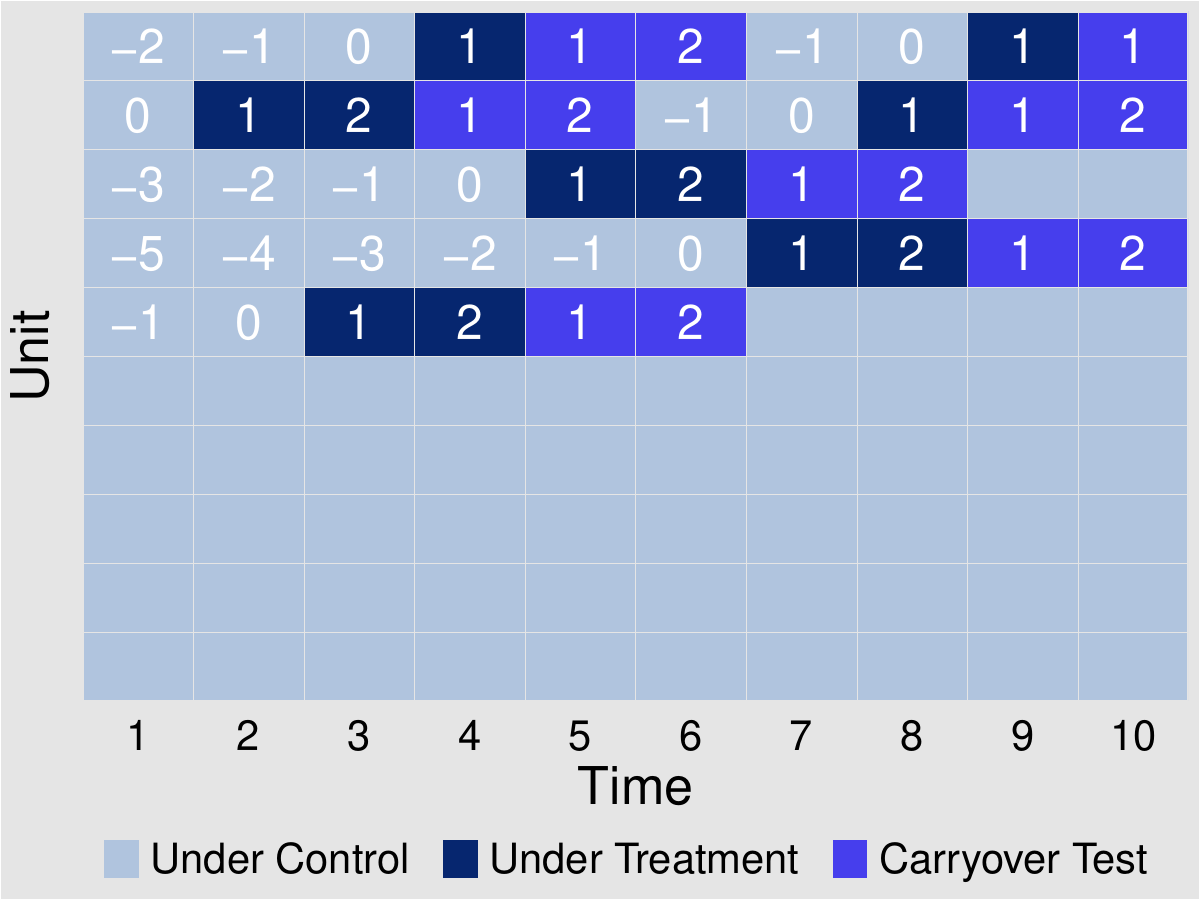}}\\
\end{center}
\footnotesize\textbf{Note:} The above figures show how observations in periods relative to the onset or ending of the treatment are being used in the diagnostic tests. 
\end{minipage}
\vspace{-1em}
\end{figure}

We describe the procedures for the diagnostic tests below. Figure~\ref{fg:tests}(a) and (b) illustrate how the placebo test and test for no carryover effects are performed, respectively.

\paragraph{Placebo test.} 
\begin{enumerate}
    \item If the equivalence approach is being used, specify a significant level $\alpha$, like $\alpha = 0.05$, the length of placebo period $l$, and two equivalence thresholds $-\theta_2$ and $\theta_1$ such that 
    $-\theta_2 < 0 < \theta_1$. In in Figure~\ref{fg:tests}(a), $l = 2$ (Periods 0 and -1).
    
    \item Regard $l$ observations immediately before the onset of the treatment for each treated unit ($C_{i} = 1$) as observations under the placebo, remove them when fitting the model, and obtain an estimate for average ``treatment effect'' in placebo periods denoted as $\widehat{ATT}^P$.
    
    \item Use bootstrap or jackknife method to obtain a 
    $(1 - \alpha)$ or $(1 - 2\alpha)$ confidence interval for $\widehat{ATT}^P$. Denote 
    $\widehat{ATT}^P_u$ the upper bound and 
    $\widehat{ATT}^P_l$ the lower bound.
    
    \item Test $\widehat{ATT}^P$ against the null hypothesis using either the DIM approach or the equivalence approach.  
\end{enumerate}

\paragraph{Test for no pretrend.} The test for no pretrend is a special case of the placebo test when $l = 1$. It conducts the placebo test repeatedly by removing the $j$'th period before the treatment starts, $j = 1, 2, \cdots$ (in Figure~\ref{fg:tests}(a), the periods marked with 0, -1, -2, -3).

\paragraph{Test for no carryover effects.} 
\begin{enumerate}
    \item If the equivalence approach is being used, specify a significant level $\alpha$, like $\alpha = 0.05$, the length of carryover effect period $l$, and two equivalence thresholds $-\theta_2$ and $\theta_1$ such that $-\theta_2 < 0 < \theta_1$. In in Figure~\ref{fg:tests}(b), $l = 2$ (Periods 1 and 2 in purple).
    
    \item Regard the first $l$ observations after the ending of 
    the treatment as periods potentially under carryover effects, remove them when fitting the model, and obtain an estimate for average treatment 
    effect in carryover effect periods denoted as $\widehat{ATT}^C$.
    
    \item Use bootstrap or jackknife method to obtain a 
    $(1 - \alpha)$ or $(1 - 2\alpha)$ confidence interval for $\widehat{ATT}^C$. Denote 
    $\widehat{ATT}^C_u$ the upper bound and 
    $\widehat{ATT}^C_l$ the lower bound.

    \item Test $\widehat{ATT}^C_u$ against the null hypothesis using either the DIM approach or the equivalence approach.  
\end{enumerate}

\begin{figure}[!ht]
\caption{Illustrating the Diagnostic Tests\\Allowing Limited Carryover Effects (Up to Two Periods)}\label{fg:tests2}
\centering
\begin{minipage}{0.95\linewidth}
\begin{center}
\hspace{-2em}
\subfigure[Placebo Test]{\includegraphics[width = 0.45\textwidth]{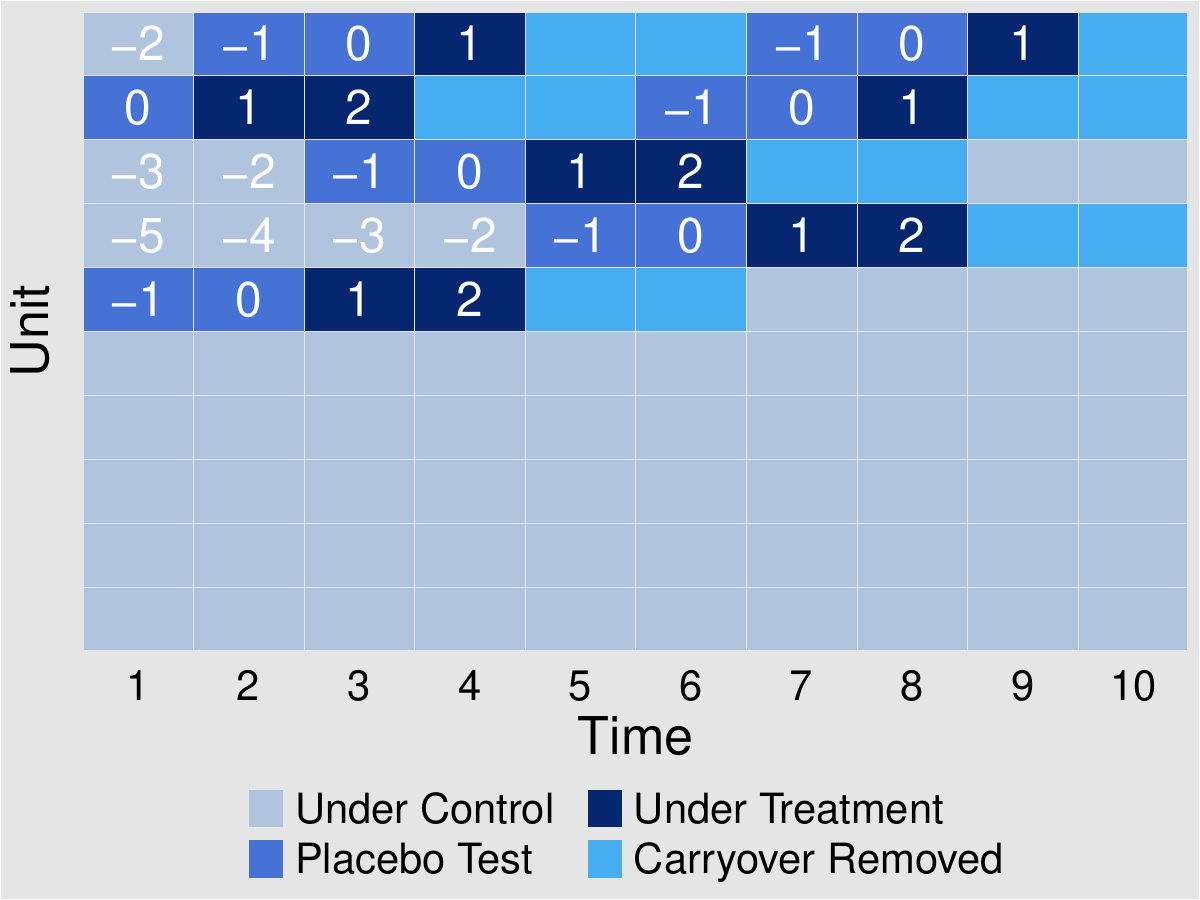}}\hspace{2em}
\subfigure[Test for No Carryover Effects]{\includegraphics[width = 0.45\textwidth]{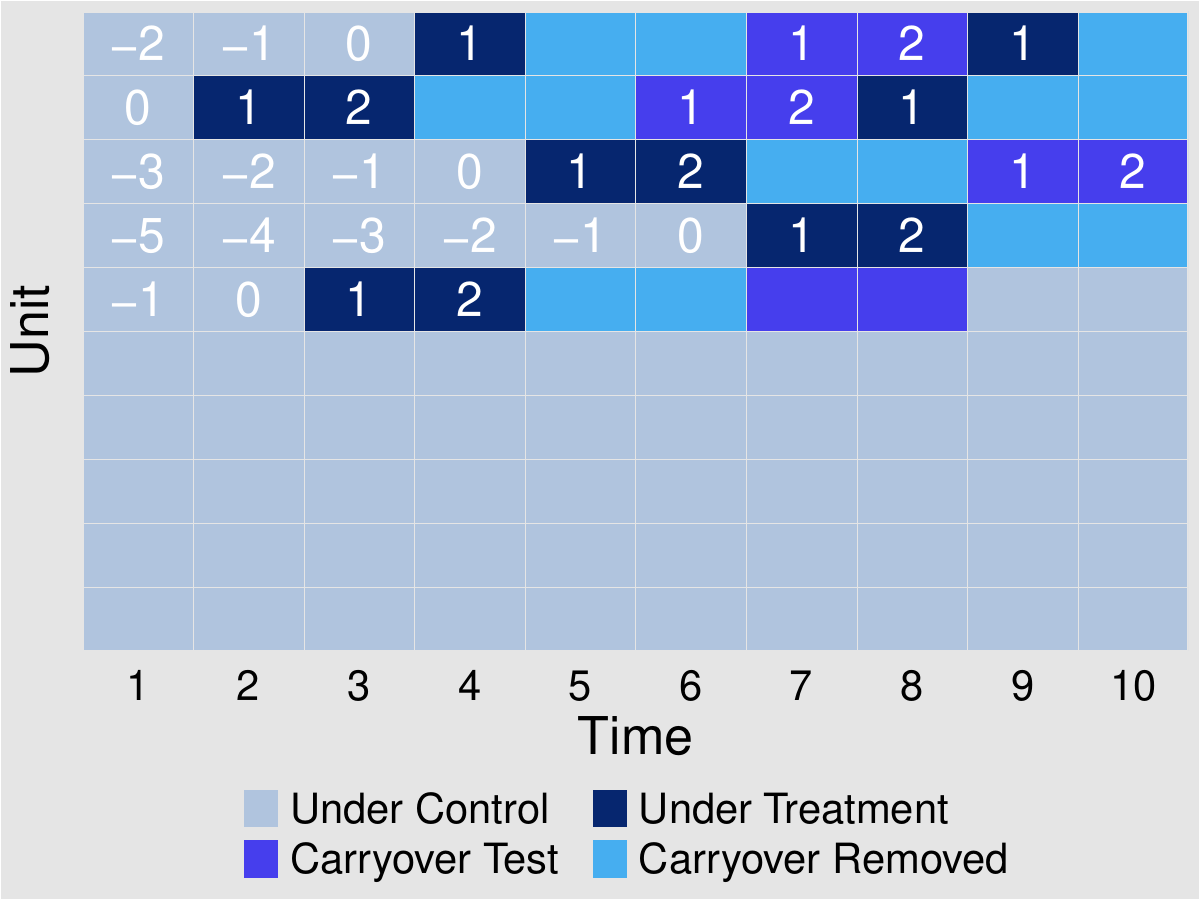}}\\
\end{center}
\footnotesize\textbf{Note:} The above figures show how observations in periods relative to the onset and ending of the treatment are being used in the diagnostic tests. 
\end{minipage}
\vspace{-1em}
\end{figure}

\paragraph{Allowing limited carryover effects.} Sometimes, as in \citet{FM2015-yy}, researchers may find evidence for limited carryover effects. Researchers can thus specify the number periods after the treatment's ending that allow carryover effects can take place. After removing those periods, researchers can then proceed to re-estimate the ATT and re-conduct the diagnostic tests. Figure~\ref{fg:tests2} illustrate the placebo test (a) and test for no carryover effects (b) when two periods after the treatment's ending (in sky blue) are removed. We apply this method in drawing Figure 10 in the main text. 

\clearpage

\clearpage

\section{Proofs}

\subsection{Unbiasedness and Consistency of FEct and IFEct}

Denote the number of all observations, the number of observations with $D_{it} = 1$, and the number of observations with $D_{it} = 0$ as $n$, $n_{\mathcal{M}}$, and $n_{\mathcal{O}}$, respectively. Under FEct, our Assumptions 1, 2, and 3 lead to the following model specification:
  \begin{align}
   Y_{it} = \mathbf{X}_{it}'\beta & + \alpha_{i} + \xi_{t} +
  \varepsilon_{it},  \hspace{1mm} (i, t) \in \mathcal{O},   \label{aeq.sp}\\
  & \sum_{D_{it} = 0}\alpha_{i} =  \sum_{D_{it} = 0} \xi_t, \notag \\
 \varepsilon_{it} \perp \{D_{js}, & \mathbf{X}_{js} , \alpha_{j}, \xi_{s}\} \text{ for any } i,j \in \{1,2,\dots,N\} \text{ and } s,t \in \{1,2,\dots,T\}. \notag
   \end{align}   
The data we use to estimate these parameters constitute an unbalanced panel since we are not using observations whose $D_{it}=1$. Following \citet{wansbeek1989estimation}, we rearrange the observations so that data on $N$ units ``are ordered in $T$ consecutive sets," thus the index $t$ ``runs slowly" and $i$ ``runs quickly." Denote the number of untreated units in period $t$ as $N_t$, then $N_t \leq N$ and $\sum_{t=1}^{T} N_t = n_{\mathcal{O}}$, the number of untreated observations in the dataset. Similarly, denote the number of periods in which unit $i$ is untreated as $T_i$. Then $T_i \leq T$ and $\sum_{i=1}^{N} T_i = n_{\mathcal{O}}$.  Let $M_t$ be the $N_t \times N$ matrix where row $i$ equals to the corresponding row in the unit matrix $I_{N}$ if $i$ is observed in period $t$. Then we can rewrite Equation (\ref{aeq.sp}) in the matrix form:
\begin{equation*}
Y = \mathbf{X} \beta + \Delta (\alpha, \xi)' + \varepsilon
\end{equation*}
  where $\mathbf{X} = (\mathbf{x}_{11}, \mathbf{x}_{21}, \cdots, \mathbf{x}_{NT})'$ is a $n_{\mathcal{O}} \times K$ matrix, $\iota_n$ denotes the $n_{\mathcal{O}}$-dimension vector consisted of 1s, $\Delta = (\Delta_{1}, \Delta_{2})$,  $\Delta_{1} = \begin{pmatrix}
  \mathbf{M}_1 \\
  \mathbf{M}_2 \\
  \vdots \\
  \mathbf{M}_T
  \end{pmatrix}$, and $\Delta_{2} = \begin{pmatrix}
  \mathbf{M}_1 \iota_{N} & & &    \\
  & \mathbf{M}_2 \iota_{N} \\
  & & \ddots \\
  & & & \mathbf{M}_T \iota_{N}
  \end{pmatrix}$. \\

Under IFEct, the model specification that satisfies Assumptions 1, 2, and 3 has the following form:
   \begin{align}
 &  Y_{it} = \mathbf{X}_{it}'\beta + \mathbf{\lambda}_{i}^{'} f_{t} + \alpha_{i} + \xi_{t} + \varepsilon_{it},  \hspace{1mm} D_{it}=0, \\
 & \sum_{D_{it} = 0}\alpha_{i} = 0,  \sum_{D_{it} = 0} \xi_t = 0, \mathbf{\Lambda}' \mathbf{\Lambda} = diagnal, \mathbf{F}'\mathbf{F} / T = \mathbf{I}_r,  \notag \\
 & \varepsilon_{it} \perp \{D_{js}, \mathbf{X}_{js} , \alpha_{j}, \xi_{s}, \mathbf{\lambda}_{j}, f_{s}\} \text{ for any } i,j \in \{1,2,\dots,N\} \text{ and } s,t \in \{1,2,\dots,T\}. \notag
   \end{align}   
 in which $\mathbf{\Lambda} = \left[\lambda_1, \lambda_2, \hdots, \lambda_N \right]'$ and $\mathbf{F} = \left[f_1, f_2, \hdots, f_T \right]'$. 
From now on we denote the projection matrix of matrix $\mathbf{A}$ as $P_{\mathbf{A}}$  and the corresponding residual-making matrix as $Q_{\mathbf{A}}$.

Proving the ATT estimator's consistency requires some regularity conditions. First, following \citet{Bai2009} and \citet{xu2017generalized}, we assume that the error terms have weak serial dependence:

\noindent\textbf{Weak serial dependence:}\\
1. $E\left[\varepsilon_{it} \varepsilon_{is} \right] = \sigma_{i,ts}, |\sigma_{i,ts}| \leq \bar{\sigma}_i$ for all $(t,s)$ such that $\frac{1}{N}\sum_i^N \bar{\sigma}_i < M$. \\
2. For every $(t,s)$, $E\left[N^{-1/2} \sum_i^N \varepsilon_{it} \varepsilon_{is} - E\left[\varepsilon_{it} \varepsilon_{is} \right] \right]^4 \leq M$. \\
3. $\frac{1}{NT^2} \sum_{t,s,u,v}\sum_{i,j}|cov\left[\varepsilon_{it} \varepsilon_{is}, \varepsilon_{ju} \varepsilon_{jv} \right]| \leq M$ and\\
$\frac{1}{N^2T} \sum_{t,s}\sum_{i,j,k,l}|cov\left[\varepsilon_{it} \varepsilon_{jt}, \varepsilon_{ks} \varepsilon_{ls} \right]| \leq M$.\\
4. $E\left[\varepsilon_{it} \varepsilon_{js} \right] = 0$ for all $i \neq j$, $(t,s)$. \\
These assumptions imply Assumption 3 in \citet{MoonWeidner2013} that $\frac{||\varepsilon||}{NT} \rightarrow 0$ as $N, T$ go to infinity. We also need some restrictions on parameters in the models:

\noindent\textbf{Restriction on parameters:}\\
1.  For each $t$, $\frac{N_t}{N} \rightarrow p_t$ as $N \rightarrow \infty$, where $0 \leq p_t < 1$ is a constant that varies with $t$.  \\
2. All entries of the matrix $E\left[\mathbf{x_{it}}\mathbf{x_{it}}' \right]$ is bounded by $M$. \\
3. For each unit $i$, all the covariates have weak serial dependence: $\sum_{t}^{T_i}X_{it, j} \times \sum_{t}^{T_i}X_{it, k} \leq M$ for any $(k, j)$. \\
4. Define $W(\lambda)$ as $\{\frac{1}{N}tr \left(\mathbf{x}^{'}_{k_1}Q_{\lambda}\mathbf{x}_{k_2}Q_{\mathbf{F}} \right)\}_{K \times K}$ and $w(\lambda)$ as the smallest eigenvalue of $W(\lambda)$. Define $W(f)$ as $\{\frac{1}{N}tr \left(\mathbf{x}_{k_1}Q_{f}\mathbf{x}^{'}_{k_2}Q_{\mathbf{\Lambda}} \right)\}_{K \times K}$ and $w(f)$ as the smallest eigenvalue of $W(f)$. Then either $\lim_{N,T \rightarrow \infty} min_{\lambda}\text{ } w(\lambda) > 0$, or $\lim_{N,T \rightarrow \infty} min_{f}\text{ } w(f) > 0$ holds. 

The last restriction comes from \citet{MoonWeidner2013} for the consistency of the IFEct model.

\begin{lemma}
Under model specification (A1) and regularity conditions, all the following limits\footnote{All the convergences here are convergence in probability.} exist: 
$(a) \lim_{N \rightarrow \infty} \frac{\mathbf{X}'\mathbf{X}}{N}$, 
$(b) \lim_{N \rightarrow \infty} \frac{\mathbf{X}'\epsilon}{N}$, 
$(c) \lim_{N \rightarrow \infty} \frac{\mathbf{X}'\Delta_2}{N}$, 
$(d) \lim_{N \rightarrow \infty} \frac{\Delta_2'\Delta_2}{N}$, 
$(e) \lim_{N \rightarrow \infty} \frac{\Delta_1'\Delta_1}{N}$,\\ 
$(f) \lim_{N \rightarrow \infty} \frac{\mathbf{X}'\Delta_1 diag\{\frac{1}{T_i}\}\mathbf{X}\Delta_1'}{N}$, where $diag\{\frac{1}{T_i}\}$ is a diagonal matrix with $\frac{1}{T_i}$ being the ith entry on the diagonal. 
\end{lemma}
\begin{proof} We start from proving (a). When the regularities conditions are satisfied, we can apply the weak law of large numbers:
\begin{align*}
\lim_{N \rightarrow \infty} \frac{\mathbf{X}'\mathbf{X}}{N} = \lim_{N \rightarrow \infty} \frac{\sum_{i}^{N} \sum_{t}^{T_i} \mathbf{x}_{it}\mathbf{x}_{it}^{'}}{N} =  \frac{\sum_{i}^{N} \sum_{t}^{T_i}
E\left[\mathbf{x}_{it}\mathbf{x}_{it}^{'}\right]}{N} = \bar{T}_i  E\left[\mathbf{x}_{it}\mathbf{x}_{it}' \right]
\end{align*}
which is bounded by $\bar{T}_i M$.
Similarly,
\begin{align*}
\lim_{N \rightarrow \infty} \frac{\mathbf{X}'\mathbf{\varepsilon}}{N} = \bar{T}_i  E\left[\mathbf{x}_{i,t}\mathbf{\varepsilon_{i,t}}' \right] = \mathbf{0}_{NT \times 1}
\end{align*}
For (c), we know that
\begin{align*}
\lim_{N \rightarrow \infty} \frac{\mathbf{X}'\Delta_2}{N} & = \lim_{N \rightarrow \infty} \frac{\sum_{i}^{N} \sum_{t}^{T_i} \mathbf{x}_{it}\Delta_{2, it}^{'}}{N}   =  \frac{\sum_{i}^{N} E\left[\sum_{t}^{T_i} \mathbf{x}_{it}\Delta_{2, it}^{'}\right]}{N} = \frac{\sum_{i}^{N} E\left[\mathbf{A}_i \right]}{N}
\end{align*}
where $\mathbf{A}_i$ is a $K \times T$ matrix, and the $t$th column of $\mathbf{A}_i$ equals to $\mathbf{0}_{K \times 1}$ when $D_{it} = 1$ and equals to $\mathbf{x}_{it}$ when $D_{it} = 0$. Clearly the limit exists under regularity conditions.

(d) and (e) are obvious.
For (f), 
\begin{align*}
& \lim_{N \rightarrow \infty} \frac{\mathbf{X}'\Delta_1 diag\{\frac{1}{T_i}\}  \mathbf{X}\Delta_1'}{N} \\
= & \lim_{N \rightarrow \infty} \frac{\sum_i^N \{\sum_{t}^{T_i}X_{it, j} \times \sum_{t}^{T_i}X_{it, k} / T_i\}_{K\times K}}{N} \\
= & \frac{\sum_i^N E\left[\frac{\mathbf{B}_i  }{T_i}\right]}{N}
\end{align*}
where $\mathbf{B}_i$ is a $K \times K$ matrix and the $(j, k)$th entry of $\mathbf{B}_i$ is $\sum_{t}^{T_i}X_{it, j} \times \sum_{t}^{T_i}X_{it, k}$. It is bounded by $\frac{M}{T_i}$.
(g) can be similarly proven.
\end{proof}

\begin{lemma}
Under model specification (A1) and regularity conditions, a. estimates of $\beta$, $\alpha_{i}$, and $\xi_{t}$ from equations (1) to (3), i.e. $\hat{\beta}$, $\hat{\alpha}_{i}$, and $\hat{\xi}_{t}$, are unbiased, and b. $\hat{\beta}$ and $\hat{\xi}_{t}$ are consistent as $N \rightarrow \infty$.
\end{lemma}

\begin{proof}
\noindent  As shown in \citet{wansbeek1989estimation}, $\beta$ under model specification (A1) can still be estimated using the within estimator. Multiplying both sides of demeaned equation (4) with $Q_{[\Delta]}$, we have $Q_{[\Delta]} Y = Q_{[\Delta]} \mathbf{X} \beta + Q_{[\Delta]} \varepsilon$, then it is easy to show that:
\begin{align*}
\hat{\beta} & = (\mathbf{X}' Q_{[\Delta]} \mathbf{X})^{-1}\mathbf{X}' Q_{[\Delta]} Y = (\mathbf{X}' Q_{[\Delta]} \mathbf{X})^{-1}\mathbf{X}' Q_{[\Delta]} [\mathbf{X} \beta + \Delta (\alpha, \xi)' + \varepsilon] \\
 & = \beta + (\mathbf{X}' Q_{[\Delta]} \mathbf{X})^{-1}X' Q_{[\Delta]}  \varepsilon
\end{align*}   
   Hence, $E[\hat{\beta}] = \beta + E[(\mathbf{X}' Q_{[\Delta]} \mathbf{X})^{-1}\mathbf{X}' Q_{[\Delta]}  \varepsilon] = \beta$, and $E[\hat{\mu}] = E[\bar{Y} - \bar{X} \hat{\beta}] = \bar{Y} - \bar{X}\beta = \mu$. \\ \\
   Similarly, 
   \begin{align*}
   Q_{[\mathbf{X}]}Y = Q_{[\mathbf{X}]}\Delta (\alpha, \xi)' + Q_{[\mathbf{X}]} \varepsilon
   \end{align*}
  The level of fixed effects, $(\alpha, \xi)'$, can also be estimated using ordinary least squares under the two constrains (2) and (3), which is equivalent to the following constrained minimization problem:
\begin{align*}
& Min_{\gamma} \quad (Q_{[\mathbf{X}]}Y - Q_{[\mathbf{X}]}\Delta \gamma)'(Q_{[\mathbf{X}]}Y - Q_{[\mathbf{X}]}\Delta \gamma) \\
& with \quad \Pi \gamma = 0
\end{align*}
where $\gamma = (\alpha, \xi)'$, and $\Pi_{1 \times (N + T)} = \begin{pmatrix}
  T_1, T_2, \hdots, T_N, -N_1, -N_2, \hdots, -N_T 
\end{pmatrix}$. \\
The solution to the minimization problem is given by the following equation: 
\begin{align*}
\Phi \begin{pmatrix}
  \hat{\gamma} \\
  \hat{\lambda}
\end{pmatrix} = 
\begin{pmatrix}
  \Delta' Q_{[\mathbf{X}]} \Delta, & \Pi' \\
  \Pi, & 0
\end{pmatrix} \begin{pmatrix}
  \hat{\gamma} \\
  \hat{\lambda}
\end{pmatrix} = \begin{pmatrix}
  \Delta' Q_{[\mathbf{X}]} Y \\
  0
\end{pmatrix}
\end{align*}
where $\lambda$ represents the corresponding Lagrangian multipliers. Finally, $\hat{\gamma} = (\hat{\alpha}, \hat{\xi})' = \Phi^{-1}_{11} \Delta' Q_{[\mathbf{X}]} Y$. Here $\Phi^{-1}_{11}$ is the upper-left block of $\Phi^{-1}$. 
For unbiasedness of these estimates, notice that
\begin{align*}
E(\hat{\alpha}, \hat{\xi})' & = E[\Phi^{-1}_{11} \Delta' Q_{[\mathbf{X}]} Y] \\
& = E[\Phi^{-1}_{11} \Delta' Q_{[\mathbf{X}]} \Delta (\alpha, \xi)'] \\
& = E[(I - \Phi^{-1}_{12}\Pi)(\alpha, \xi)'] \\
& = (\alpha, \xi)'.
\end{align*}
\noindent The second equality uses the fact that $Q_{[\mathbf{X}]}\mathbf{X} = 0$. The third equality builts upon the definition of $\Phi^{-1}_{11}$ and $\Phi^{-1}_{12}$: $\Phi^{-1}_{11} \Delta' Q_{[\mathbf{X}]}\Delta + \Phi^{-1}_{12}\Pi = I$. The last equality exploits the constraint $\Pi (\alpha, \xi)' = 0$.

Now, for $\hat{\beta}$ and $\hat{\xi}$, it is easy to show that:
\begin{align*}
\lim_{N \rightarrow \infty} \begin{pmatrix}
\hat{\beta} \\
\hat{\xi}
\end{pmatrix}  = & \begin{pmatrix}
\beta \\
\xi
\end{pmatrix}  + \lim_{N \rightarrow \infty}\left[\begin{pmatrix}
\mathbf{X}'\\
\Delta_{2}^{'}
\end{pmatrix} Q_{[\Delta_1]} (\mathbf{X}, \Delta_{2})\right]^{-1}\left[\begin{pmatrix}
\mathbf{X}'\\
\Delta_{2}^{'}
\end{pmatrix} Q_{[\Delta_1]} \varepsilon \right] \\
& \begin{pmatrix}
\beta \\
\xi
\end{pmatrix}  + \lim_{N \rightarrow \infty}\left[\begin{pmatrix}
\mathbf{X}'\\
\Delta_{2}^{'}
\end{pmatrix} Q_{[\Delta_1]} (\mathbf{X}, \Delta_{2}) / N \right]^{-1}\left[\begin{pmatrix}
\mathbf{X}'\\
\Delta_{2}^{'}
\end{pmatrix} Q_{[\Delta_1]} \varepsilon / N \right] 
\end{align*}
And,
\begin{align*}
\begin{pmatrix}
\mathbf{X}'\\
\Delta_{2}^{'}
\end{pmatrix} & Q_{[\Delta_1]} (\mathbf{X}, \Delta_{2}) = \begin{pmatrix}
\mathbf{X}'\mathbf{X} & \mathbf{X}'\Delta_{2} \\
\Delta_{2}^{'}\mathbf{X} & \Delta_{2}^{'}\Delta_{2}
\end{pmatrix} - 
\begin{pmatrix}
\mathbf{X}'\Delta_{1} \\
\Delta_{2}^{'}\Delta_{1}
\end{pmatrix} \left(\Delta_{1}^{'}\Delta_{1}\right)^{-1} (\mathbf{X}\Delta_{1}^{'}, \Delta_{2}\Delta_{1}^{'}) \\
= & \begin{pmatrix}
\mathbf{X}'\mathbf{X} & \mathbf{X}'\Delta_{2} \\
\Delta_{2}^{'}\mathbf{X} & \Delta_{2}^{'}\Delta_{2}
\end{pmatrix} - 
\begin{pmatrix}
\mathbf{X}'\Delta_{1} \\
\Delta_{2}^{'}\Delta_{1}
\end{pmatrix} diag\{\frac{1}{T_i}\} (\mathbf{X}\Delta_{1}^{'}, \Delta_{2}\Delta_{1}^{'})
\end{align*}
Using Lemma 1, we know that as $N \rightarrow \infty$, each term in the expression above will converge in probability to a fixed matrix. Using Slutsky's theorem,  $\left[\begin{pmatrix}
\mathbf{X}'\\
\Delta_{2}^{'}
\end{pmatrix} Q_{[\Delta_1]} (\mathbf{X}, \Delta_{2}) / N\right]^{-1}$ also converges to a fixed matrix. Similarly, we can show that $\left[\begin{pmatrix}
\mathbf{X}'\\
\Delta_{2}^{'}
\end{pmatrix} Q_{[\Delta_1]} \varepsilon / N \right]$ converges to $\mathbf{0}_{N \times 1}$ as $N \rightarrow \infty$, which leads to the consistency result.

On the contrary, $\hat{\alpha}_i$ is inconsistent when only $N \rightarrow \infty$ as the number of parameters changes accordingly.
\end{proof}

\begin{lemma}
Under model specification (A2) and regularity conditions, a. estimates of $\beta$, $\alpha_{i}$, $\xi_{t}$, $\lambda_{i}$, and $f_{t}$ from equations (5) to (9), i.e. $\hat{\beta}$, $\hat{\alpha}_{i}$, $\hat{\xi}_{t}$, $\hat{\lambda}_{i}$, and $\hat{f}_{t}$ are a. unbiased, and b. consistent as $N, T  \rightarrow \infty$.
\end{lemma}

\begin{proof}
\citet{MoonWeidner2013} show that all the coefficients of an IFE model can be estimated via a quasi maximum likelihood estimator and the estimates are unbiased as well as consistent when both $N$ and $T$ increase to infinity. We also know that estimates obtained from the EM algorithm converge to the quasi-MLE solution since it is the unique extrema. Hence the lemma holds due to properties of QMLE.
\end{proof}

\begin{proposition}[Unbiasedness and Consistency of FEct]: Under model specification (A1), as well as regularity conditions,
\begin{center}
    $\E[\widehat{ATT}_{s}] = ATT_{s}$; $\E[\widehat{ATT}] = ATT$;\\
    $\widehat{ATT}_{s} - ATT_{s} \overset{p}{\to} 0$; and $\widehat{ATT} - ATT \overset{p}{\to} 0$ as $N\to\infty$.
\end{center}\vspace{-1ex}
\end{proposition}

\begin{proof}
We only show the unbiasedness and consistency of $\widehat{ATT}_{s}$. For $\widehat{ATT}$, the proof is similar and omitted.
  \begin{align*}
    \widehat{ATT_{s}} & = \frac{1}{|\mathcal{S}|}
                        \sum_{(i,t) \in \mathcal{S}} \left( Y_{it} - \mathbf{X}_{it}'\hat\beta
                        - \hat\alpha_{i} - \hat\xi_{t} \right)\\
                      & = \frac{1}{|\mathcal{S}|}
                        \sum_{(i,t) \in \mathcal{S}} \left\lbrace
                        \mathbf{X}_{it}'(\beta-\hat{\beta}) +
                        (\alpha_{i} -\hat\alpha_{i})
                        + (\xi_{t} - \hat\xi_{t}) + \delta_{it} \right\rbrace  
  \end{align*}
  Using lemma 1, we know that
  \begin{align*}
   \E[\widehat{ATT_{s}}]  &= \frac{1}{|\mathcal{S}|}
                        \sum_{(i,t) \in \mathcal{S}} \left\lbrace
                        E[\mathbf{X}_{it}'(\beta-\hat{\beta})] +
                        E[\alpha_{i} -\hat\alpha_{i}]
                        + E[\xi_{t} - \hat\xi_{t}] + \delta_{it} \right\rbrace \\
                        & = \frac{1}{|\mathcal{S}|}
                        \sum_{(i,t) \in \mathcal{S}} 
                        \delta_{it} \\
                        & = ATT_{s}
  \end{align*}
Therefore, unbiasedness holds. For consistency, we know from the proof of lemma 2 that:
  \begin{align*}
 \lim_{N \rightarrow \infty}   \widehat{ATT_{s}} & =  \lim_{N \rightarrow \infty}\frac{1}{|\mathcal{S}|}
                        \sum_{(i,t) \in \mathcal{S}} \left( Y_{it} - \mathbf{X}_{it}'\hat\beta
                         - \hat\alpha_{i} - \hat\xi_{t} \right) \\
                      & = \lim_{N \rightarrow \infty} \frac{1}{|\mathcal{S}|}
                        \sum_{(i,t) \in \mathcal{S}}  \left(\delta_{it} + \alpha_{i} - \hat\alpha_{i}\right) + \bar{X}_{it}'(\beta - \hat\beta)  + (\xi_{t} - \hat\xi_{t})
  \end{align*}
Lemma 2 indicates that  as $N \rightarrow \infty$, $\hat\beta$ and $\hat\xi_{t}$ converge to $\beta$, and $\xi_t$, respectively. The only thing to be shown is $\lim_{N \rightarrow \infty} \frac{1}{\sum_{i} D_{it}}
                        \sum_{i,D_{it}=1}  \left(\alpha_{i} - \hat\alpha_{i}\right) = 0$. This is true since $E\left[\alpha_{i} - \hat\alpha_{i}\right] = 0$ and $\mathrm{Var}\left[\alpha_{i} - \hat\alpha_{i}\right]$ is bounded by the regularity conditions. Therefore $\lim_{N \rightarrow \infty}   (\widehat{ATT_{s}} - \frac{1}{|\mathcal{S}|}
                        \sum_{(i,t) \in \mathcal{S}}  \delta_{it}) = \lim_{N \rightarrow \infty}   (\widehat{ATT_{s}} - ATT_s) = 0$, consistency holds.
\end{proof}

\begin{proposition}[Unbiasedness and Consistency of IFEct]: Under model specification (A2), as well as regularity conditions,
 \begin{align*}
     \E[\widehat{ATT}_{s}] = ATT_{s} &\text{ and } \E[\widehat{ATT}] = ATT;\\
    \widehat{ATT}_{s} \overset{p}{\to}  ATT_{s} &\text{ and } 
    \widehat{ATT} \overset{p}{\to}  ATT \text{ as } N, T \to\infty.
 \end{align*}
\end{proposition}

\begin{proof} From lemma 3, we know that estimates for $\beta$, $\alpha_{i}$, $\xi_{t}$, $\lambda_{i}$, and $f_{t}$  are unbiased and consistent as $N, T  \rightarrow \infty$. Hence, $\widehat{ATT_{t}}$ and $\widehat{ATT}$ are also unbiased and consistent, following the same logic in the proof of Proposition 1.
\end{proof}

\subsection{FEct as a Weighting Estimator}

\begin{proposition}[FEct as a weighting estimator]: Under model specification (A1), and when there is no covariate,
  \begin{center}
    $\widehat{ATT}_{s} = \frac{1}{|\mathcal{S}|}
                        \sum_{(i,t) \in \mathcal{S}} [Y_{it} - \mathbf{W}^{(it)'} \mathbf{Y}_{\mathcal{O}}]$,
  \end{center} 
  where $\mathbf{W}^{(it)'} = \left(\dots, W_{js}^{(it)}, \dots\right)_{(j,s) \in \mathcal{O}}$ is a vector of weights that satisfy
  \begin{align*}
  \sum_{(s: (i, s)\in\mathcal{O})}W_{is}^{(it)} = 1, \sum_{(j: (j, t)\in\mathcal{O})}W_{jt}^{(it)} = 1, \sum_{(j: s \neq t, (j, s)\in\mathcal{O})}W_{js}^{(it)} = \sum_{(s: (j \neq i, (j, s)\in\mathcal{O})}W_{js}^{(it)} = 0.
  \end{align*}
\end{proposition}

\begin{proof} When there is no covariate,
\begin{align*}
\hat\alpha_{i} + \hat\xi_{t} & = \nu_{it}' \begin{pmatrix}
  \hat{\alpha} \\
  \hat{\xi}
\end{pmatrix} = \nu_{it}' \Phi^{-1}_{11} \Delta' \mathbf{Y}_{\mathcal{O}} = \mathbf{W}^{(it)'} \mathbf{Y}_{\mathcal{O}},
\end{align*}
where $\mathbf{W}^{(it)'}  = \nu_{it}' \Phi^{-1}_{11} \Delta' = \left(\dots, W_{js}^{(it)}, \dots\right)_{(j,s) \in \mathcal{O}}$.
Therefore,
\begin{align*}
    \widehat{ATT_{s}} & = \frac{1}{|\mathcal{S}|}
                        \sum_{(i,t) \in \mathcal{S}} \left( Y_{it} - \hat       
                       \alpha_{i} - \hat\xi_{t} \right) \\
                      & = \frac{1}{|\mathcal{S}|}
                        \sum_{(i,t) \in \mathcal{S}} \left( Y_{it} - \nu_{it}' \Phi^{-1}_{11} \Delta' \mathbf{Y}_{\mathcal{O}} \right) \\
                        & = \frac{1}{|\mathcal{S}|}
                        \sum_{(i,t) \in \mathcal{S}} (Y_{it} -  \mathbf{W}^{(it)'} \mathbf{Y}_{\mathcal{O}}).
  \end{align*}
Note that 
\begin{align*}
\hat\alpha_{i} + \hat\xi_{t} & = \mathbf{W}^{(it)'} \hat{\mathbf{Y}}_{D_{it} = 0} =\mathbf{W}^{(it)'} \begin{pmatrix}
 \vdots \\
  \hat{\alpha}_j + \hat{\xi}_s \\
  \vdots
\end{pmatrix}_{(j,s) \in \mathcal{S}}.
\end{align*}
Note that under the constraints, the fixed effects are independent to each other. Hence, for the equation to hold, we must have $\sum_{(s: (i, s)\in\mathcal{O})}W_{is}^{(it)} = 1$,  $\sum_{(j: (j, t)\in\mathcal{O})}W_{jt}^{(it)} = 1$, $\sum_{(j: s \neq t, (j, s)\in\mathcal{O})}W_{js}^{(it)} = \sum_{(s: (j \neq i, (j, s)\in\mathcal{O})}W_{js}^{(it)} = 0$, as on the left-hand side there are only $\hat\alpha_{i}$ and $\hat\xi_{t}$. We can see that the weight for each untreated observation $(j, s)$ is larger if there are fewer untreated observations in unit $j$ or period $s$. 
\end{proof} \\

We now use a simple example to compare weights under the FEct estimator and those under the classic FE estimator. Consider a dataset with 3 units and 4 periods and the treatment status is as follows: 

\begin{table}[!th]
\centering
\caption{Treatment Status}\label{atb:assign}
\setlength{\extrarowheight}{2pt}
\begin{tabular}{cc|cccc|c|}
  & \multicolumn{1}{c}{} & \multicolumn{5}{c}{Periods} \\
  & \multicolumn{1}{c}{} & \multicolumn{1}{c}{$1$}  & \multicolumn{1}{c}{$2$}  & \multicolumn{1}{c}{$3$} & \multicolumn{1}{c}{$4$}  & \multicolumn{1}{c}{$\bar{D}_{i.}$} \\\cline{3-7}
            & $1$ & $0$ & $0$ & $0$ & $0$ & $0$ \\ 
Units  & $2$ & $0$ & $0$ & $0$ & $1$ & $1/4$ \\\
            & $3$ & $0$ & $1$ & $1$  & $1$ & $3/4$ \\\cline{3-7}
            & $\bar{D}_{.t}$ & $0$ & $1/3$ & $1/3$  & $2/3$ & $5/6$\\\cline{3-7}
\end{tabular}
\end{table}
where $\bar{D}_{i.}$ is the average treatment status of unit $i$ and $\bar{D}_{.t}$ is the average treatment status of period $t$. We have 4 treated observations, $(2,4)$, $(3,2)$, $(3,3)$, and $(3,4)$, and 12 untreated ones. According to \citealt{de_Chaisemartin2018-iw}, the FE estimator, $\widehat{\delta}_{FE}$, equals to $\frac{1}{2}\widehat{\delta}_{24} + \frac{3}{10}\widehat{\delta}_{32} + \frac{3}{10}\widehat{\delta}_{33} - \frac{1}{10}\widehat{\delta}_{34}$, where each $w_{it} = \frac{\epsilon_{it}}{\sum_{it: (i, t) \in \mathcal{M}} \epsilon_{it}}$ and $\epsilon_{it} = D_{it} - \bar{D}_{i.} - \bar{D}_{.t} + \bar{D}$. The last estimate contributes negatively to the ATT estimate. Meanwhile, the FEct estimator, $\widehat{\delta}_{FEct}$, equals to $\frac{1}{4}\widehat{\delta}_{24} + \frac{1}{4}\widehat{\delta}_{32} + \frac{1}{4}\widehat{\delta}_{33} + \frac{1}{4}\widehat{\delta}_{34}$. All the four weights are equal and positive. Under both FE and FEct, $\widehat{\delta}_{it} = \mathbf{W}^{(it)'} \mathbf{Y}_{\mathcal{O}}$, where the weights satisfy constraints in Proposition 3 (and conditions 2 and 3 in \citealt{arkhangelsky2021double}) although they specific values may vary across the two estimators.

\clearpage
\section{Inferential Methods}

We rely on nonparametric block bootstrap and jackknife---both clustered at the unit level---to obtain uncertainty estimates for the treatment effect estimates. In the bootstrap procedure, we resample, with replacement, an equal number of units from the original sample. When a unit is drawn, its entire time series of data, including the outcomes, treatment status, and covariates, are replicated.  We obtain standard errors and confidence intervals of treatment effect estimates using conventional standard deviation and percentiles methods, respectively \citep{Efron1993}.

When the number of treated units is small (but bigger than one), jackknife resampling is an appealing alternative to bootstrapping \citep{miller1974, efron1981}. In each run, the procedure drops one unit (again, with its entire time series) and re-estimates the treatment effects. The variance of the ATT estimate is produced by $\widehat{\mathrm{Var}}(\widehat{ATT}) = \frac{N-1}{N}\sum_{i = 1}^{N} (\widehat{ATT}^{-i} - \overline{\widehat{ATT}})^{2}$, in which $\widehat{ATT}^{-i}$ is the ATT estimate from the sample in which the $i$'th unit is dropped and $\overline{\widehat{ATT}}$ is the average of jackknife estimates. We obtain confidence internals and $p$-values using a standard normal distribution. To address \citet{howmuch2004}'s well-known critique, both methods allow the error terms to be serially correlated but assume homoscedasticity of the errors across units. Both methods require the number of units $N$ to be large. 

We study the finite sample properties of the bootstrap and jackknife variance estimators using simulations. We simulate samples with $T = 20$ and $N = 50, 100$ and a staggered adoption treatment assignment mechanism. We assume that no time-varying confounders exist while the treatment effects are heterogeneous, hence, FEct is consistent for the ATT while the twoway fixed effects model is not. Following \citet{Arkhangelsky2019-lz}, we plot the quantiles of the distribution for standardized errors of the ATT estimates, i.e., $(\widehat{ATT} - ATT)/\widehat{\mathrm{Var}}(\widehat{ATT})^{1/2}$, based on 1,000 simulated samples against the quantiles of the standard normal distribution---a Quantile-Quantile plot (QQ plot)---using three combinations of estimators and inferential methods: (1) twoway fixed effects with block bootstrapped standard errors; (2) FEct with block bootstrapped standard errors; and (3) FEct with jackknife standard errors. If the ATT estimator is consistent and asymptotically normal and the chosen variance estimator precisely estimates its variance, the QQ plot should be very close to 45-degree line. 

\begin{figure}[!ht]
\caption{QQ Plots for Bootstrapped and Jackknife Standard Errors}\label{fg:sim.inference}
\centering
\begin{minipage}{0.95\linewidth}
\begin{center}
\hspace{-2em}
\subfigure[$N = 50$]{\includegraphics[width = 0.95\textwidth]{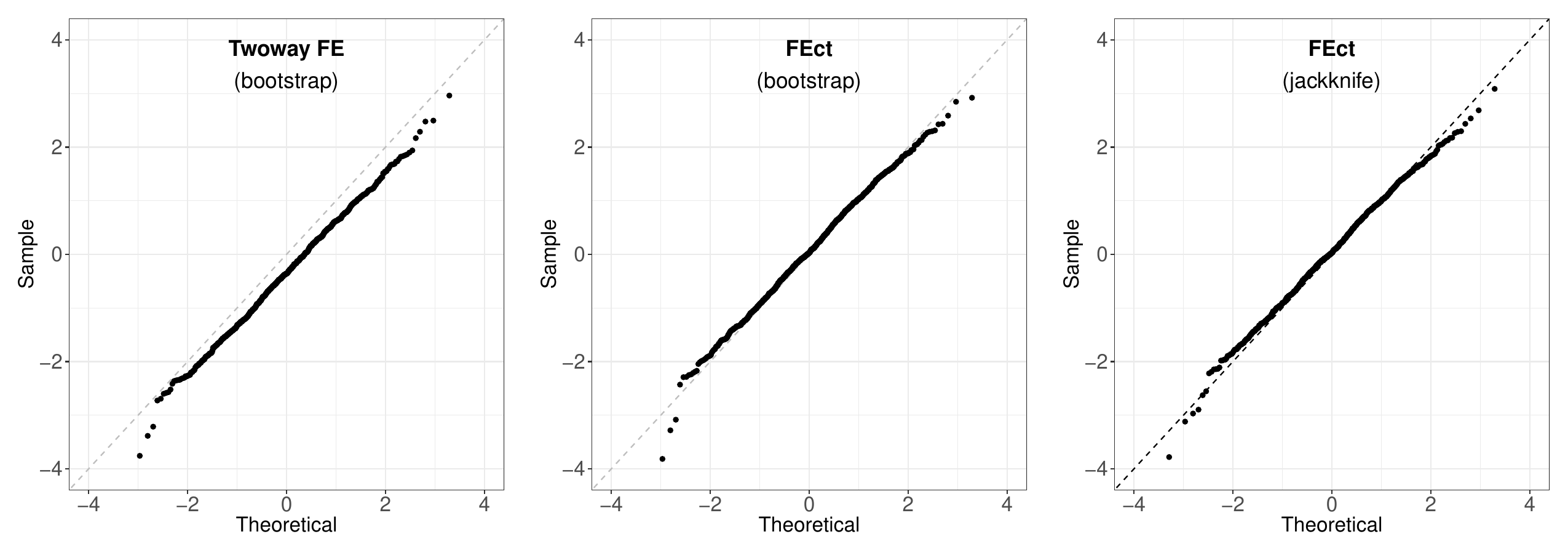}}\\
\subfigure[$N = 100$]{\includegraphics[width = 0.95\textwidth]{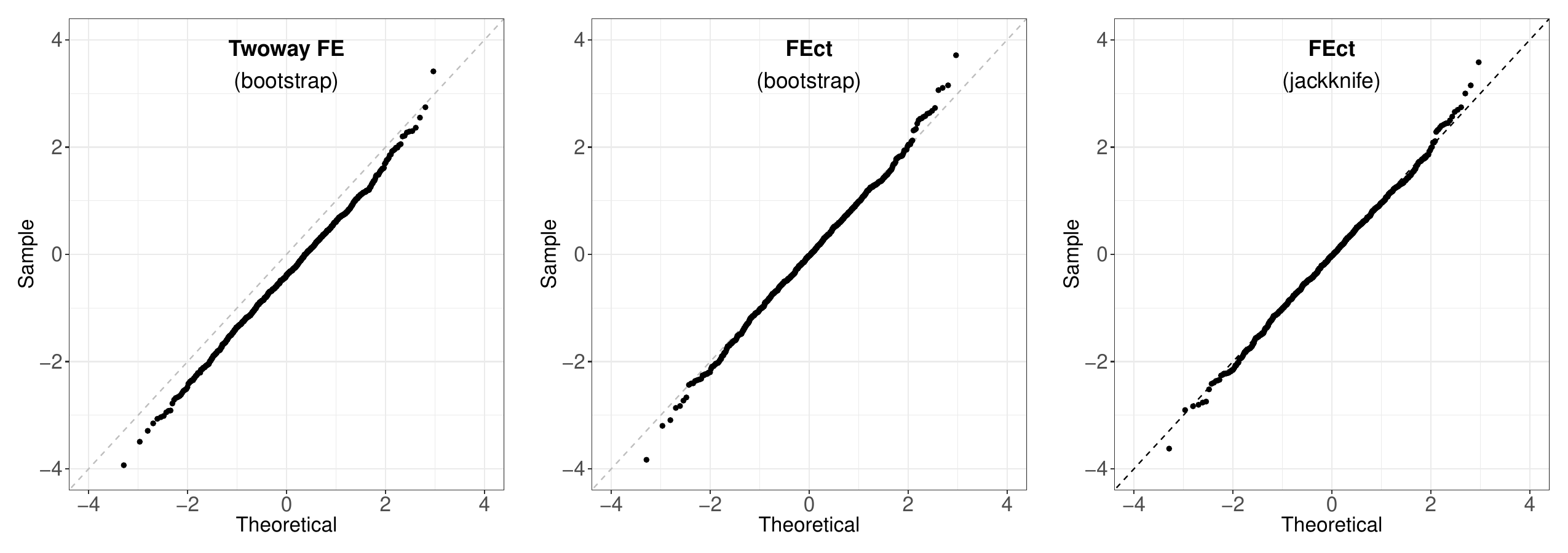}}
\end{center}
\footnotesize\textbf{Note:} The above figures show the standard Gaussian QQ plot of the standardize errors $(\widehat{ATT} - ATT)/(\widehat{\mathrm{Var}}(\widehat{ATT}))^{1/2}$ for the following combination of estimators and inferential methods: (1) twoway fixed effects with block bootstrapped standard errors; (2) FEct with block bootstrapped standard errors; and (3) FEct with jackknife standard errors, each aggregated from 1,000 simulations using samples with dimensions $T = 20$ and $N = 30$, $N = 50$ or $N = 100$. The 45-degree indicates the benchmark: consistent point estimates with perfectly calibrated Gaussian standard errors. 
\end{minipage}
\end{figure}

Figure~\ref{fg:sim.inference} presents the results. First, because the twoway fixed effects estimator is inconsistent, the QQ plot does not pass point $(0, 0)$. Second, when we apply the FEct estimator, both bootstrap and jackknife procedures precisely estimate the variance of an ATT estimate: both QQ plots are almost exactly on the 45-degree lines. Our finding suggests that the literature's recommendation to use block bootstrap or jackknife for variance estimation for panel models \citep{howmuch2004,cameron2015} can be extended to counterfactual estimators, such as FEct, IFEct, and MC.
\clearpage

\clearpage 

\section{Additional Monte Carlo Evidence}\label{sc:sim}

In this section, we report results of three sets of Monte Carlo exercises to demonstrate (1) the finite sample properties of the proposed inferential methods; (2) the differences between the IFEct and MC estimators, and (3) the main advantages of the equivalence test over the $F$ test. Before doing so, we first describe the data generating processes (DGP) of the simulated sample.

\subsection{Describing the DGP of the Simulated Example}

\begin{figure}[!th]
\caption{Treatment Status: The Simulated Samples}\label{fg:sim.treat}
\centering
\begin{minipage}{0.85\linewidth}{
\begin{center}
\includegraphics[width = 0.8\textwidth]{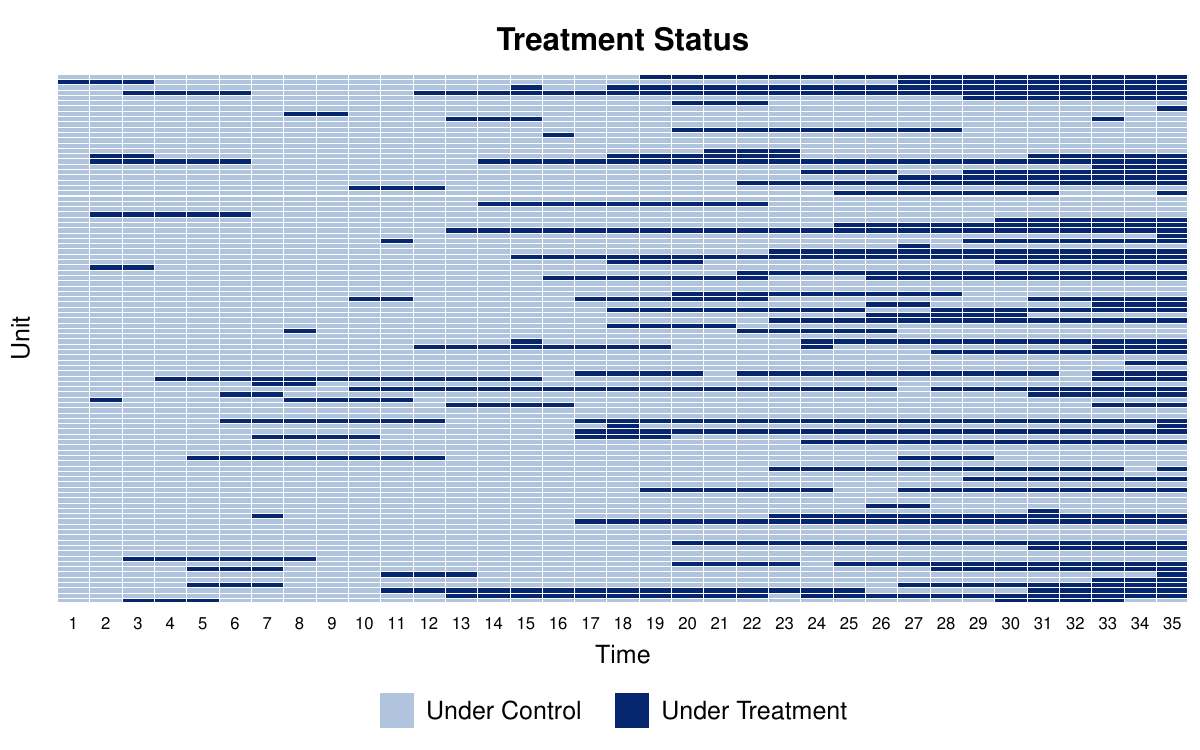}\\
\end{center}
}
\footnotesize\textbf{Note:} The above figure plots the treatment status of the simulated example, in which treatment reversal is allowed. The plot is made by the \texttt{panelView} package.
\end{minipage}
\end{figure}

\begin{figure}[!th]
\caption{Outcome Variable: The Simulated Samples}\label{fg:sim.outcome}
\centering
\begin{minipage}{0.85\linewidth}{
\begin{center}
\includegraphics[width = 0.8\textwidth]{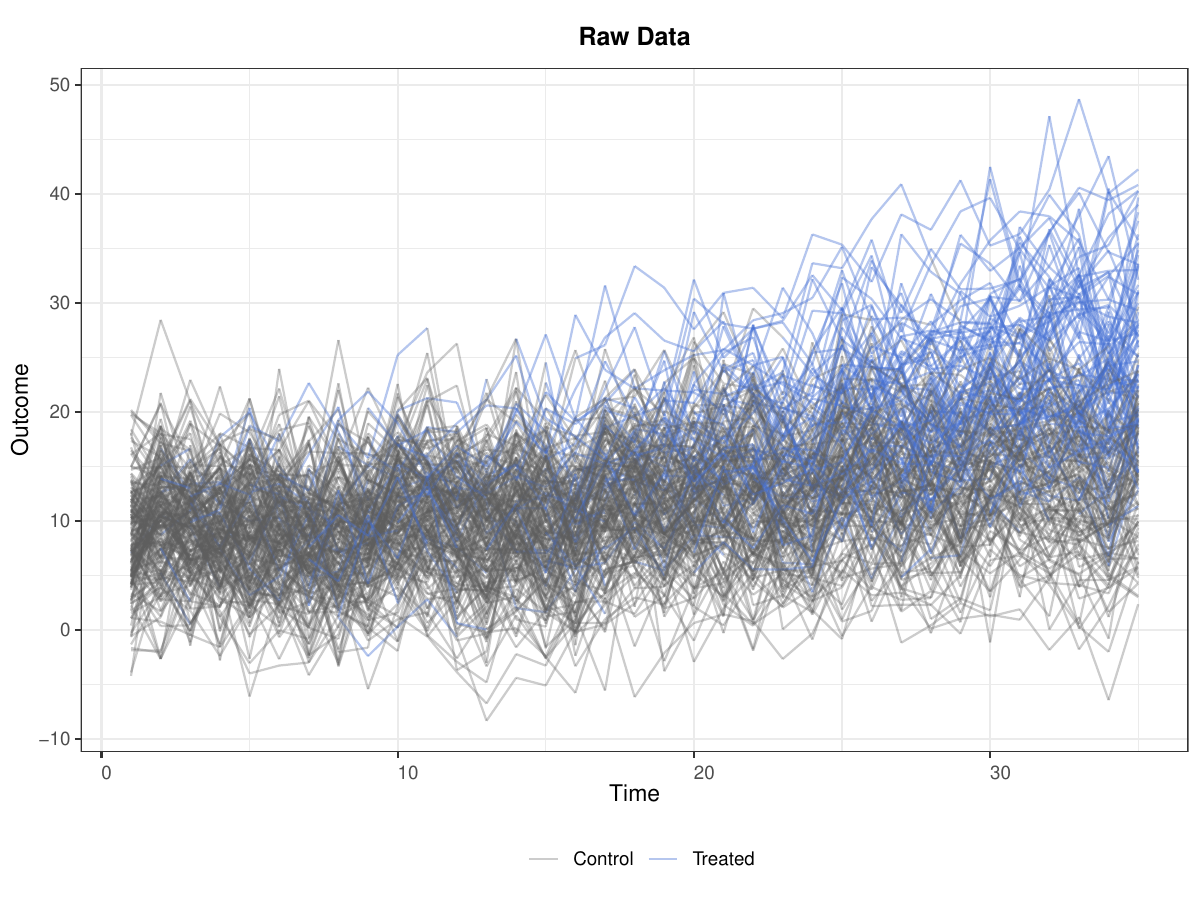}\\
\end{center}
}
\footnotesize\textbf{Note:} The above figure plots the outcome variable in the simulated example. The plot is made by the \texttt{panelView} package.
\end{minipage}
\end{figure}

We describe the DGP of the simulated example as follows. 
\begin{itemize}
    \item {\bf Outcome model}: $Y_{it}(0) = \delta_{it} D_{it} + 5 + 1 \cdot X_{it,1} + 3 \cdot X_{it,2} + 1.5 \lambda_{i1}\cdot f_{1t} + \lambda_{i2}\cdot f_{2t}+ \alpha_{i} + \xi_{t} + \varepsilon_{it}$, in which $f_{1t}$ is a linear trend plus a white noise: $f_{1t} = t + \nu_{t}$, and $\nu_{t} \overset{i.i.d}{\sim} N(0,1)$, then it is normalized to have variance 1. $f_{2t}$ is an i.i.d $N(0,1)$ white noise. Both $\lambda_{i1}$ and $\lambda_{i2}$ are i.i.d $N(0,1)$. Two covariates $X_{1,it}$ and $X_{2,it}$ are included in the model. They are both i.i.d. $N(0,1)$. Unit fixed effects $\alpha_{i}\sim N(0,1)$. Time fixed effects $\xi_{t}$ follows a stochastic drift. The error term $\varepsilon_{it}$ is also i.i.d. $N(0,2)$.
    \item {\bf Treatment effects}: $\delta_{it}= 0.4 s_{t} + e_{it}$, in which $s_{t}$ represents the number of periods since the latest treatment's onset and $e_{it}$ is i.i.d. $N(0,0.2^2)$. This means the expected value of the treatment effect gradually increases as a unit takes up the treatment, e.g. from 0.4 in the first period after receiving the treatment to 2.0 in the fifth period. 
    \item {\bf Treatment assignment}: denote $p_{it}$ the probability of getting treated for unit $i$ in period $t$: $logit(p_{it}) = -1 + 0.5 D_{t-1} + 0.5\lambda_{i}'f_{t} + 0.2\alpha_{i} + 0.2\xi_{t} + \mu_{it}$, in which $\mu_{it} \overset{i.i.d}{\sim} N(0,0.1^2)$.
\end{itemize}
Note that this DGP satisfies Assumptions~1-3. Figures~\ref{fg:sim.treat} and \ref{fg:sim.outcome} show the treatment status and outcome variable in the simulated example. 

\clearpage

\subsection{IFEct versus MC}

We compare the performance of the IFEct and MC estimators using DGPs similar to that of the simulated example: $Y_{it} = \delta_{it} D_{it} + 5 + \frac{1}{\sqrt{r}}\sum_{m=1}^{r} \lambda_{im} \cdot f_{mt} + \alpha_{i} + \xi_{t} + \varepsilon_{it}.$ We simulate samples of 200 units and 30 time periods, and all treated units receive the treatment at period 21 ($T_{0}= 20$). Following \citet{li2018inference}, we vary the number of factors $r$ from 1 to 9 and adjust a scaling parameter $\frac{1}{\sqrt{r}}$ such that the total contribution of all factors (and their loadings) to the variance of $Y$ remains constant. Our intuition is that IFEct (i.e., hard impute) performs better than MC (i.e., soft impute) when only a small number of factors are present and each of them exhibits relatively strong signals while MC outperforms IFEct when a large set of weak factors exist. In other words, MC should handle sparsely distributed factors better than parametric models like IFEct. 
\begin{figure}[!ht]
\caption{Monte Carlo Exercises: IFEct vs. MC}\label{fg:sim.estimators}
\centering
\begin{minipage}{0.95\linewidth}
\begin{center}
\includegraphics[width = 0.8\textwidth]{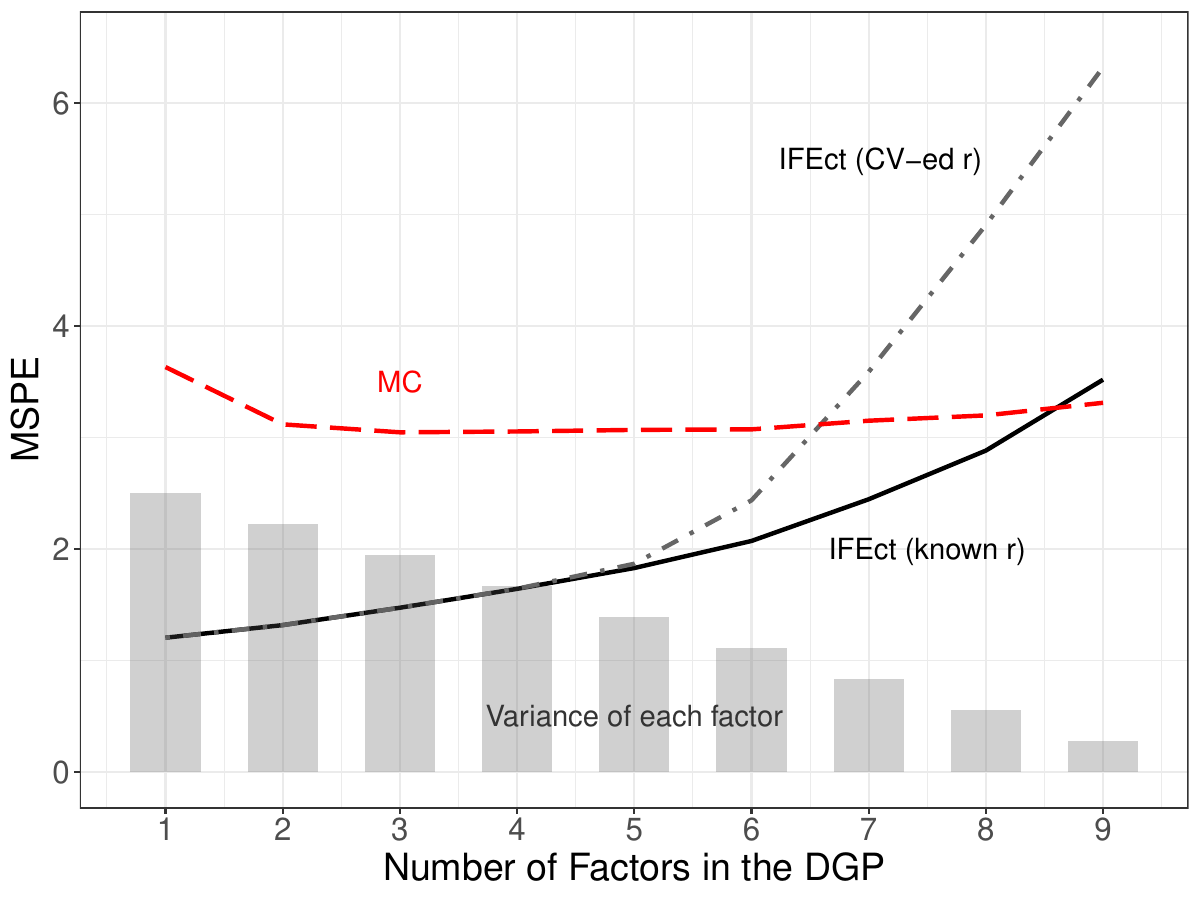}
\end{center}
\footnotesize\textbf{Note:} The above figure compares the mean squared prediction errors (MSPEs) for treated counterfactuals using the IFEct and MC estimators with different DGPs in which the total variance of all factors are kept constant. 
\end{minipage}
\end{figure}

The results are shown in Figure~\ref{fg:sim.estimators}, which depicts the MSPE of treated counterfactuals, i.e., $\frac{1}{\#\mathbf{1}\{(i,t)| D_{it}=1\}}\sum_{D_{it} = 1} [Y(0) - \hat{Y}_{it}(0)]^{2}$ , from 500 simulations using these two methods. The black solid line and gray dashed line represent the MSPE of IFEct with the correct number of factors ($r$) and with cross-validated $r$'s, respectively, while the red dot-dashed line marks the MSPEs of the MC estimator with a crossed validated tuning parameter $\lambda$. The result shows that MC gradually catches up with, and eventually beats, IFEct (with correctly specified $r$) as the number of factor grows and each factor produces weaker signals. It also suggests that, when factors become weaker, it is more difficult for the cross-validation scheme to pick them up, resulting in worse predictive performance, while the MC estimator is robust to a large number of factors because factors and loadings are not directly estimated. 
\bigskip

\subsection{$F$ Test versus the Equivalence test.} 

As explained in the paper, there is a trade-off between the $F$ test and the equivalence test (TOST) for testing no pre-trend: (1) when the sample size is small, the $F$ test has a power issue while TOST does not; and (2) when the sample size is relatively large, TOST are more liberal than the $F$ test in front of small biases. To illustrate these, we simulate data using the following DGP similar to that in the previous section but with only one factor: $Y_{it} = \delta_{it} D_{it} + 5 + k \cdot \lambda_{i}  f_{t} + \alpha_{i} + \xi_{t} + \varepsilon_{it},$ in which we vary $k$ to adjust the influence of a potential confounder $U_{it} = \lambda_{i} f_{t}$, which is correlated with $D_{it}$. For each $k$, we run 600 simulations. In each simulation, we first generate a sample of $N = 100$ units (50 treated and 50 controls) of 40 periods. We estimate a FEct model without taking into account the time-varying confounder. We then expand the sample size such that $N = 300$ and re-do the analysis.

\begin{figure}[!ht]
\caption{Monte Carlo Exercises: $F$ vs. equivalence tests}\label{fg:sim.tests}
\begin{center}
\hspace{-1em}
\subfigure[$N = 100$]{\includegraphics[width = 0.45\textwidth]{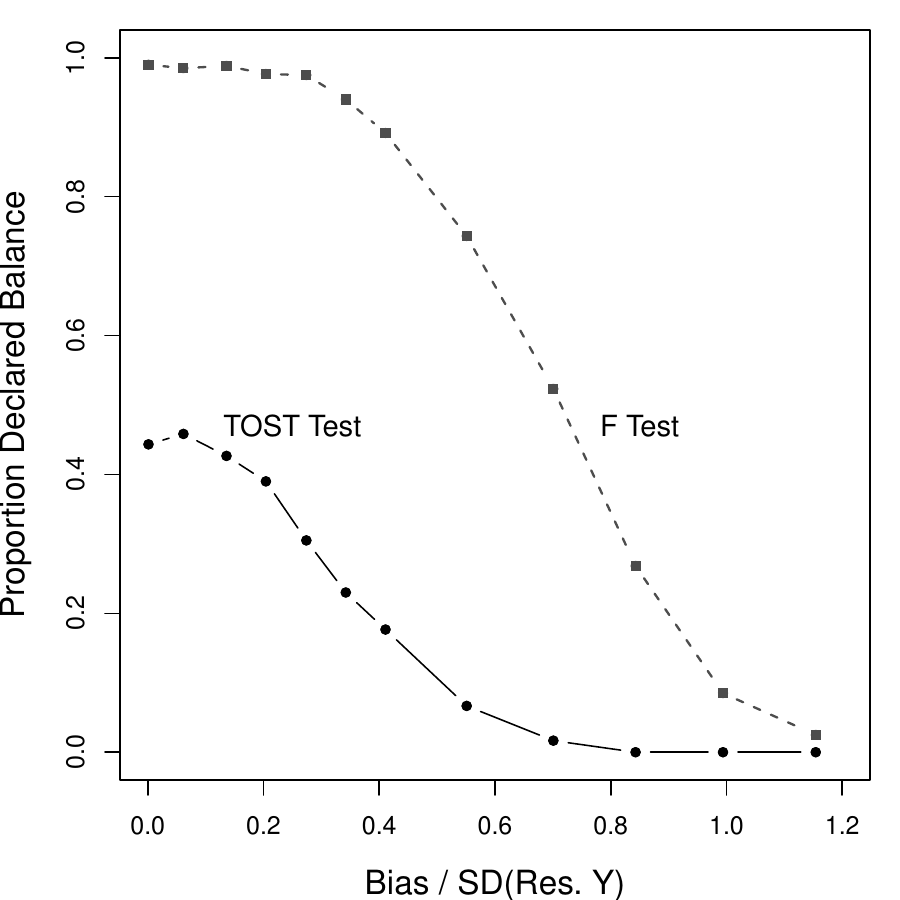}}\hspace{0.5em}
\subfigure[$N = 300$]{\includegraphics[width = 0.45\textwidth]{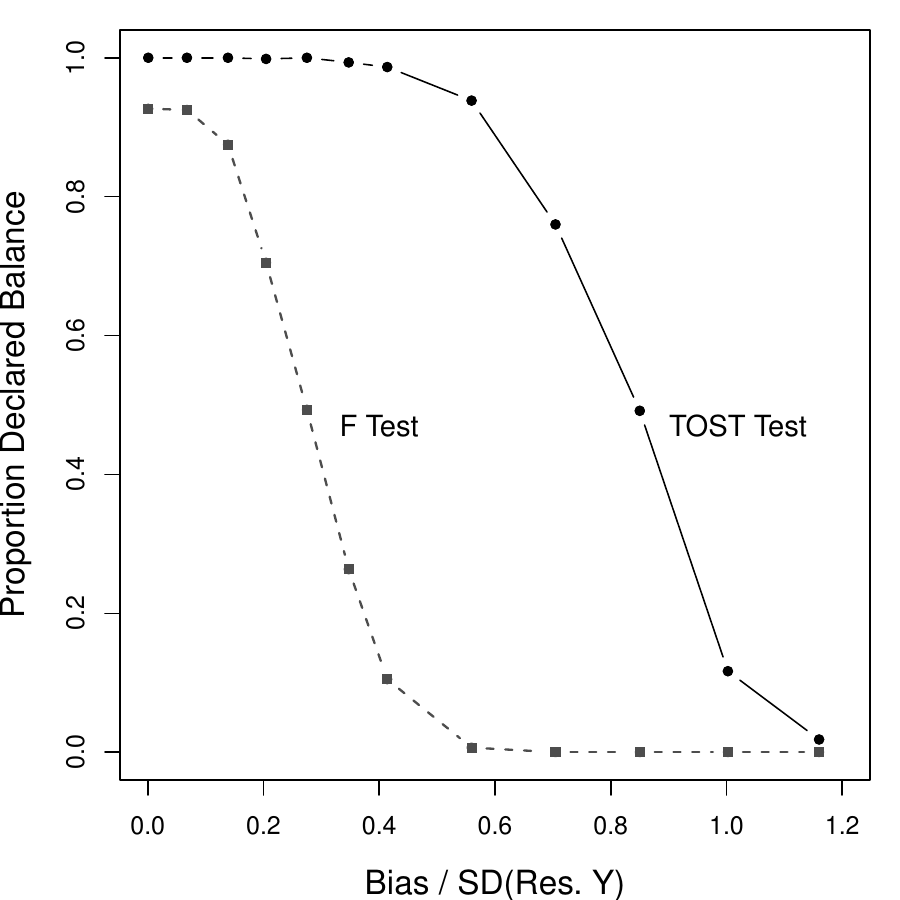}}\hspace{0.5em}
\end{center}
\footnotesize\textbf{Note:} The above figures show the results from Monte Carlo exercises that compare the $F$ test and the equivalence test when an unobserved confounder exists. In plot (a), $N = 100$; in plot (b), $N = 300$. Each dot is based on results from 600 simulations. 
\end{figure}

In Figure~\ref{fg:sim.tests}(a), we plot the proportion of times the equivalence test (solid line) or the $F$ test (dashed line) backs the strict exogeneity assumption against the normalized bias induced by the confounder when $N = 100$. It shows that it is highly likely that an $F $test cannot reject the null of zero residual average due to lack of power, even when the biases are large. In contrast, the probability that the equivalence test rejects inequivalence (hence, declaring equivalence) drops quickly as the bias increases. In other words, the equivalence test is more powerful in detecting imbalances than the conventional $F$ test. Figure~\ref{fg:sim.tests}(b) shows that, when the sample size is relatively large ($N = 300$), the non-rejection rate of the $F$ test declines quickly as the influence of the confounder grows. In comparison, the equivalence test rejects inequivalence (hence, declaring equivalence) when a relatively inconsequential confounder is at present; as the confounder becomes more influential, it starts to sound the alarm. These patterns are similar to what \citet{hartman2018equivalence} report in a cross-sectional setting. 
\clearpage

\clearpage



\section{Additional Information on the Empirical Examples}

\subsection{Replicating Hainmueller and Hangartner (2015)}

\begin{figure}[!th]
\caption{Treatment Status: Indirect Democracy and\\Naturalization Rate}\label{fg:XY2015}
\centering
\begin{minipage}{0.85\linewidth}{
\begin{center}
\includegraphics[width = 0.8\textwidth]{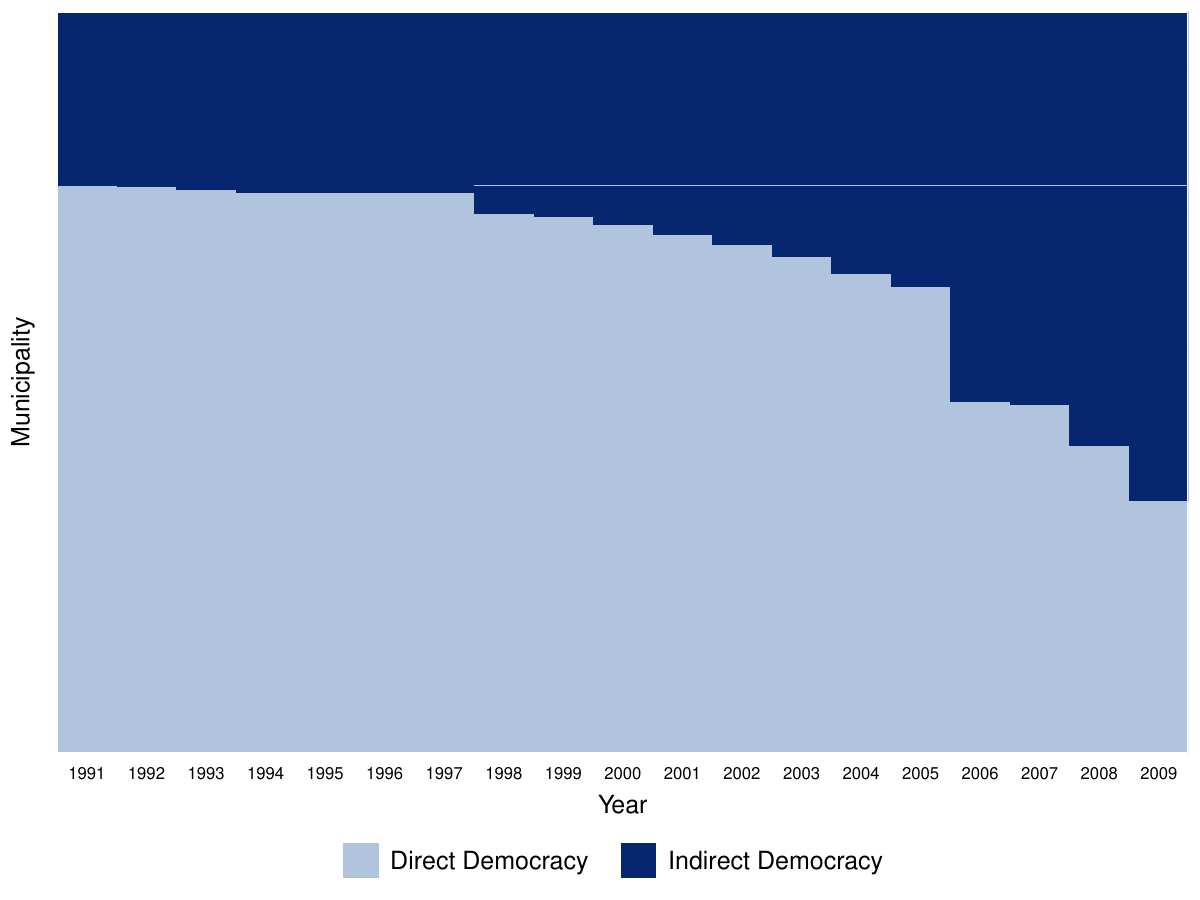}\\
\end{center}
}
\footnotesize\textbf{Note:} The above figure plots the treatment status for the first 50 units using data from \citet{hainmueller2015does}. The pattern of treatment assignment follows staggered adoption. Municipalities are ordered based on the timing when they started to adopt indirect democracy to make naturalization decisions. The plot is made by the \texttt{panelView} package.
\end{minipage}
\end{figure}

\clearpage

\subsection{Replicating Fouirnaies and Mutlu-Eren (2015)}

\begin{figure}[!th]
\caption{Original Treatment Effect Plot}\label{fg:FM2015.original}
\centering
\begin{minipage}{1\linewidth}{
\begin{center}
\includegraphics[width = 0.75\textwidth]{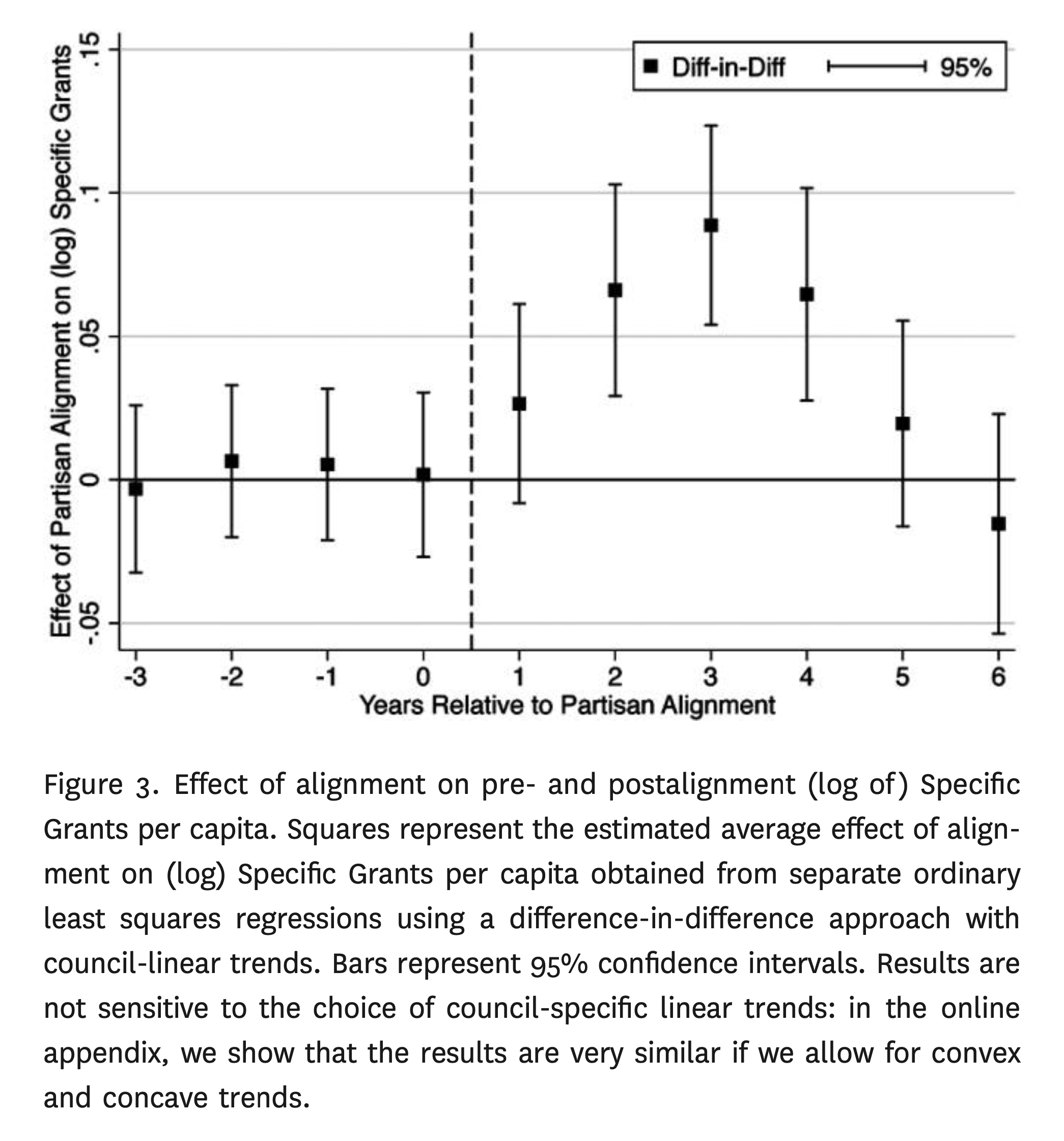}\\
\footnotesize\textbf{Note:} The above figure is adapted from \citet{FM2015-yy}.
\end{center}
} 
\end{minipage}
\end{figure}
\clearpage

\begin{figure}[!th]
\caption{Treatment Status: Partisan Alignment and Specific Grants}\label{fg:FM2015.treat}
\centering
\begin{minipage}{1\linewidth}{
\begin{center}
\includegraphics[width = 0.55\textwidth]{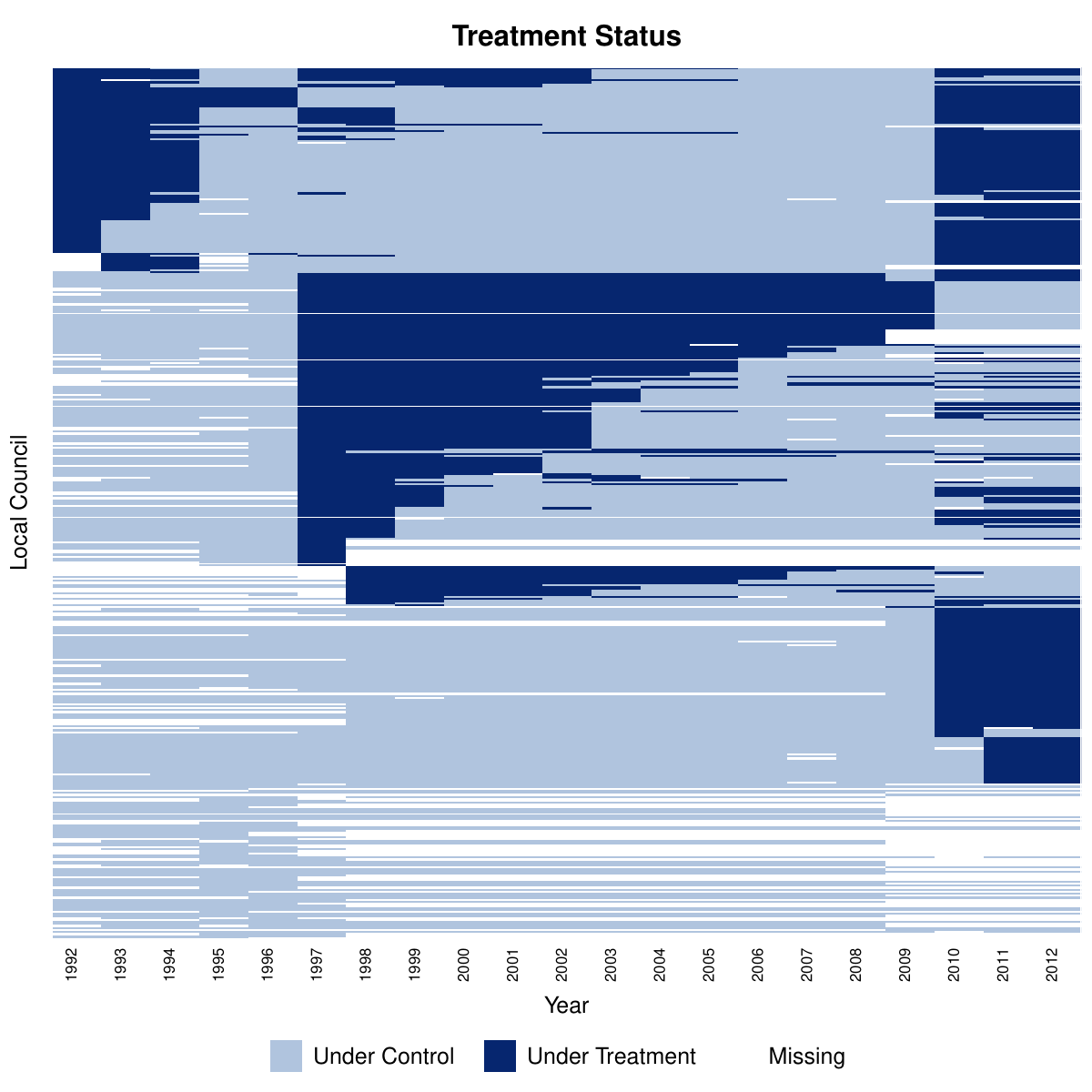}
\end{center}
}
\footnotesize\textbf{Note:} The above figure plots the treatment status  using data from \citet{FM2015-yy}. Local councils in England are ordered based on the timing when they are politically aligned with the government party. The plot is made by the \texttt{panelView} package.
\end{minipage}
\end{figure}

\begin{figure}[!ht]
\caption{The Effect of Partisan Alignment on Specific Grants\\Testing No Pre-Trend}\label{fg:FM2015.equiv}
\centering
\begin{minipage}{1\linewidth}{
\centering
\includegraphics[width = 1\textwidth]{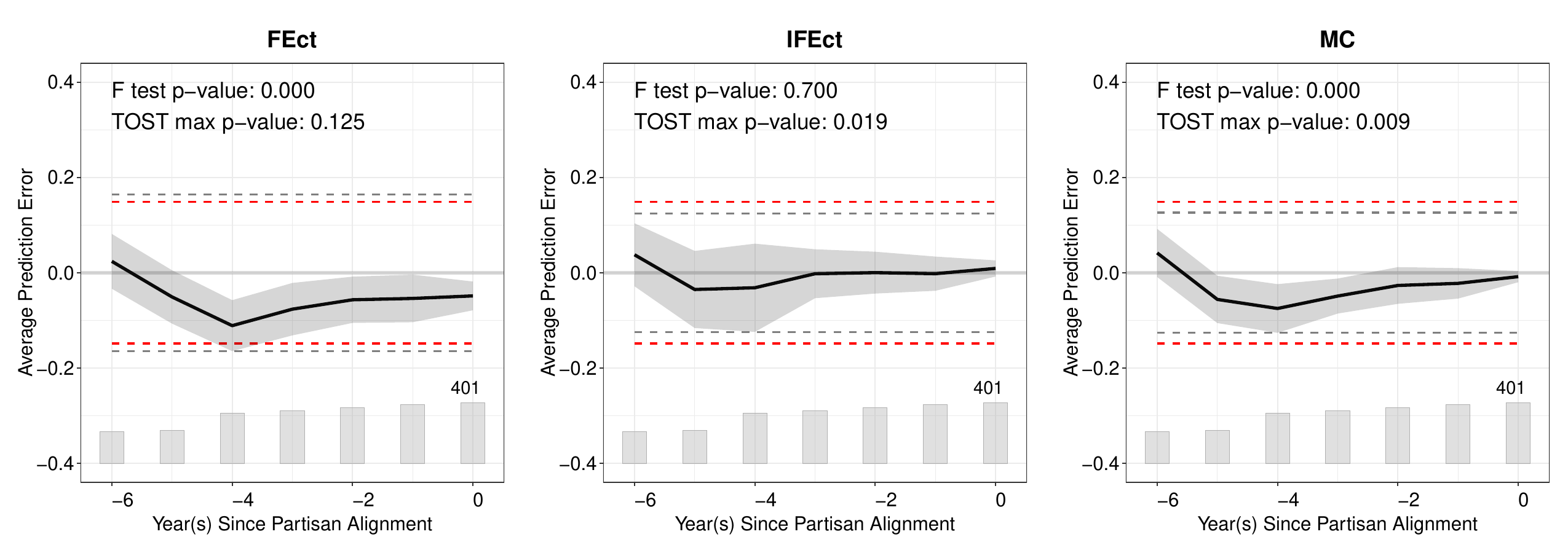}}
\footnotesize\textbf{Note:} The above figures show the results from tests for no pre-trend in the application of partisan alignment on specific grants in the UK. The bar plot at the bottom of each figure illustrates the number of treated units at a given time period relative to the onset of the treatment. Pretreatment residual averages and their 90\% confidence intervals are drawn; the red and gray dashed lines mark the equivalence range and the minimum range, respectively. With the equivalence threshold set at $0.36\hat\sigma^2$, all three models pass the equivalence test. 
\end{minipage}
\end{figure}
\clearpage

\begin{figure}[!ht]
\caption{The Effect of Partisan Alignment on Grant Allocation\\Three Years after the Treatment Ends as the ``Carryover Period''}\label{fg:FM2015b}
\centering
\begin{minipage}{0.9\linewidth}{
\centering
\subfigure[Dynamic Treatment Effects]{\includegraphics[width = 1\textwidth]{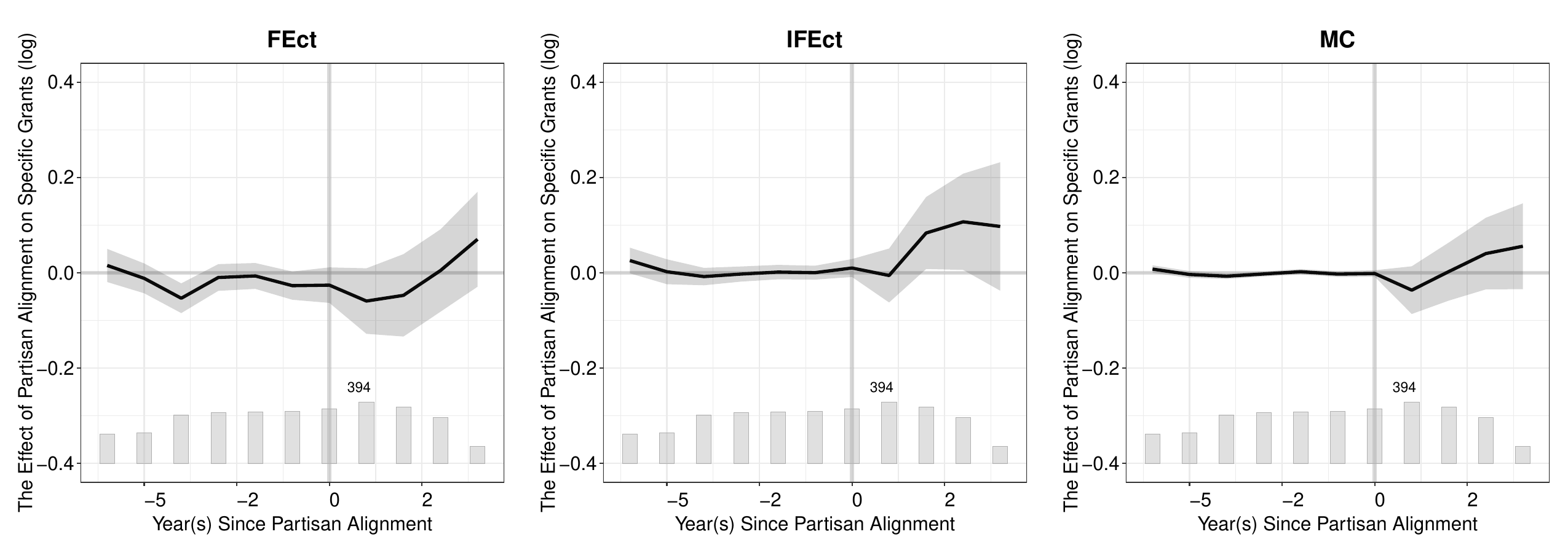}}\\
\subfigure[Placebo Test]{\includegraphics[width = 1\textwidth]{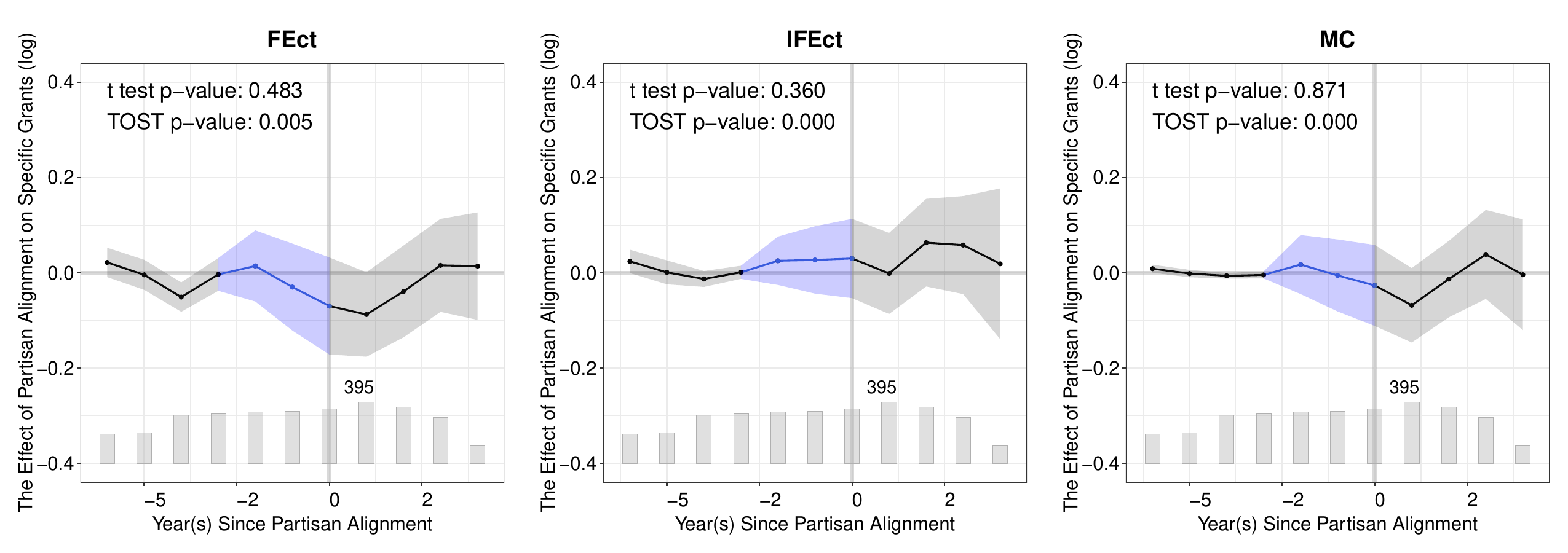}}
\subfigure[Test for Carryover Effects]{\includegraphics[width = 1\textwidth]{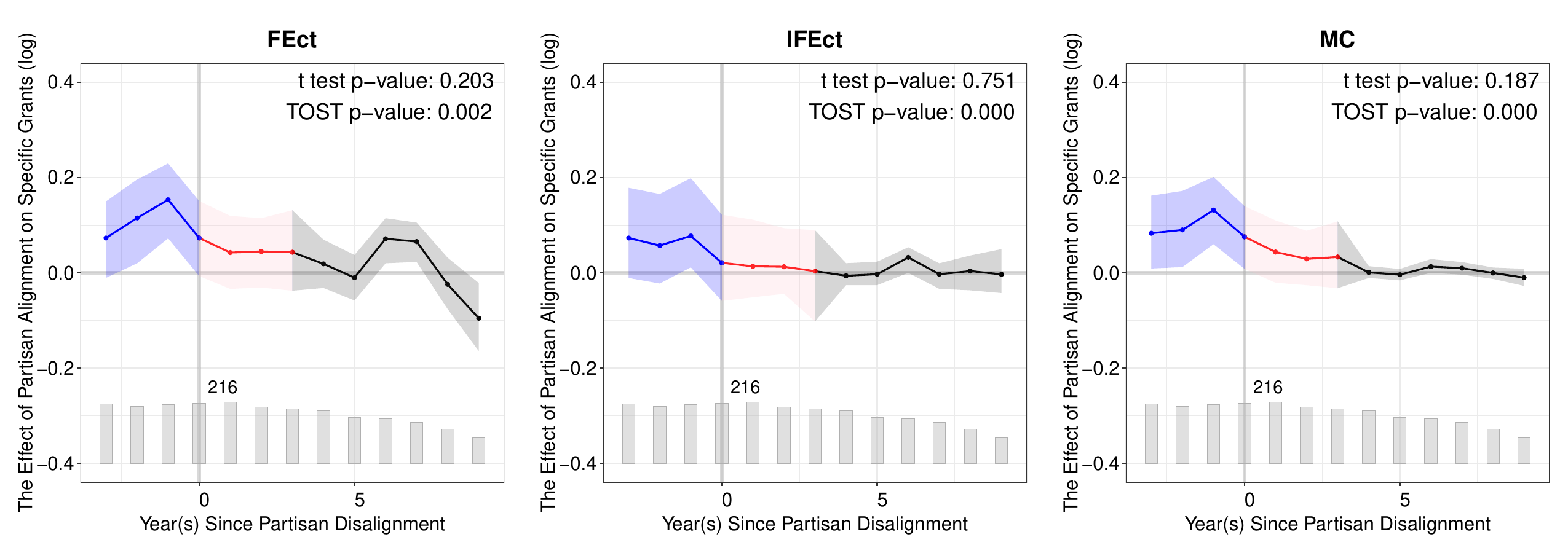}}
}
\footnotesize\textbf{Note:} The blue dots in (b) represent the periods used in the placebo tests. In (c), the blue and red dots represent the periods removed from the model-modeling stage (Step 1) and the periods used in the tests for no carryover effects, respectively. 
\end{minipage}
\end{figure}
\clearpage

\begin{figure}[!ht]
\caption{The Effect of Partisan Alignment on Specific Grants:\\by Cohort}\label{fg:FM2015.cohort}
\centering
\begin{minipage}{1\linewidth}{
\centering
\subfigure[FEct]{\includegraphics[width = 1\textwidth]{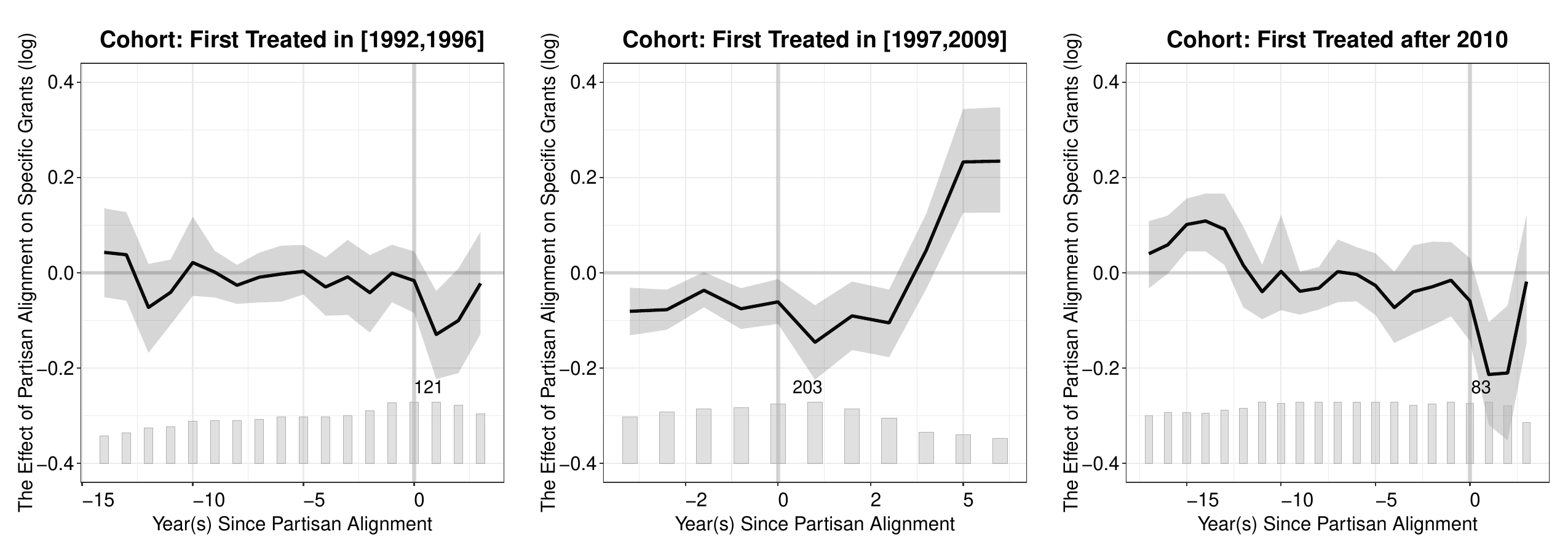}}
\subfigure[IFEct]{\includegraphics[width = 1\textwidth]{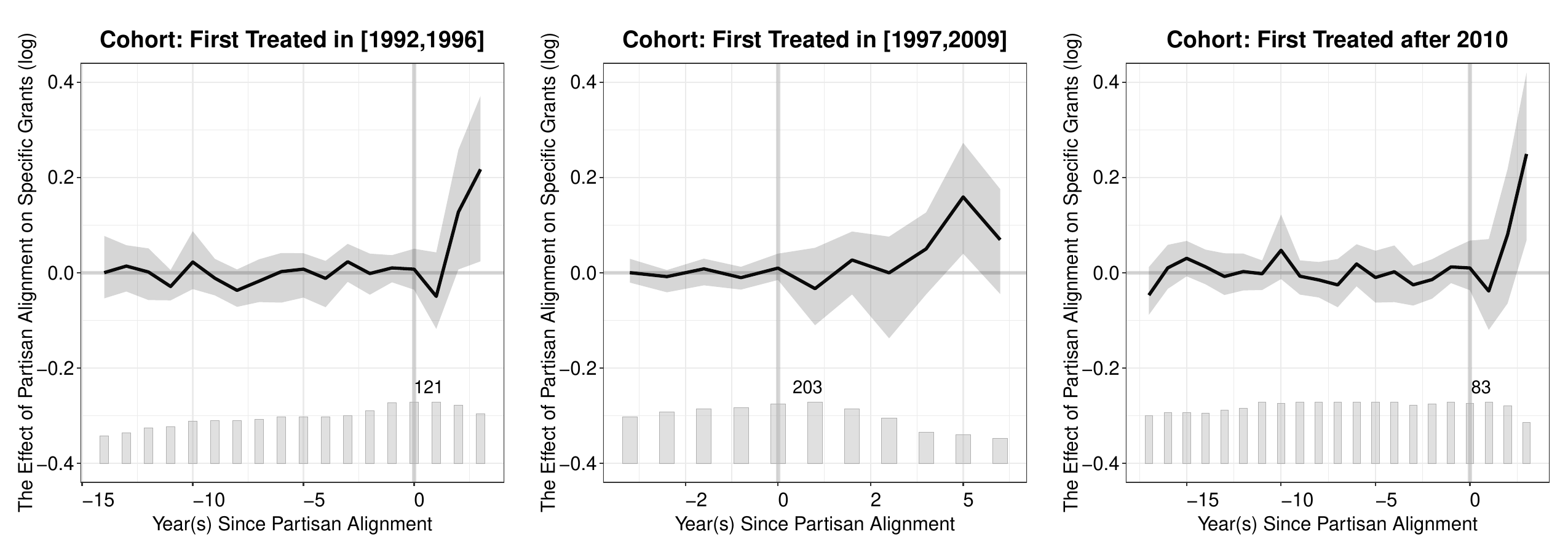}}}
\footnotesize\textbf{Note:} The above figures show the effect of partisan alignment on specific grants using FEct (a) and IFEct (b). Cohorts are defined based on the timing when a local council in England is first aligned with the government party. They broadly correspond to the three blocks of units in the treatment status plot (Figure~\ref{fg:FM2015.treat}).
\end{minipage}
\end{figure}
\clearpage

\newpage

\bibliographystyle{apsr}
\bibliography{tscs}